\documentclass[aps,pre,showkeys,
reprint,   
twocolumn, 
nofootinbib,
floatfix]{revtex4-2}
\usepackage[T1]{fontenc}
\usepackage{graphicx}
\usepackage{amsmath}
\usepackage{amsfonts}
\usepackage{subcaption}
\usepackage{xcolor}
\usepackage{float}
\makeatletter
\let\newfloat\newfloat@ltx
\makeatother
\usepackage{hyperref}
\hypersetup{
    colorlinks=true,
    urlcolor=blue,
}
\usepackage{cleveref}

\begin{document}
\title{Vanishing opinions in Latan\'e model of opinion formation}
\author{Maciej Dworak}
\author{Krzysztof Malarz}
\thanks{\includegraphics[width=10pt]{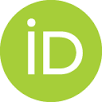}~\href{https://orcid.org/0000-0001-9980-0363}{0000-0001-9980-0363}}
\email{malarz@agh.edu.pl}
\affiliation{\mbox{AGH University of Science and Technology, Faculty of Physics and Applied Computer Science},\\
al. Mickiewicza 30, 30-059 Krak\'ow, Poland}
\begin{abstract}
In this paper, the results of computer simulations based on Nowak--Szamrej--Latan\'e model with multiple (from two to five) opinions available in the system are presented. 
We introduce the noise discrimination level (which says how small the clusters of agents could be considered as negligible) as a quite useful quantity that allows qualitative characterization of the system. 
We show that depending on the introduced noise discrimination level, the range of actors' interactions (controlled indirectly by an exponent in distance scaling function, the larger the exponent the more influential the nearest neighbors are) and the information noise level (modeled as social temperature, which increases results in increase of randomness in taking the opinion by the agents), the ultimate number of the opinions (measured as the number of clusters of actors sharing the same opinion in clusters greater than the noise discrimination level) may be smaller than the number of opinions available in the system. 
These are observed in small and large information noise limits but result in either unanimity, or polarization, or randomization of opinions.
\end{abstract}
\date{\today}

\maketitle

\section{\label{sec:intro}Introduction}

The formation and dynamics of opinions \cite{Galam_2022,Kozitsin_2022,Weron_Szwabinski_2022,PhysRevE.105.044112,Muslim_2022,Lian_2022,Su_2022,Zachary_2022,Nguyen_2020,Galam_2020a,Galam_2020b} and its spread and propagation \cite{Szvetelszky2006,Choi_2020} seem to be a vivid section of sociophysics \cite{Castellano-2009,Stauffer_2013,Galam_2017,Ishii_2018,Schweitzer_Sociophysics,Sobkowicz_2019,Jusup_2022}.
Existing models \cite{Fortunato-Stauffer-2005,Grabisch_2020} may be grouped into two families: with discrete or continuous opinions. 
The latter are represented by Hegselmann--Krause model \cite{Hegselmann-2002,Schawe_2020a,Schawe_2020b}, Deffuant {\em et al.} model \cite{Deffuant-2000,Deffuant-2006,Malarz2006b,Mathias-2016,000399951000005} (in a one-dimensional opinion space), the Zaller--Deffuant model \cite{Kulakowski2009469,Gronek2011,Kulakowski2012,Kulakowski2014} (in a two-dimensional opinion space), compromise model \cite{cond-mat_0111494,Ben-Naim_2003_PhysicaD,Ben-Naim_2003_PhysicaA} or others \cite{Weisbuch_2002,Laguna_2004}.
In the family of discrete models, a particular role is played by toy models dealing with binary opinions and simplified rules of opinion formation, 
with majority \cite{Galam_2002,Oliveira_2019}, 
voter \cite{Holley_1975,Lima2006,Fernandez-Gracia_2014}, 
Sznajd \cite{Sznajd-2000,Sznajd-2005,Sznajd-2005a,Kulakowski2008,Sznajd-Sznajd-Sznajd}, 
Galam \cite{cond-mat_0409484,GalamReview} models, among others.

For example, in the voter model \cite{Holley_1975}, 
the opinions of any given actor on some issue change at random times under the influence of the opinions of his/her neighbors. 
An actor's opinion at any given time can take one of two values. At random times, a random individual is selected, and that actor's opinion is changed according to a stochastic rule. Specifically, for one of the chosen actor's neighbors, one is chosen according to a given set of probabilities, and that individual's opinion is transferred to the chosen actor. 

In the majority model \cite{Galam_2002}, at each time step, a group of $r$ actors is selected, where $r$ can be constant or changed in each successive
step. All randomly selected actors adopt the opinion that dominates the group.
If the size $r$ of a group of neighbors is even, in case of a tie, either the group adopts an arbitrarily determined biased opinion or maintains the {\em status quo}.

In the original one-dimensional version of the Sznajd model \cite{Sznajd-2000} agent in position $i$ adopts the opinion of the actor sitting in position $i+2$ and the actor in position $i+1$ adopts the opinion of the actor sitting in position $i-1$.
These rules ultimately lead system to one of three (stable and fixed) attracting points: either two states of {\em unanimity} or one state of alternately opposite opinions (`antiferromagnetic' state). 

These models may be particularly useful for modeling the thinking dichotomy, that is, binary thinking that involves only two extreme attitudes\footnote{Typical answers (measuring opinions) for dichotomy-like questionnaires are: ‘No’ and ‘Yes’.}.
Such a situation occurs for voters in countries with two-parties systems (like in USA), or for actors answering fundamental or simple questions.
For example, people usually well know if they like chicken livers with onion (or not), people usually well know if they believe that our Earth is flat (or not), people usually well know if they are pro or contra abortion, etc.

Somewhere on the border between two (discrete/continuous) families of models, discrete opinion models allow multiple opinions to appear \cite{Gekle-2005,Kulakowski2010,Ozturk_2013,1902.03454,2002.05451,Martins_2020,Zubillaga_2022,Li_2022,Doniec_2022,000409112600017,000316891200004,000432967700004}.
These models still allow us to observe geometrical {\em clusterization} of opinions, but also their {\em polarization}, which is naturally forced (assumed) in the case of models with binary opinions.
Such models are particularly attractive for modeling indifferents as an interface between pro and contra, modeling responses to Likert-scale questionnaires\footnote{Typical answers (measuring opinions) for Likert-like questionnaires are: ‘Strongly disagree’, ‘Disagree’, ‘Neither agree nor disagree’, ‘Agree’ and ‘Strongly agree’.}, or modeling voter decisions in multiparty systems.

Here, we use a discrete multi-choice opinion model based on computerized version \cite{Nowak-1990} of opinion formation based on Latan\'e theory of social impact \cite{Darley1968,Latane-1976,Latane-1981} (see References~\onlinecite{Kacperski-2000,Holyst-2000,1902.03454,2002.05451,2010.15736} for examples of model applications and Reference~\onlinecite{ARCPIX253} for a comprehensive review).

In Reference~\onlinecite{1902.03454} Nowak--Szamrej--Latan\'e model \cite{Nowak-1990} was modified to allow multiple (more than two) opinions. 
It was shown that in the presence of information noise (modeled as social temperature $T$) the signatures of order/disorder phase transition were observed: in the average fraction of actors sharing the $i$-th opinion; its variation; average number of clusters of actors with the same opinion and the average size of the largest cluster of actors who share the same opinion.
The social temperature $T$ played a role as a standard Boltzmann distribution parameter that contains the social impact as the equivalent of energy.
The order and disordered phases were observed for low ($T<T_C$) and high ($T>T_C$), respectively.
For a homogeneous society (with identical actors’ supportiveness and persuasiveness) the critical social temperature $T_C$ decreased with increasing number of available opinions $K$. 

The authors of Reference~\onlinecite{2002.05451} showed that opinion formation and spread were influenced by both: $i$) flow of information between actors (effective range of interactions between actors) and $ii$) randomness in adopting opinions (noise level).
Noise not only leads to opinions disorder, but also promotes consensus under certain conditions.
In the disordered phase and when the exchange of information is spatially effectively limited, various faces of disorder were observed, including system states, where the signatures of self-organized criticality manifested themselves as a scale-free probability distribution function for sizes of cluster of actors sharing the same opinion.
Then increasing the noise level leads the system to a disordered random state. 
The critical noise level $T_C$ above which the histograms of the sizes of the opinion groups lost their scale-free character increases with an increase in the ease of information flow.

In this paper, we continue the studies presented in References~\onlinecite{1902.03454,2002.05451}.
Namely, with computer simulation based on Nowak--Szamrej--Latan\'e model \cite{Nowak-1990} we check: 
  $i$) how influential are the nearest neighbors with respect to the entire population;
 $ii$) the opinion clusterization (including the distribution of these cluster numbers and their sizes);
$iii$) and distribution of surviving opinions.

The rest of the paper is organized as follows. 
In \Cref{sec:model} a detailed description of the model is presented.
\Cref{sec:results} contains the results of simulations.
The results obtained are discussed in \Cref{{sec:discussion}} and summarized in \Cref{sec:conclusion}.
The list of references and three appendixes---presenting detailed results on: 
examples of final spatial opinion distribution (\Cref{apx:example_opinions});
average number of clusters (\Cref{apx:num_clu});
the number of surviving opinions (\Cref{apx:num_sur})---close the manuscript.

\section{\label{sec:model}Model}

The model is based on previous attempts \cite{ThesisBancerowski,1902.03454,2002.05451,2010.15736,ThesisDworak} to describe the dynamics of opinion in the context of the theory of social impact \cite{Darley1968,Latane-1976,Latane-1981} in its computerized version \cite{Nowak-1990}.
The system contains $N$ actors labeled with $i=0,\cdots,N-1$.
Every actor $i$ at time $t$ has an opinion $\xi_i(t)\in\mathbf{\Xi}$.
The set $\mathbf{\Xi}$ of available opinions consists of $K$ different opinions $\{\Xi_1,\cdots,\Xi_K\}$.
The social impact $\mathcal{I}_{i,k}(t)$ exerted in time $t$ on an actor $i$ by all actors who share opinions $\Xi_k$ is calculated as
\begin{subequations}
\label{eq:szamrej}
\begin{equation}
\label{eq:szamrej_sum_same}
\mathcal{I}_{i,k}(t) = \sum^{N-1}_{j=0}{\frac{4s_j}{g(d_{i,j})} \cdot \delta(\Xi_k, \xi_j(t)) \cdot \delta(\xi_j(t),\xi_i(t))}
\end{equation}
or
\begin{equation}
\label{eq:szamrej_sum_diff}
\mathcal{I}_{i,k}(t) = \sum^{N-1}_{j=0}{\frac{4p_j}{g(d_{i,j})} \cdot \delta(\Xi_k, \xi_j(t)) \cdot [1-\delta(\xi_j(t),\xi_i(t))}], 
\end{equation}
where Kronecker delta $\delta(x,y)=0$ when $x\ne y$ and $\delta(x,y)=1$ when $x=y$. 
\end{subequations}
The term $\delta(\Xi_k,\xi_j(t))$ in \Cref{eq:szamrej} indicates that the impact $\mathcal{I}_{i,k}(t)$ on the $i$-th agent in time $t$ is exerted only by agents $j$ who at time $t$ believe in the opinion $\Xi_k$ ($\xi_j(t)=\Xi_k$).
The term $\delta(\xi_j(t),\xi_i(t))$ in \Cref{eq:szamrej_sum_same} vanishes when $\xi_i(t)\ne\xi_j(t)$, i.e., it produces a non-zero contribution of the impact $\mathcal{I}_{i,k}(t)$ on agent $i$ only when agent $j$ shares the opinion of agent $i$. Thus, therm $s_j$ is considered to be the {\em supportiveness} of the $j$-th actor.
On the contrary, the term $[1-\delta(\xi_j(t),\xi_i(t))]$ resets the impact when agents $i$ and $j$ share the same opinion.
It means that the components of the sum \eqref{eq:szamrej_sum_diff} can be non-zero only when interacting in time $t$ agents have different opinions $\xi_i(t)\ne \xi_j(t)$ and thus $p_j$ play a role of {\em persuasiveness} of the $j$-th agent.
The supportiveness $s_i$ and persuasiveness $p_i$ are taken randomly from the interval $[0,1]$.  
$d_{i,j}$ stands for the Euclidean distance between agents $i$ and $j$.
The distance scaling function $g(\cdot)$ should be a non-decreasing function that ensures a decreasing influence from more and more distant actors.
Here, we assume that
\begin{equation}
\label{eq:fg}
g(x) = 1+x^\alpha,
\end{equation}
where the exponent $\alpha$ is a model control parameter. 

After calculating impacts \eqref{eq:szamrej} for each actor $i$ and every opinion $\Xi_k$ available in the system, the temporal evolution of $i$-th actor opinion $\xi_i$ can be predicted based on either deterministic (in absence of information noise) or non-deterministic (in presence of information noise) way.

In the deterministic version (without information noise), the actor $i$ in the next time step $(t+1)$ takes the opinion $\Xi_k$ that the believers exerted the largest impact on him/her:
\begin{equation}
\label{eq:deterministic}
\begin{split}
    \xi_i(t + 1) = \Xi_k \Longleftrightarrow\\ \mathcal{I}_{i,k}(t) = \max(\mathcal{I}_{i,1}(t), \mathcal{I}_{i,2}(t),\dots,\mathcal{I}_{i,K}(t)).
\end{split}
\end{equation}

\begin{figure}[htbp]
\subcaptionbox{\label{fig:example_map_K2}$K=2$}{\includegraphics[width=0.26\textwidth]{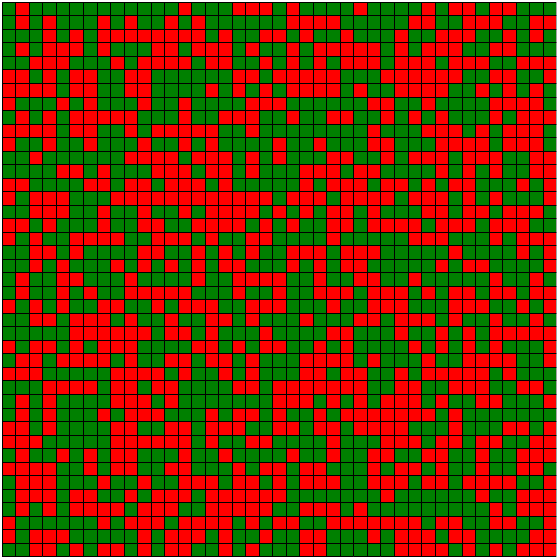}}
\hfill
\subcaptionbox{\label{fig:example_map_K4}$K=4$}{\includegraphics[width=0.26\textwidth]{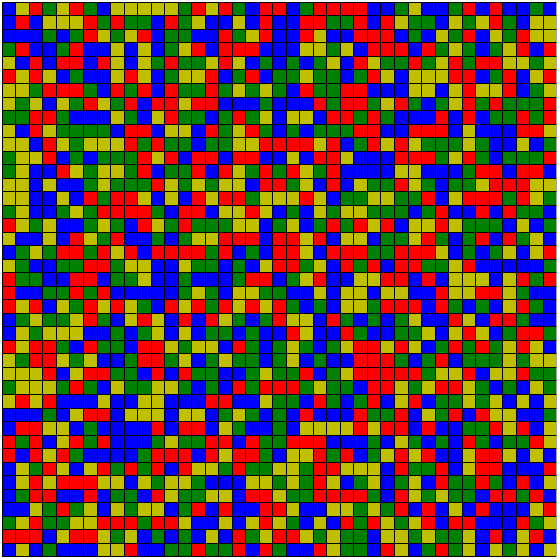}}
\caption{\label{fig:example_map}Example of random initial state of the system for (\subref{fig:example_map_K2}) $K=2$ and (\subref{fig:example_map_K4}) $K=4$.
Various colors correspond to various opinions.}
\end{figure}

When information noise is present in the system, the social impact $\mathcal{I}_{i,k}(t)$ \eqref{eq:szamrej} determines the probability $P_{i,k}(t)$
of accepting opinion $\Xi_k$ in the next time step $(t+1)$ by $i$-th actor.
To that end, we introduce a (temperature-like) information noise parameter $T$ \cite{social_temperature} and a Boltzmann-like factor
\begin{subequations}
    \label{eq:probability_eq}
    \begin{equation}
    p_{i,k}(t) = \exp\left(\frac{\mathcal{I}_{i,k}(t)}{T} \right), \label{eq:probability_p}
    \end{equation}
which allow us to define the above-mentioned probability
    \begin{equation}
    P_{i,k}(t) = \frac{p_{i,k}(t)}{\sum^K_{j=1} p_{i,j}(t)}. \label{eq:probability_P}
    \end{equation}    
\end{subequations}
Then, $i$-th actor accepts in the next time step $(t+1)$ opinion $\Xi_k$
\begin{equation}
\label{eq:probabilistic}
    \xi_i(t+1) = \Xi_k, \text{ with probability } P_{i,k}(t).
\end{equation}

We assume that the actors occupy nodes of the square grid 
\[
\mathcal{G}=\{(x,y): 0\le x,y< L,\quad x,y\in\mathbb{Z} \}
\]
and agent's label $i=Lx+y$. 
The open boundary conditions are assumed.
Initially (at $t=0$), the agents take a random opinions.
The examples of the initial system states are presented in \Cref{fig:example_map} for $K=2$ (\Cref{fig:example_map_K2}) and for $K=4$ (\Cref{fig:example_map_K4}). 
Various opinions are marked by various colors.
The algorithm of performed simulations is presented in Algorithm 1 \cite{ThesisDworak}.
The source code of program (written in C) is available in Reference~\onlinecite{app_Dworak}. 

\section{\label{sec:results}Results}

In this Section we describe the results of computer simulations carried for square lattice with $L^2=41^2$ actors. If not stated otherwise the results are gathered after $t=1000$ time steps and averaged over $R=100$ independent system realizations (for various random initial spatial distribution of opinions $\xi_i(t=0)$, supportiveness $s_i$ and persuasiveness $p_i$ values). 

\subsection{\label{sec:beta}How influential are the nearest-neighbours in respect to the entire population?}

\begin{figure*}[htbp]
\centering
\subcaptionbox{ $n=1$, $r=0$\label{fig:neighbour_n01}}{\includegraphics[width=.24\textwidth]{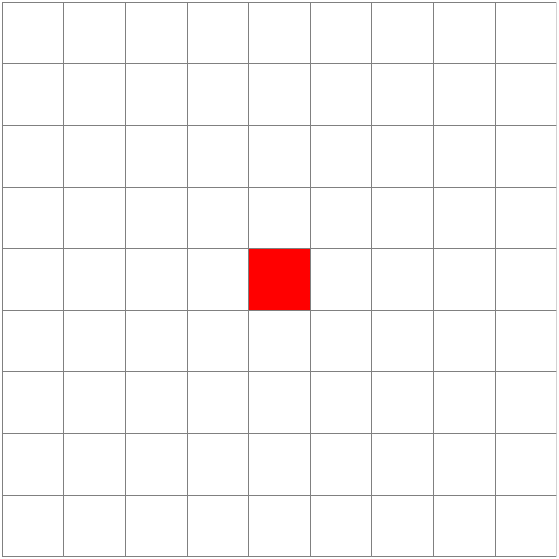}}\hfill
\subcaptionbox{ $n=9$, $r=1$\label{fig:neighbour_n09}}{\includegraphics[width=.24\textwidth]{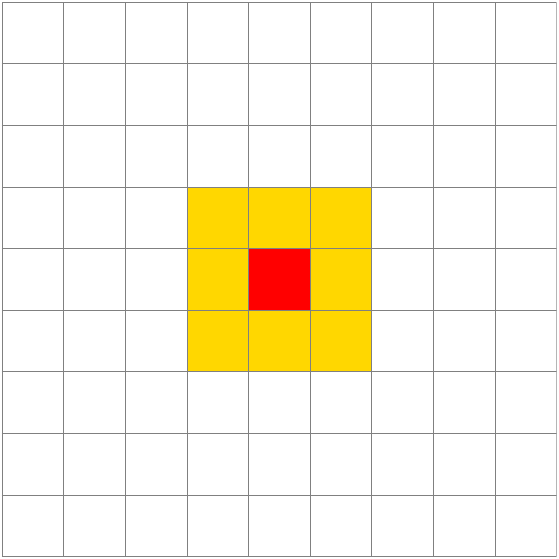}}\hfill
\subcaptionbox{$n=25$, $r=2$\label{fig:neighbour_n25}}{\includegraphics[width=.24\textwidth]{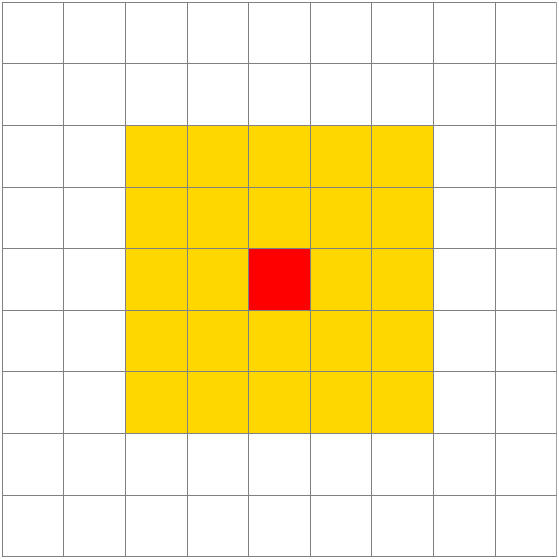}}\hfill
\subcaptionbox{$n=49$, $r=3$\label{fig:neighbour_n49}}{\includegraphics[width=.24\textwidth]{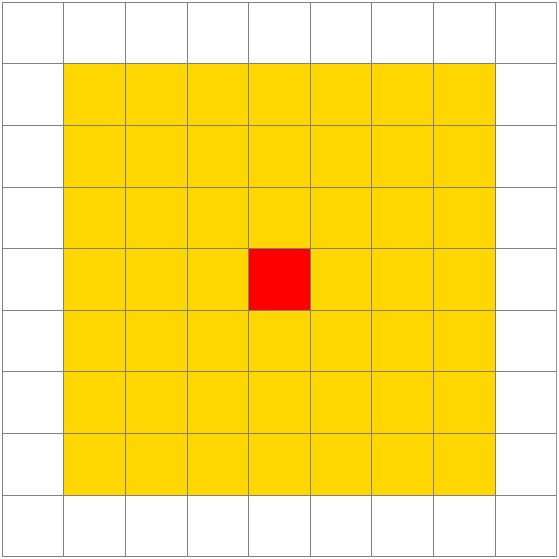}}
\caption{\label{fig:neighbourhoods}The sketches of shapes of the neighborhoods closest to the sites (\subref{fig:neighbour_n01}) $n=1$, (\subref{fig:neighbour_n09}) $n=9$, (\subref{fig:neighbour_n25}) $n=25$, (\subref{fig:neighbour_n49}) $n=49$ sites. The values of the $r$ parameters indicated in the figures in the headline influence summation limits in the nominator of \Cref{eq:beta}.}
\end{figure*}

To better understand the role played by the $\alpha$ parameter, we check the ratio
\begin{equation}
\label{eq:beta}
\beta(n) = \dfrac{{(L-2r)^{-2}}\cdot {\sum_{x=r}^{(L-r)} \sum_{y=r}^{(L-r)} \sum_{k=1}^K \mathcal{I}_{i,k}^n(t\to\infty)}}
{L^{-2} \cdot\sum_{i=1}^{L^2} \sum_{k=1}^K \mathcal{I}_{i,k}(t\to\infty)},
\end{equation}
which describes the opinion-independent relative influence of $n$ geometrically nearest neighbors with respect to the total impact coming from all actors.
Examples of shapes of these nearest neighborhoods containing $n=1$, 9, 25, 49 actors are sketched in \Cref{fig:neighbourhoods}.
The measured influence ratio $\beta(n)$ is averaged over $(L-2r)^2$ actors with $r=0$ for $n=1$, $r=1$ for $n=9$, $r=2$ for $n=25$, $r=3$ for $n=49$, etc., reflecting the possibility of placing the yellow square from \Cref{fig:neighbourhoods} in the square grid $\mathcal{G}$ without protruding beyond the boundaries of the system.
The term $\mathcal{I}_{i,k}^n$ stands for social impact calculated according to \Cref{eq:szamrej} but with an upper summation index replaced by $(n-1)$ instead of $(N-1)$. 
The impacts $\mathcal{I}_{i,k}^n$ and $\mathcal{I}_{i,k}$ are measured at the long-term simulation limit $(t\to\infty)$.
The results of the simulations of $\beta(n)$ are presented in \Cref{tab:In-to-I}.

\begin{table}[htbp]
\caption{\label{tab:In-to-I}Average ratio $\beta(n)$ [defined in \Cref{eq:beta}] of the influence of the neighborhood with $n$ sites (presented in \Cref{fig:neighbourhoods}) to the total influence of the entire network with $L^2$ sites for various values of $K$ and $\alpha$.}
\centering
\begin{ruledtabular}
\begin{tabular}{ r  llll } 
$\alpha$ & 2 & 3 & 4 & 6 \\
\hline
$n$ & \multicolumn{4}{c}{$K=2$} \\
\hline
  1 & 0.05987(13) & 0.14973(63) & 0.2209(13)   & 0.2902(16)\\
  9 & 0.25269(45) & 0.58795(79) & 0.80513(74)  & 0.95820(21)\\
 25 & 0.39642(56) & 0.76437(75) & 0.92761(38)  & 0.993687(41)\\ 
 49 & 0.49898(57) & 0.84573(64) & 0.96450(21)  & 0.998328(12)\\
 81 & 0.57600(53) & 0.89073(54) & 0.97971(13)  & 0.9994007(45)\\
121 & 0.63641(46) & 0.91866(44) & 0.987262(83) & 0.9997408(20)\\
169 & 0.68530(41) & 0.93739(36) & 0.991477(58) & 0.9998727(10)\\
225 & 0.72578(35) & 0.95064(29) & 0.994033(42) & 0.99993158(54)\\
289 & 0.75989(31) & 0.96039(24) & 0.995680(31) & 0.99996061(31)\\
361 & 0.78903(28) & 0.96778(20) & 0.996790(23) & 0.99997609(18)\\
\hline
$n$ & \multicolumn{4}{c}{$K=3$} \\
\hline
  1 & 0.05990(17) & 0.15080(92) & 0.2232(16)   & 0.2937(22)\\
  9 & 0.25275(62) & 0.5873(15)  & 0.8041(10)   & 0.95793(26)\\
 25 & 0.39649(85) & 0.7635(13)  & 0.92698(53)  & 0.993625(52)\\
 49 & 0.49906(96) & 0.8449(11)  & 0.96414(29)  & 0.998311(15)\\
 81 & 0.5761(10)  & 0.89006(90) & 0.97950(18)  & 0.9993947(53)\\
121 & 0.6365(10)  & 0.91812(72) & 0.98712(12)  & 0.9997382(22)\\
169 & 0.6854(10)  & 0.93694(59) & 0.991385(81) & 0.9998715(10)\\
225 & 0.72586(96) & 0.95027(48) & 0.993969(57) & 0.99993091(62)\\
289 & 0.75996(93) & 0.96008(39) & 0.995633(41) & 0.99996022(35)\\
361 & 0.78909(88) & 0.96753(32) & 0.996755(31) & 0.99997585(21)\\
\hline
$n$ & \multicolumn{4}{c}{$K=4$} \\
\hline
  1 & 0.05990(16) & 0.15095(98) & 0.2247(20)   & 0.2962(27)\\
  9 & 0.25275(50) & 0.5871(15)  & 0.80338(98)  & 0.95757(28)\\
 25 & 0.39649(65) & 0.7633(14)  & 0.92657(51)  & 0.993549(51)\\
 49 & 0.49906(70) & 0.8448(11)  & 0.96393(29)  & 0.998291(15)\\
 81 & 0.57609(70) & 0.88999(90) & 0.97938(18)  & 0.9993876(54)\\
121 & 0.63649(69) & 0.91807(74) & 0.98705(12)  & 0.9997353(22)\\
169 & 0.68538(66) & 0.93690(60) & 0.991331(79) & 0.9998701(12)\\
225 & 0.72586(63) & 0.95024(49) & 0.993931(56) & 0.99993012(63)\\
289 & 0.75997(59) & 0.96006(41) & 0.995605(40) & 0.99995976(36)\\
361 & 0.78911(56) & 0.96751(34) & 0.996735(30) & 0.99997557(21)\\
\hline
$n$ & \multicolumn{4}{c}{$K=5$} \\
\hline
  1 & 0.05988(15) & 0.1511(10)  & 0.2248(21)   & 0.2976(26)\\
  9 & 0.25267(48) & 0.5867(15)  & 0.8029(11)   & 0.95741(29)\\
 25 & 0.39638(58) & 0.7629(13)  & 0.92634(61)  & 0.993525(55)\\
 49 & 0.49893(59) & 0.8445(11)  & 0.96380(34)  & 0.998284(15)\\
 81 & 0.57595(57) & 0.88970(90) & 0.97930(21)  & 0.9993847(57)\\
121 & 0.63634(55) & 0.91784(73) & 0.98700(14)  & 0.9997340(25)\\
169 & 0.68523(53) & 0.93673(60) & 0.991302(92) & 0.9998694(12)\\
225 & 0.72571(51) & 0.95010(49) & 0.993909(65) & 0.99992979(66)\\
289 & 0.75982(49) & 0.95995(40) & 0.995590(46) & 0.99995958(37)\\
361 & 0.78896(49) & 0.96742(33) & 0.996723(35) & 0.99997546(24)\\
\end{tabular}
\end{ruledtabular}
\end{table}

Within the estimated uncertainties, the ratio $\beta(n)$ does not depend on the number $K$ of opinions available in the system and appears to be a purely geometric characteristic of the model.
Of course, we expected an observed increase of $\beta(n)$ with an increase of $n$ independently on $K$ and $\alpha$.
Much more interesting is the observed monotonic increase of $\beta(n)$ with the increase of the distance scaling function exponent $\alpha$.
For $\alpha=2$ roughly 25\% of the impact comes from $n=9$ nearest-neighbors.
This ratio increases to $\beta(9)\approx 59\%$ for $\alpha=3$, $\beta(9)\approx 80\%$ for $\alpha=4$ and $\beta(9)\approx 96\%$ for $\alpha=6$.
For $n=25$ roughly $\beta(25)\approx 39\%$, 76\%, 92\% and 99\% of the social impact exerted comes from only those twenty five neighbors for $\alpha=2$, 3, 4 and 6, respectively.
In other words, the $\alpha$ parameter says how influential the nearest neighbors are with respect to the entire population: the larger $\alpha$ the more influential the nearest neighbors are.

\subsection{\label{sec:opinionmaps}The final opinions distributions}

\begin{figure}[htbp]
\subcaptionbox{\label{sfig:a3K4T0-1}$T=0$}{\includegraphics[width=.21\textwidth]{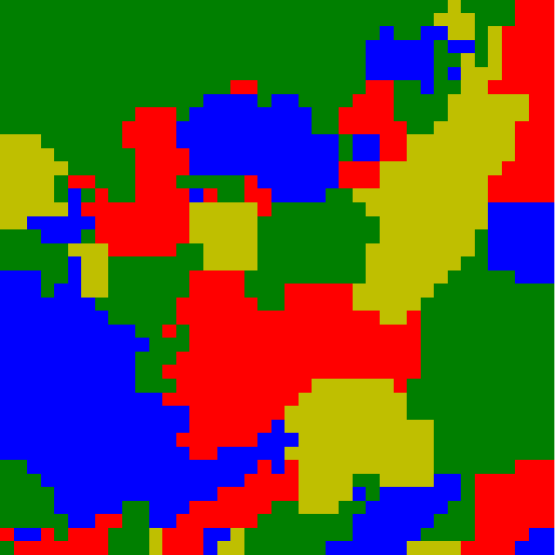}}\hfill
\subcaptionbox{\label{sfig:a3K4T0-2}$T=0$}{\includegraphics[width=.21\textwidth]{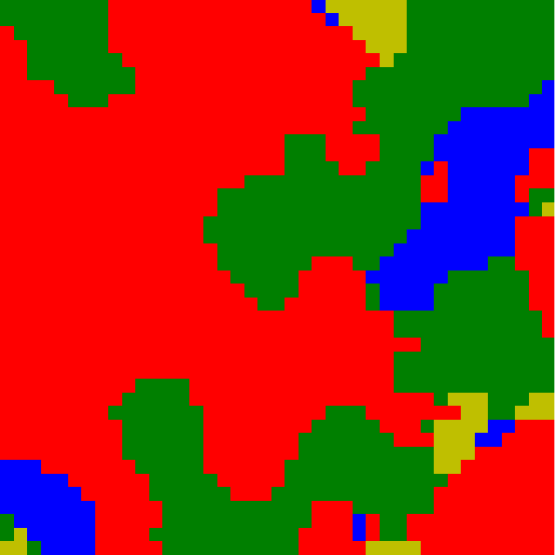}}\\
\subcaptionbox{\label{sfig:a3K4T1-1}$T=1$}{\includegraphics[width=.21\textwidth]{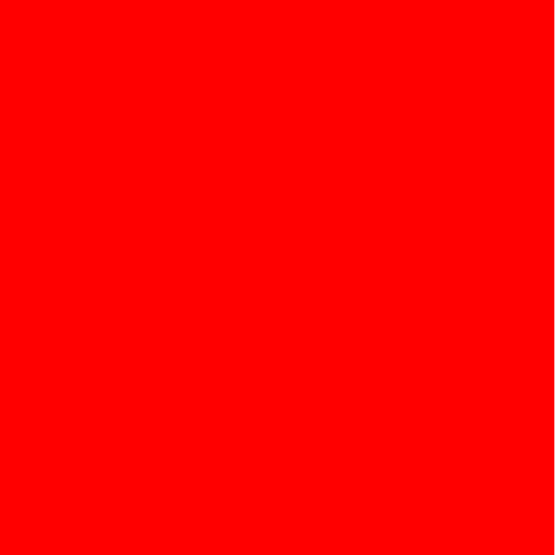}}\hfill
\subcaptionbox{\label{sfig:a3K4T1-2}$T=1$}{\includegraphics[width=.21\textwidth]{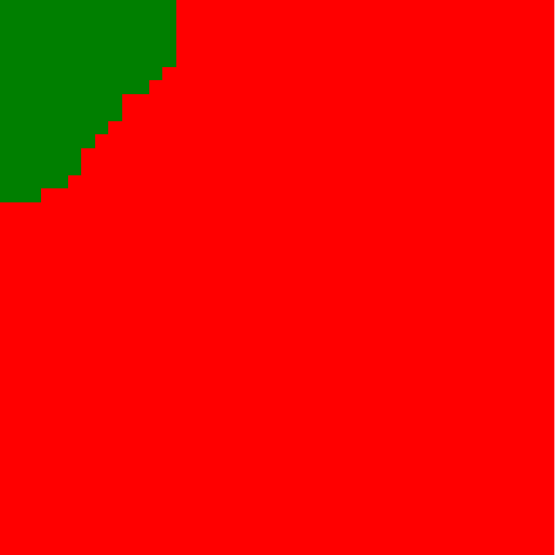}}\\
\subcaptionbox{\label{sfig:a3K4T2-1}$T=2$}{\includegraphics[width=.21\textwidth]{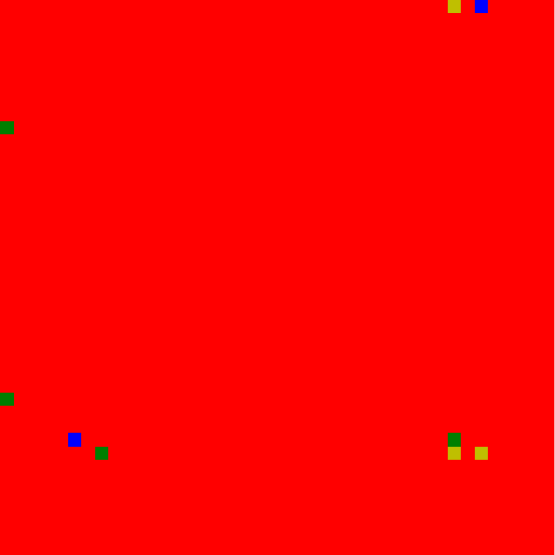}}\hfill
\subcaptionbox{\label{sfig:a3K4T2-2}$T=2$}{\includegraphics[width=.21\textwidth]{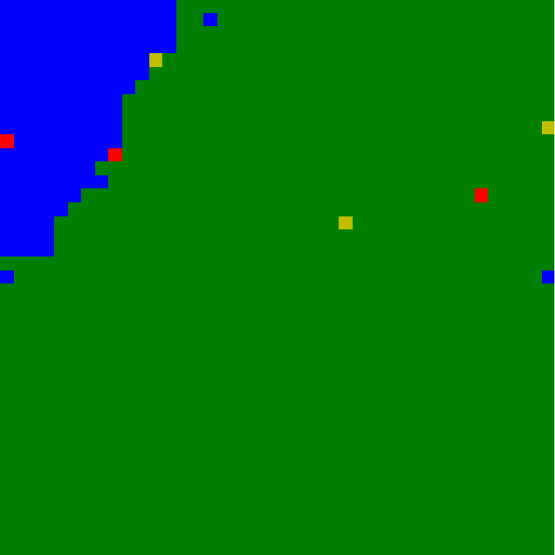}}\\
\subcaptionbox{\label{sfig:a3K4T3-1}$T=3$}{\includegraphics[width=.21\textwidth]{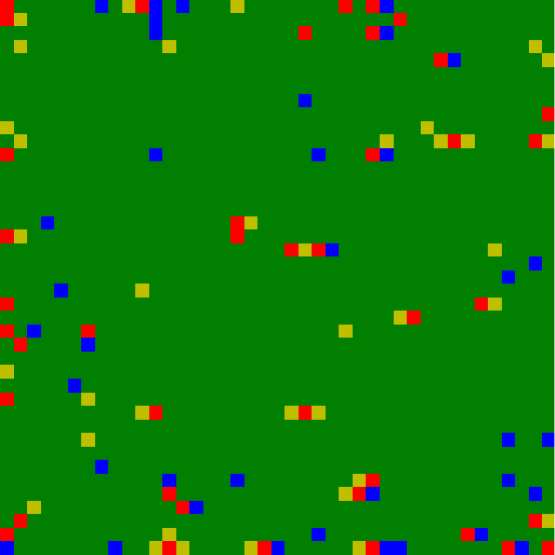}}\hfill
\subcaptionbox{\label{sfig:a3K4T3-2}$T=3$}{\includegraphics[width=.21\textwidth]{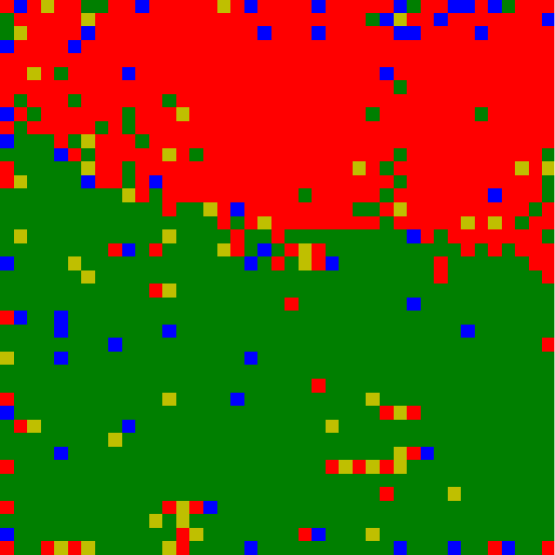}}\\
\subcaptionbox{\label{sfig:a3K4T4-1}$T=4$}{\includegraphics[width=.21\textwidth]{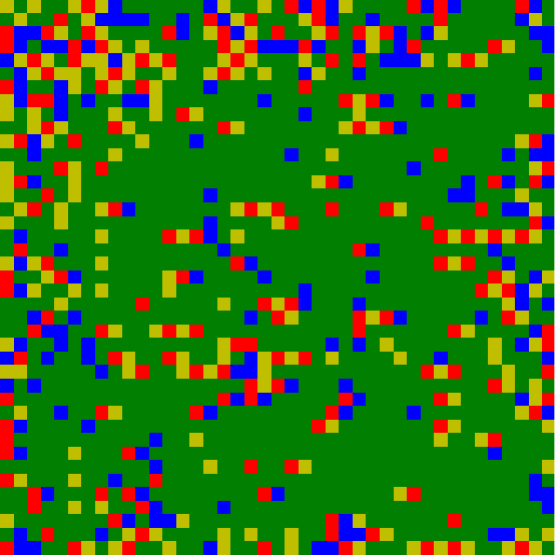}}\hfill
\subcaptionbox{\label{sfig:a3K4T4-2}$T=4$}{\includegraphics[width=.21\textwidth]{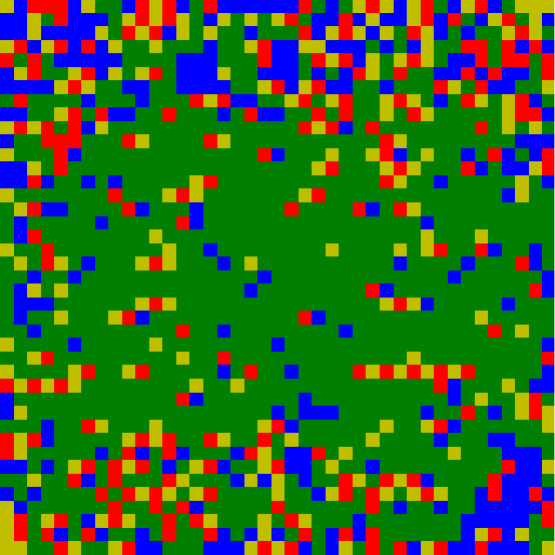}}\\
\caption{\label{fig:opinions_a3K4}Examples of two most probable spatial distributions of the final opinion after $10^3$ time steps. $L=41$, $\alpha=3$, $K=4$ and various levels of noise $T$.}
\end{figure}

The initial random opinions presented in \Cref{fig:example_map} evolve according to \Cref{eq:deterministic} (in the absence of information noise $T=0$) or \Cref{eq:probabilistic} (for $T>0$).
This temporal evolution subsequently changes the spatial opinion distribution. 
In \Cref{fig:opinions_a3K4} examples of two most probable final opinion spatial distributions for various noise levels $T$ after $10^3$ time steps are presented.
The exponent in the distance scaling function is assumed to be $\alpha=3$.
The system contains $L^2=41^2$ actors and $K=4$ possible opinions.

For a deterministic version of algorithm ($T=0$, see \Cref{sfig:a3K4T0-1,sfig:a3K4T0-2}) all $K$ opinions initially present in the systems survive, however, the clustering of actors who share the same opinions is observed. 
Slight increase of temperature ($T=1$) `melts' the `frozen' state leading either to consensus (the same opinion shared by all actors, see \Cref{sfig:a3K4T1-1}) or polarization (two, well separated, clusters of opinions, see \Cref{sfig:a3K4T1-2}).
As a cluster of opinions---or more precisely actors---we consider a group of actors who share the same opinions and connected by the nearest-neighbor interaction (sitting in the von Neumann neighborhood, as for random site percolation problem).
The number of actors who share the same opinion and belong to the same cluster defines the cluster size $\mathcal S$.
The increase of noise level to $T=2$ allows a small number of actors to appear with other available but short-lived opinions (appearing at time $t$ and immediately disappearing at $t+1$) (see \Cref{sfig:a3K4T2-1,sfig:a3K4T2-2}) as the temperature increases $T$---according to \Cref{eq:probability_eq}---favorites the appearance of less probable opinions (exerting less impact).
The above-mentioned increase of probability \eqref{eq:probability_eq} with $T$ leads to an increase of the number of single actors or even pairs of actors with minority opinions destroying locally either consensus (see \Cref{sfig:a3K4T3-1}) or system polarization (see \Cref{sfig:a3K4T3-2}).
The further increase in $T$ also allows for the appearance of larger (but still relatively small) clusters of opinions (\Cref{sfig:a3K4T4-1,sfig:a3K4T4-2}).
Finally, for a high noise level, all opinions become equiprobable as 
\[ \lim_{T\to\infty} P_{i,k}(t)=1/K \]
in every time step $t$ for every actor $i$ and for every opinion $\Xi_k$. 
The latter leads to the system blinking with all $K$ available `colors of opinion at every time step $t$ and at every site $i$---the snapshot of the system does not differ much from the one presented in \Cref{fig:example_map_K4}.

Examples of the spatial distributions of the final opinion for $\alpha=3$ and $K=2$, 3 and 5 (\Cref{fig:opinions_a3K2,fig:opinions_a3K3,fig:opinions_a3K5})
and for $\alpha=4$ and $K=2$, 3, 4 and 5
(\Cref{fig:opinions_a4K2,fig:opinions_a4K3,fig:opinions_a4K4,fig:opinions_a4K5})
are collected in \Cref{apx:example_opinions}.

\subsection{\label{sec:clustering}Opinion clustering}

As the most common observed phenomenon in the system is opinion clustering, we check the distribution of these cluster numbers and sizes.
To this end, we utilize the Hoshen--Kopelman algorithm \cite[pp.~59--60]{Guide_to_Monte_Carlo_Simulations_2009}, \cite{Hoshen1976a,Frijters_2015,1803.09504}. With Hoshen--Kopelman algorithm, one can label every site in such a way that sites (actors sharing the same opinions) in various clusters are labeled with various labels and sites belonging to a given cluster are labeled with the same label.

Let us look again at \Cref{sfig:a3K4T1-1,sfig:a3K4T1-2} obtained for $\alpha=3$, $K=4$ and $T=1$. In \Cref{sfig:a3K4T1-1} consensus takes place and we observe a single cluster (the number of clusters $\mathcal C=1$) and all actors belong to this cluster (the size of the cluster $\mathcal S=L^2$). In \Cref{sfig:a3K4T1-2} the system polarization is observed, thus the number of observed clusters is two ($\mathcal C=2$), but most of the actors are in a `red' cluster ($\mathcal S_1\approx 0.92L^2$) while actors with minority opinion (marked with `green') are occupying upper left corner of the system ($\mathcal S_2\approx 0.08L^2$).

As for larger noise level single sites with minority opinions appear from time to time (cf. for example \Cref{sfig:a3K4T2-1,sfig:a3K4T2-2,sfig:a3K4T3-1,sfig:a3K4T3-2}) but the main picture behind remains the same (i.e. in principle we still deal either with consensuses or system polarization), it would be useful to introduce the noise discrimination level $\theta$.
For example, setting $\theta=5$ and neglecting appearance clusters with sizes $\mathcal S$ smaller than $\theta$ is sufficient to keep the picture of the number $\mathcal C$ of clusters as for those presented in \Cref{sfig:a3K4T1-1,sfig:a3K4T1-2} also for systems presented in \Cref{sfig:a3K4T2-1,sfig:a3K4T2-2,sfig:a3K4T3-1,sfig:a3K4T3-2}.

The results presented below are based on assuming various levels of discrimination $\theta$ in the spirit described above.
In other words, the $\theta$ parameter arbitrarily says how small the clusters of agents sharing the same opinion could be considered as negligible.

\subsubsection{\label{ssec:ave_C}Average number of opinion clusters}

\begin{figure}[htbp]
\centering
\includegraphics[width=.75\columnwidth]{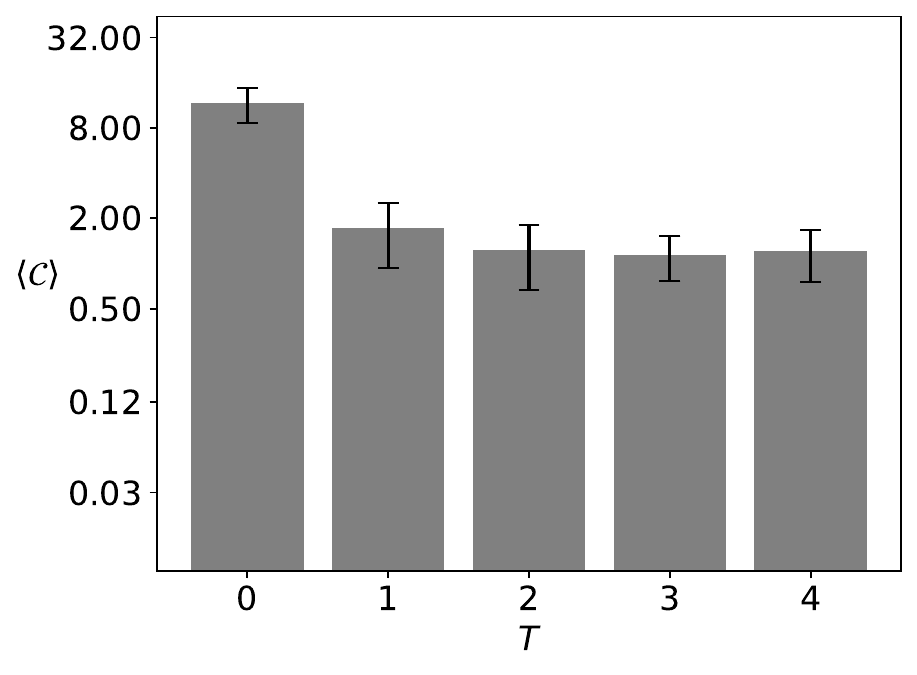}
\caption{\label{fig:clust_hist_a3K4theta25}Average number $\langle\mathcal C\rangle$ of opinion clusters after $t=10^3$ time steps for the exponent of the distance scaling function $\alpha=3$, the number $K=4$ of opinions available in the system, and the noise discrimination threshold $\theta=25$. The system contains $L^2=41^2$ actors. The results are averaged over $R=100$ independent system realizations.}
\end{figure}

In \Cref{fig:clust_hist_a3K4theta25} the average number $\mathcal C$ of opinion groups is presented for $\alpha=3$ and $K=4$. 
Statistics are based on $R=100$ replications of the system with $L^2=41^2$ actors measured after $t=10^3$ time steps of evolution.
We assume the discrimination threshold $\theta=25$. 

For $T=1$ roughly half among $R=100$ simulations end in consensus ($\mathcal C=1$) or system polarization ($\mathcal C=2$) leading to the average number of clusters $\langle\mathcal C\rangle\approx 1.73(80)$. 
The symbol $\langle\cdots\rangle$ stands for the averaging procedure on $R=100$ independent system realizations (simulations).
The increase in the level of noise $T\ge 2$ with the assumed discrimination threshold $\theta=25$ does not change the average number of clusters $\langle\mathcal C\rangle$ to much: $\langle\mathcal C\rangle=1.24(57)$, $1.14(38)$ and $1.22(46)$ for $T=2$, 3 and 4, respectively.

However, for $T=0$ this number $\langle\mathcal C\rangle\approx 11.6$ (with uncertainty 3.0) is much higher than for $T\ne 0$ (please note the logarithmic scale on the $\langle\mathcal C\rangle$ axis).
We should stress that the number of clusters $\mathcal C=17$ (\Cref{sfig:a3K4T0-1}) and $\mathcal C=8$ (\Cref{sfig:a3K4T0-2}) is higher than the number of opinions available $K=4$ in the systems. 
In other words, several different clusters of the same opinion are counted for the number $\mathcal C$. For instance, in \Cref{sfig:a3K4T0-2} we observe four clusters (of sizes $\mathcal C$ larger than $\theta=25$) of `green' opinions, two of `blue' opinions, two of `red' opinions, and none of `yellow' opinions.

The average number $\langle\mathcal C\rangle$ of clusters for various values of the distance scaling function exponent $\alpha=2$, 3, 4 and 6, number of available opinions $K=2$, 3, 4 and 5, information noise level $T=0$, 1, 2, 3 and 4 and noise discrimination levels $\theta=12$, 25 and 50 are presented in \Cref{fig:th12_clust,fig:th25_clust,fig:th50_clust} in \Cref{apx:num_clu}.

\subsubsection{\label{ssec:Smax}The sizes of the largest clusters}

\begin{figure}
\subcaptionbox{$K=2$\label{sfig:biggest_cluster_K2}}{\includegraphics[width=.34\textwidth]{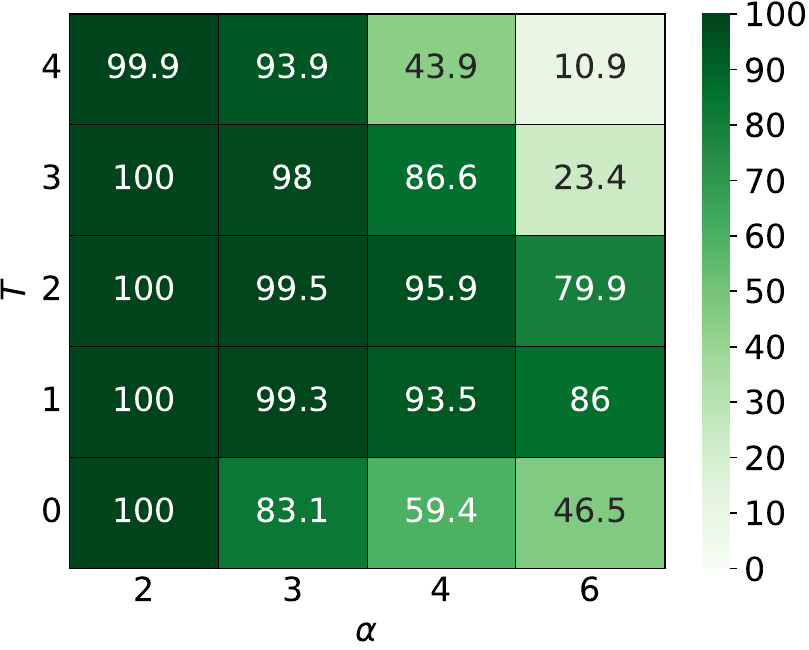}}\\
\subcaptionbox{$K=3$\label{sfig:biggest_cluster_K3}}{\includegraphics[width=.34\textwidth]{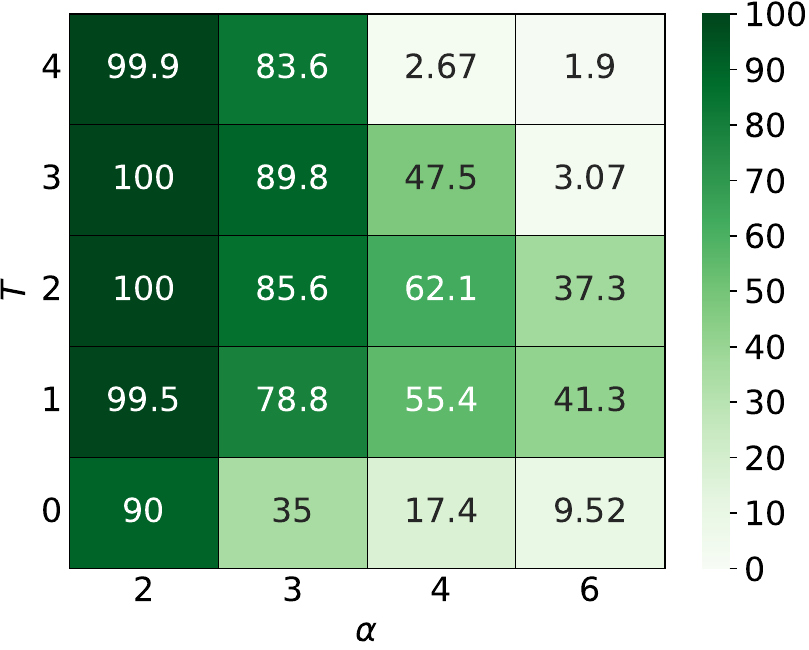}}\\
\subcaptionbox{$K=4$\label{sfig:biggest_cluster_K4}}{\includegraphics[width=.34\textwidth]{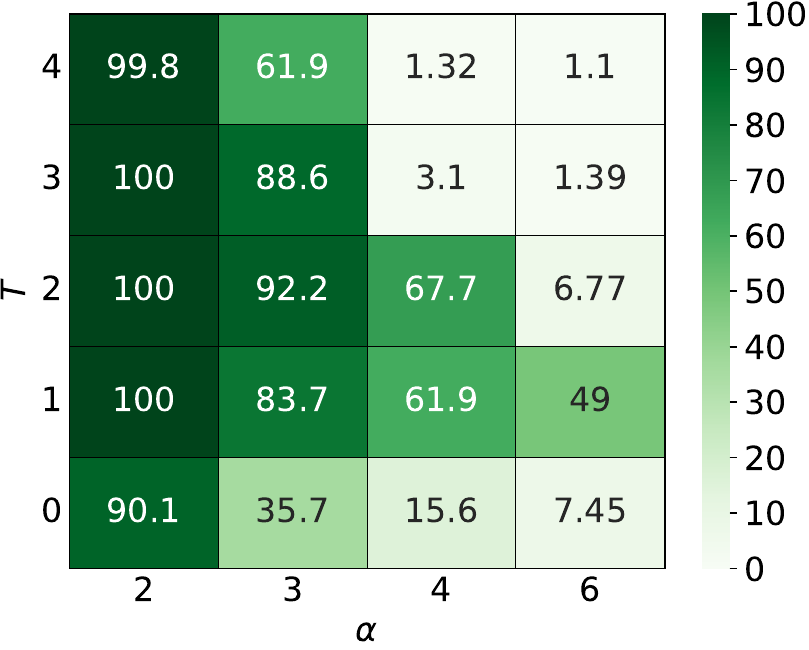}}\\
\subcaptionbox{$K=5$\label{sfig:biggest_cluster_K5}}{\includegraphics[width=.34\textwidth]{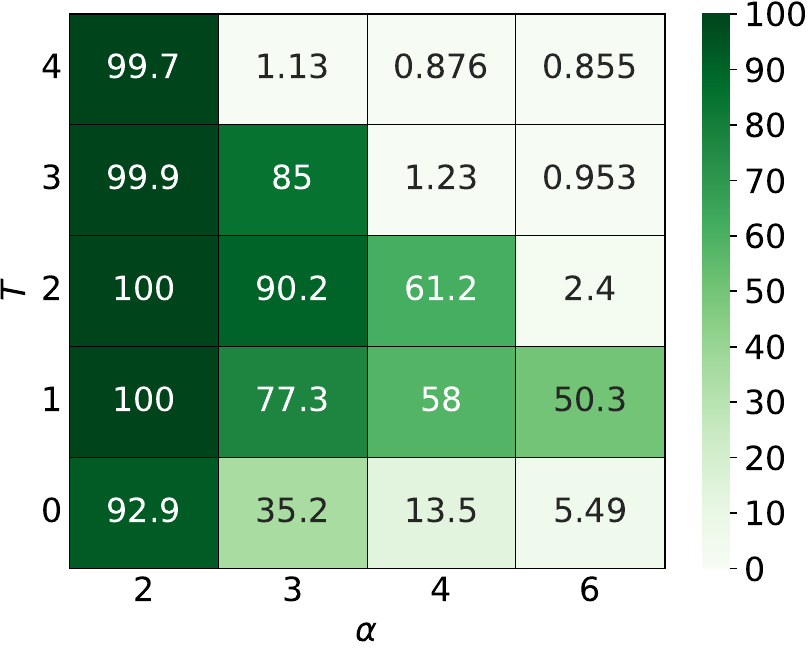}}
\caption{\label{fig:biggest_cluster_maps}The average ratio (in percents) of the size of the largest cluster $\langle\mathcal S_\text{max}\rangle$ to the size of the entire system $L^2$ depending on the parameters $\alpha$ and $T$. $L=41$, $t=10^3$, $R=100$.}
\end{figure}

In Reference~\onlinecite{2002.05451} average largest cluster size $\langle\mathcal S_\text{max}\rangle$ (normalized to the system size $L^2$) for $K=2$ and $K=3$ and various values of the noise level $T$ and the interaction range $\alpha$ were presented in Figures \href{https://doi.org/10.1371/journal.pone.0235313.g006}{6a} and \href{https://doi.org/10.1371/journal.pone.0235313.g007}{7a}, respectively.
Here, we also extend this study to a larger number $K$ of opinions available in the system, namely for $K=4$ and $K=5$.
The results are presented in \Cref{fig:biggest_cluster_maps}.

Let us again look at the thermal evolution of $\mathcal S_\text{max}$ of the system presented in \Cref{fig:opinions_a3K4}. 
Due to the freezing system for $T=0$ (as presented in \Cref{sfig:a3K4T0-1,sfig:a3K4T0-2}) the largest cluster sizes are around $\mathcal S_\text{max}=267$ and $\mathcal S_\text{max}=794$ (cluster of `green' opinion in the upper left corner and cluster of `red' opinion in the left side of \Cref{sfig:a3K4T0-1,sfig:a3K4T0-2}, respectively).
The increase in noise level to $T=1$ increases the sizes of the largest cluster to  
$\mathcal S_\text{max}=L^2$ and $\mathcal S_\text{max}=1540$ for \Cref{sfig:a3K4T1-1,sfig:a3K4T1-2}, respectively.
Then, the subsequent increase in $T$ only reduces the size of the largest cluster.

\subsection{Distribution of surviving opinions}

The methodology of clusters counting allowing for construction of histograms $\langle\mathcal C(T)\rangle$ presented in \Cref{fig:clust_hist_a3K4theta25,fig:th12_clust,fig:th25_clust,fig:th50_clust}---as mentioned in \Cref{ssec:ave_C}---neglects the clusters colors. Thus, the information provided there is insufficient to determine whether all $K$ opinions available in the system persisted until the assumed time $t=10^3$.
Now, we are interested in checking the number $1\le\Phi\le K$ of surviving opinions for various values of the parameters $K$, $\alpha$, and $T$.

As mentioned above, the system presented in \Cref{sfig:a3K4T0-2} for $K=4$, $\alpha=3$, $T=0$ has eight clusters larger than $\theta=25$, and thus the number of clusters $\mathcal C$ is eight.
As three opinions available in the system are observed, then $\Phi=3$.
In contrast, for $T=1$ (see \Cref{sfig:a3K4T1-2}) only $\Phi=2$ opinions (`red' and `green') survived. There, due to the polarization of the system, the number of clusters $\mathcal C$ and the number of surviving opinions $\Phi$ are equal.

\subsubsection{\label{ssec:Phi}Histograms of surviving opinions}

The opinion that survives in the system is the opinion that, at the end of the simulation, it is represented by at least one cluster with a size $\mathcal S$ not smaller than $\theta$.

\begin{figure}[htbp]
\centering
\includegraphics[width=.75\columnwidth]{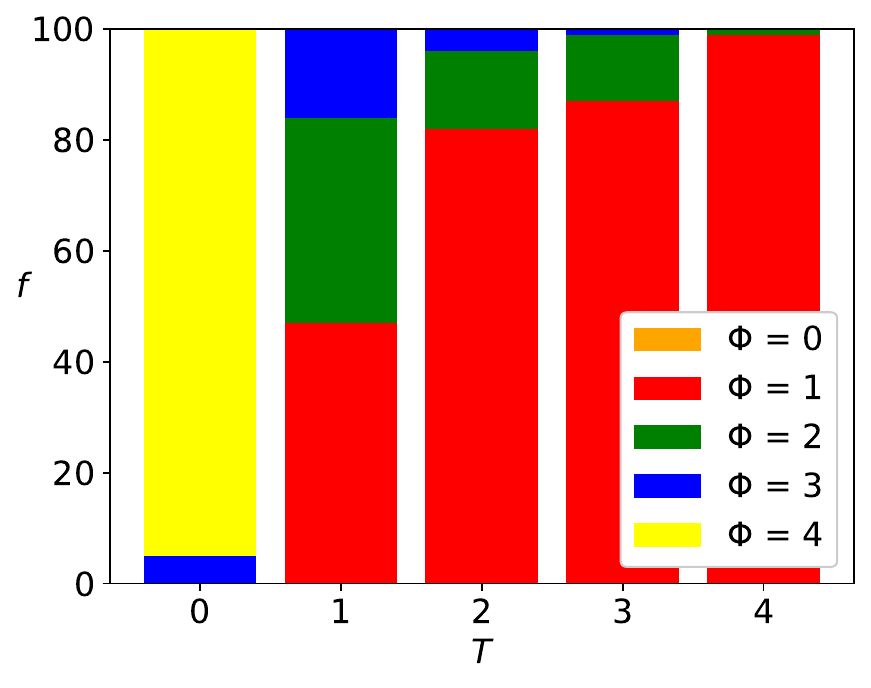}
\caption{\label{fig:surv_histogram_a3K4theta25}The histogram of frequencies $f$ of the number $\Phi$ of surviving opinions for $\alpha=3$, $K=4$ and the level of noise discrimination $\theta=25$.}
\end{figure}

In \Cref{fig:surv_histogram_a3K4theta25} the histogram of the number $\Phi(T)$ of surviving opinions for $\alpha=3$, $K=4$ and the level of noise discrimination $\theta=25$ are presented.

The results are collected again after $t=10^3$ time steps and for $R=100$ system realizations.

For $T=0$, 95\% of these $R$ simulations ended with $\Phi=4$ [$f(\Phi=4)=95\%$, yellow rectangle in the first bar of \Cref{fig:surv_histogram_a3K4theta25}] surviving opinions, and 5\% of the simulations ended with $\Phi=3$ surviving opinions [$f(\Phi=3)=5\%$, blue rectangle in the first bar of \Cref{fig:surv_histogram_a3K4theta25}].
Situations with consensus ($\Phi=1$) or system polarization ($\Phi=2$) were not observed: $f(\Phi=1)=f(\Phi=2)=0\%$ [absence of green and red rectangles in the first bar of \Cref{fig:surv_histogram_a3K4theta25}].
Finally, the orange color is also absent [$f(\Phi=0)=0\%$] in the first bar of \Cref{fig:surv_histogram_a3K4theta25}] which means that the situation of all opinions disappearing was not observed.
Of course, the rules of the game do not allow for vanishing all opinions: the case $f(\Phi=0)>0$ means that the fraction $f(\Phi=0)$ of system realizations ended with a lot of very small clusters, each of them smaller than the assumed noise discrimination level $\theta$. 

For $T=1$, 47\% of these $R$ simulations ended with $\Phi=1$ [$f(\Phi=1)=47\%$, red rectangle on the second bar of \Cref{fig:surv_histogram_a3K4theta25}] surviving opinions,  37\% of the simulations ended with $\Phi=2$ surviving opinions [$f(\Phi=2)=37\%$, green rectangle in the second bar of \Cref{fig:surv_histogram_a3K4theta25}] and 16\% of the simulations ended with $\Phi=3$ surviving opinions [$f(\Phi=3)=16\%$, blue rectangle in the second bar of \Cref{fig:surv_histogram_a3K4theta25}], etc.

For the highest noise level investigated ($T=4$) we have $f(\Phi=1)\approx 99\%$ (red rectangle in the fifth bar in \Cref{fig:surv_histogram_a3K4theta25}) and $f(\Phi=2)\approx 1\%$ (green rectangle in the fifth bar in \Cref{fig:surv_histogram_a3K4theta25}).

Histograms of frequencies $f(\Phi)$ of the numbers $\Phi$ of the surviving opinions for various values of $K$, $\alpha$, $T$ and three values of noise discrimination level $\theta=12$, 25, 50 are presented in \Cref{fig:th12_sur,fig:th25_sur,fig:th50_sur} in \Cref{apx:num_sur}.

\subsubsection{\label{ssec:Phi-star}The most probable number of surviving opinions}

\begin{figure*}[htbp]
\centering
\subcaptionbox{$K=2$, $\theta=12$\label{fig:probable_th12K2}}{\includegraphics[width=.32\textwidth,]{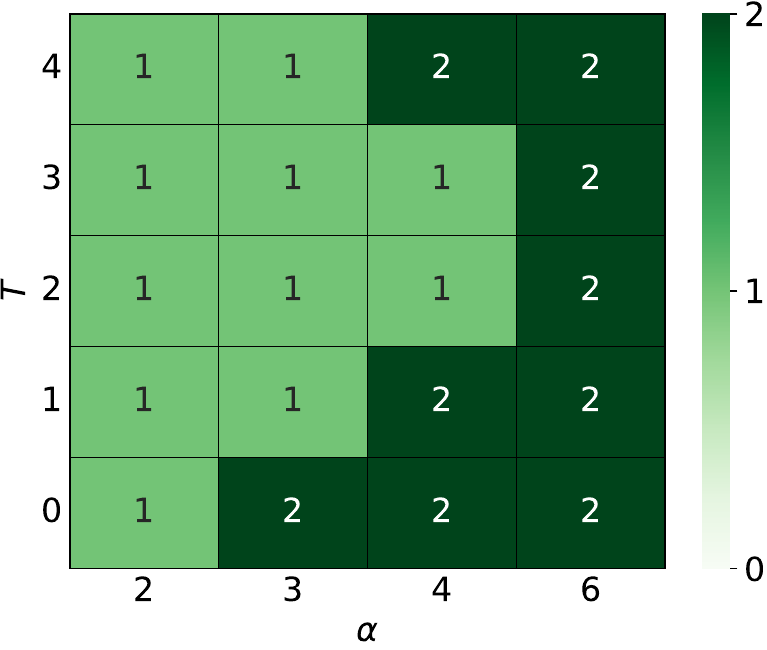}}\hfill
\subcaptionbox{$K=2$, $\theta=25$\label{fig:probable_th25K2}}{\includegraphics[width=.32\textwidth,]{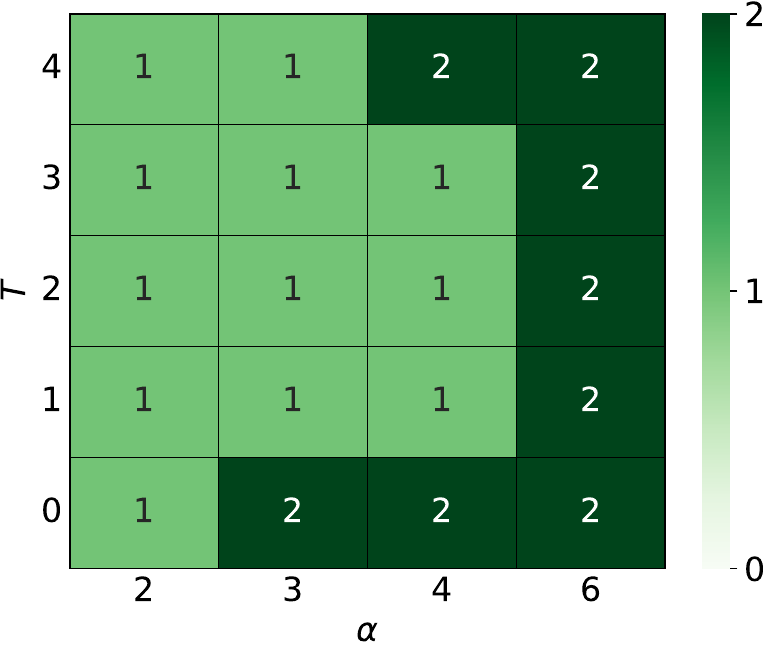}}\hfill
\subcaptionbox{$K=2$, $\theta=50$\label{fig:probable_th50K2}}{\includegraphics[width=.32\textwidth,]{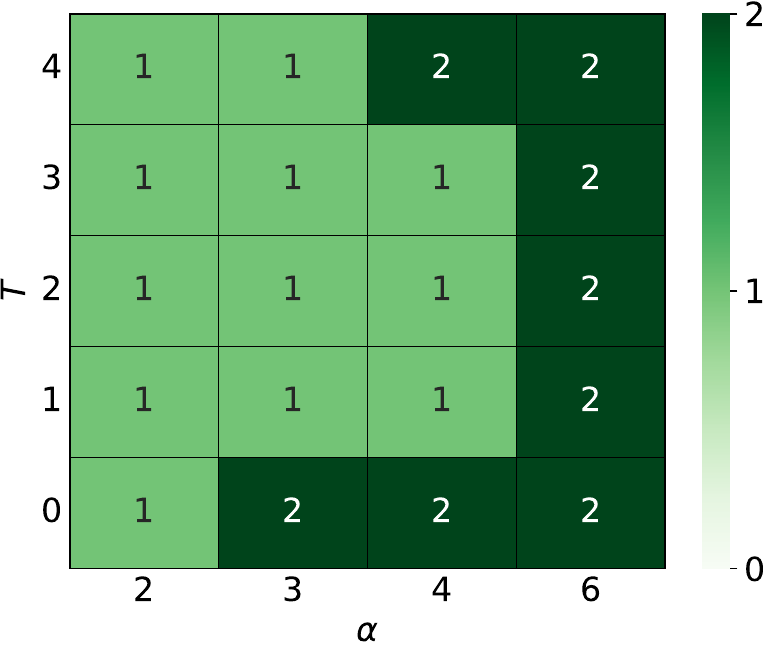}}\\
\subcaptionbox{$K=3$, $\theta=12$\label{fig:probable_th12K3}}{\includegraphics[width=.32\textwidth,]{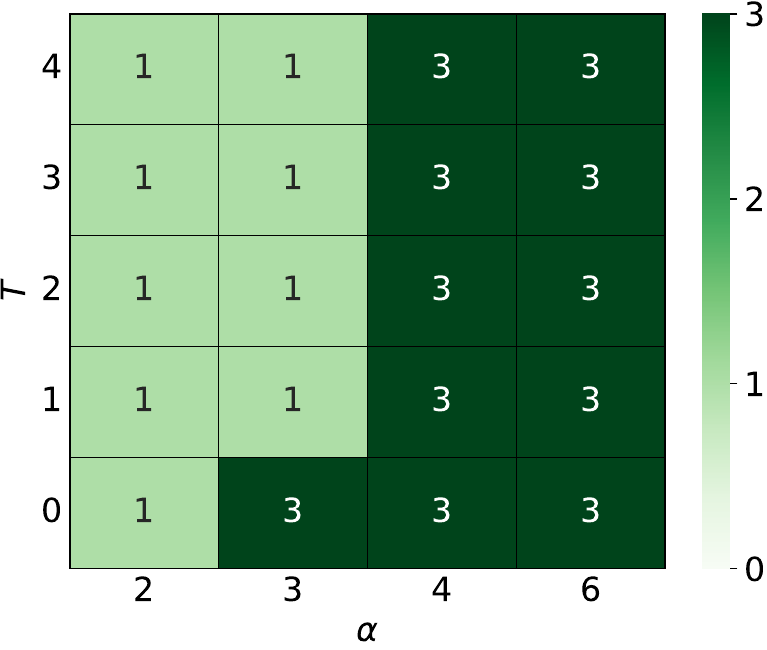}}\hfill
\subcaptionbox{$K=3$, $\theta=25$\label{fig:probable_th25K3}}{\includegraphics[width=.32\textwidth,]{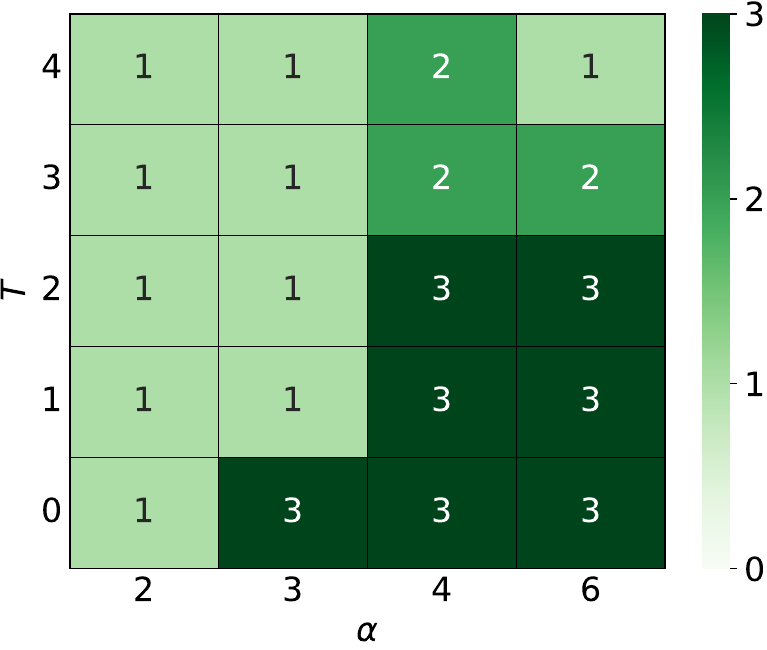}}\hfill
\subcaptionbox{$K=3$, $\theta=50$\label{fig:probable_th50K3}}{\includegraphics[width=.32\textwidth,]{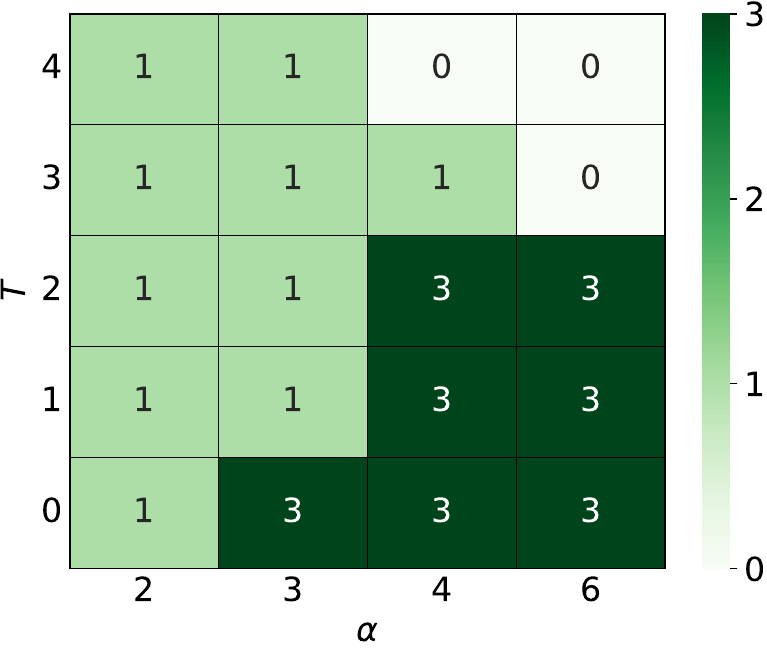}}\\
\subcaptionbox{$K=4$, $\theta=12$\label{fig:probable_th12K4}}{\includegraphics[width=.32\textwidth,]{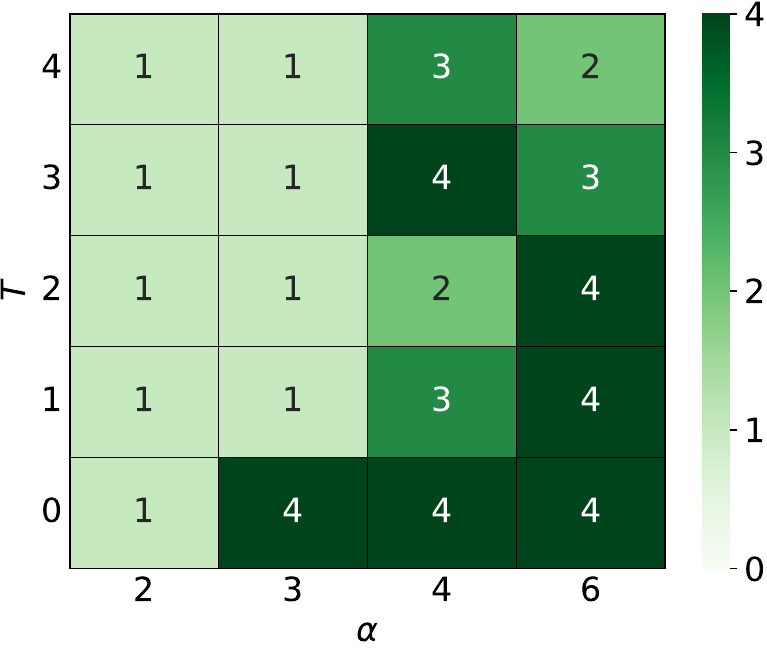}}\hfill
\subcaptionbox{$K=4$, $\theta=25$\label{fig:probable_th25K4}}{\includegraphics[width=.32\textwidth,]{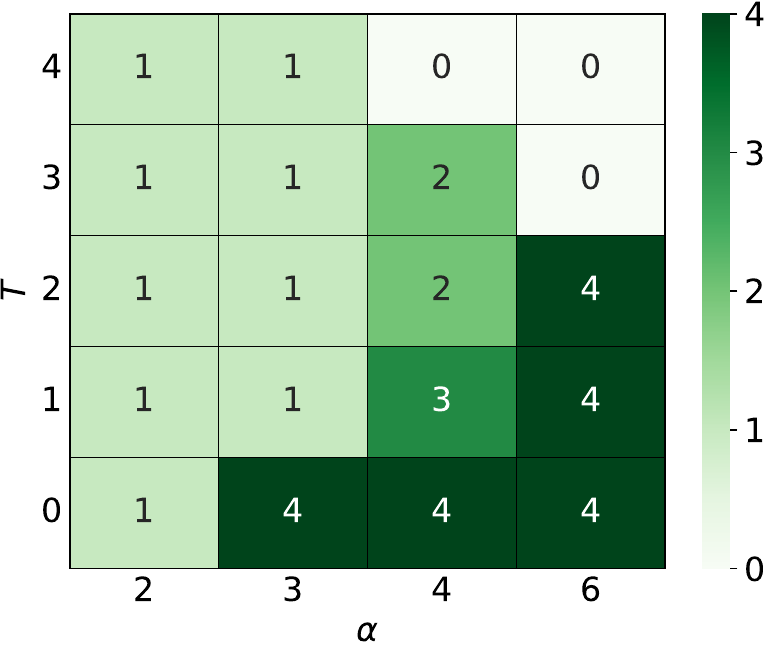}}\hfill
\subcaptionbox{$K=4$, $\theta=50$\label{fig:probable_th50K4}}{\includegraphics[width=.32\textwidth,]{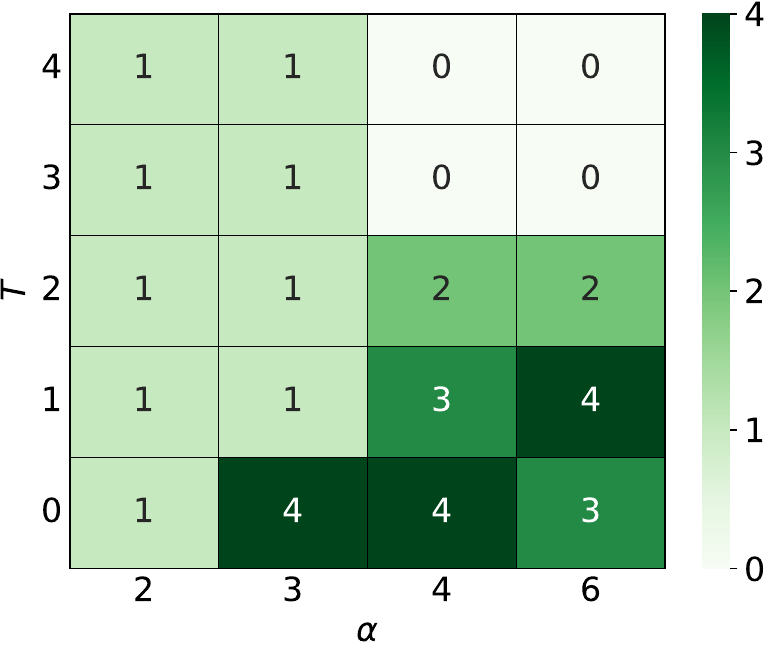}}\\
\subcaptionbox{$K=5$, $\theta=12$\label{fig:probable_th12K5}}{\includegraphics[width=.32\textwidth,]{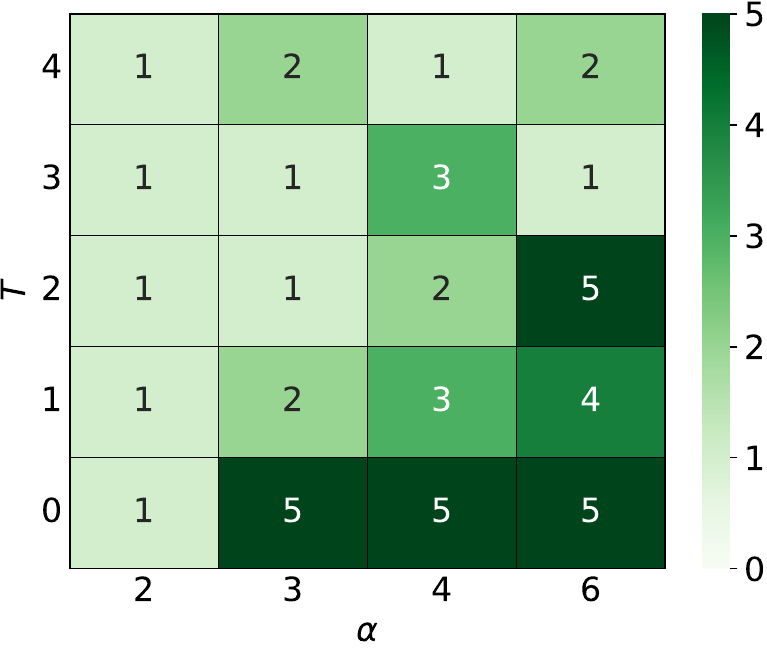}}\hfill
\subcaptionbox{$K=5$, $\theta=25$\label{fig:probable_th25K5}}{\includegraphics[width=.32\textwidth,]{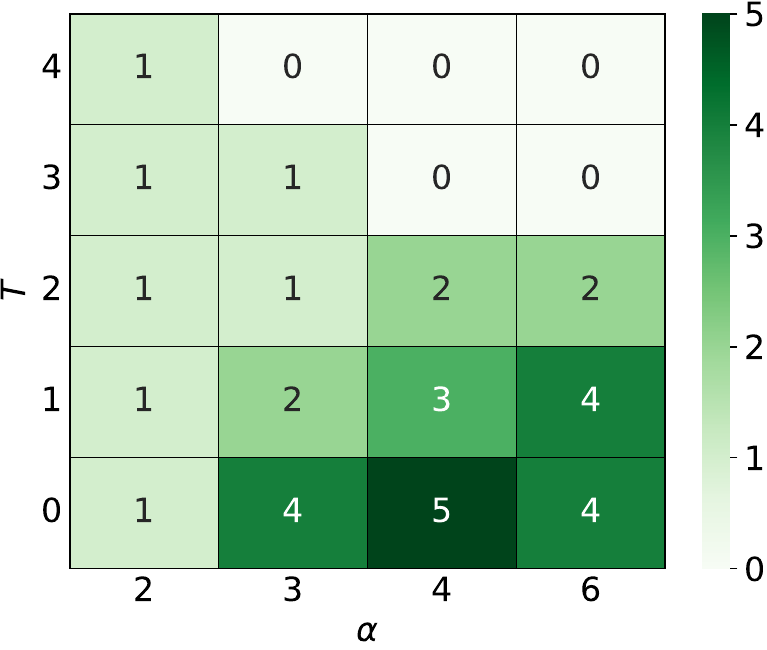}}\hfill
\subcaptionbox{$K=5$, $\theta=50$\label{fig:probable_th50K5}}{\includegraphics[width=.32\textwidth,]{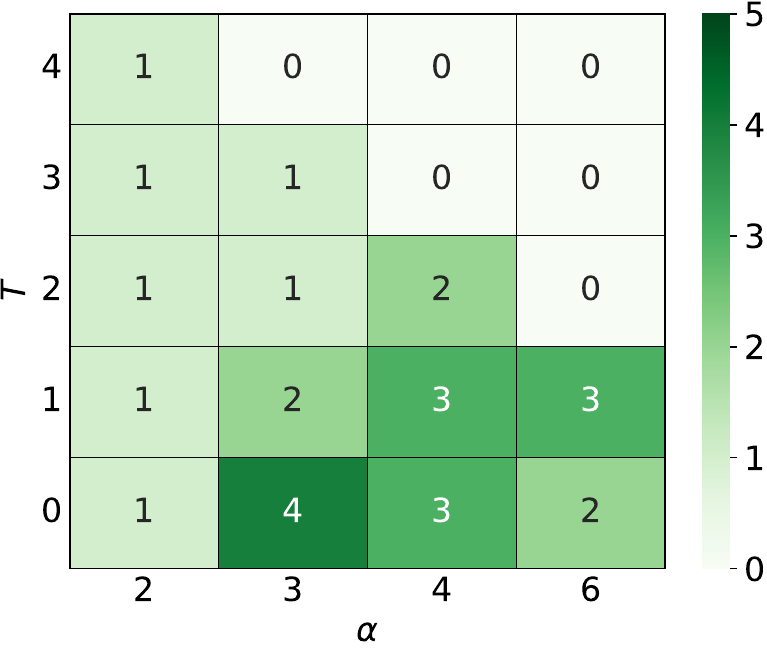}}\\
\caption{\label{fig:most}The most probable final number $\Phi^\star$ of surviving opinions for various numbers $K$ of opinions available in the system and noise discrimination thresholds $\theta$ depending on the level of information noise $T$ and the range of interaction $\alpha$.}
\end{figure*}

We finalize the presentation of the results with heat maps of the most probable final number of surviving opinions $\Phi^\star$ (see \Cref{fig:most}). We define the most probable number of surviving opinions $\Phi^\star$ as this value of $\Phi$ for which the fraction $f(\Phi)$ is the largest (for fixed values of the noise discrimination level $\theta$, the noise level of information $T$ and the effective range of interaction $\alpha$).

For example, for $K=4$, $\alpha=3$, $\theta=25$ and 
\begin{itemize} 
\item for $T=0$ (see the first bar of \Cref{fig:surv_histogram_a3K4theta25}) $\Phi^\star=4$ as $95\%=f(\Phi=4) > f(\Phi=3)=5\%$,  
\item for $T=1$, 2, 3 (see the second, third, and fourth bar of \Cref{fig:surv_histogram_a3K4theta25}) $\Phi^\star=1$ as $f(\Phi=1) > f(\Phi=2) > f(\Phi=3)$,
\item for $T=4$ (see the fifth bar of \Cref{fig:surv_histogram_a3K4theta25}) $\Phi^\star=1$ as $99\%=f(\Phi=1) > f(\Phi=2)=1\%$.
\end{itemize}

\section{\label{sec:discussion}Discussion}

\subsection{\label{disc:ave_C}Average number of opinion clusters}

For a low value of the noise discrimination level ($\theta=12$, \Cref{fig:th12_clust}) and $\alpha=2$ (see \Cref{sfig:clust_th12a2K2,sfig:clust_th12a2K3,sfig:clust_th12a2K4,sfig:clust_th12a2K5}) for the nondeterministic version of the algorithm ($T>0$), only one cluster exceeds the threshold size, regardless of the number $K$ of opinions available in the system. Therefore, the system is dominated by a single group of opinions, and consensus takes place.

Reducing the impact of distant actors ($\alpha=3$, \Cref{sfig:clust_th12a3K2,sfig:clust_th12a3K3,sfig:clust_th12a3K4,sfig:clust_th12a3K5}) allows additional clusters of size $S$ greater than $\theta=12$. Their number $\langle\mathcal C\rangle$ most often does not exceed two, except for the simulation of a high number of opinions available ($K> 3$) and high social temperature ($T=4$). For such parameter settings, we can observe on average more than two clusters, at the same time with a greater standard deviation of this number --- the number of clusters, depending on the simulation, ranges from $\langle\mathcal C\rangle=1$ to about $\langle\mathcal C\rangle=5\div 6$. Independently of the number of $K$ the deterministic case ($T=0$) produces a relatively high average number $\langle\mathcal C\rangle$ of clusters ($\langle\mathcal C\rangle=4$ for $K=2$ opinions, $\langle\mathcal C\rangle=16$ and for $K>2$).

An increased exponent ($\alpha=4$, \Cref{sfig:clust_th12a4K2,sfig:clust_th12a4K3,sfig:clust_th12a4K4,sfig:clust_th12a4K5}) results in a clear increase in the average number $\langle\mathcal C\rangle$ of clusters in the system up to $\langle\mathcal C\rangle= 32$ for $T=0$.

For the largest value considered of $\alpha=6$ (\Cref{sfig:clust_th12a6K2,sfig:clust_th12a6K3,sfig:clust_th12a6K4,sfig:clust_th12a6K5})
the most numerous sets of clusters with a size $S$ exceeding $\theta = 12$ are observed. With two opinions in the system (\Cref{sfig:clust_th12a6K2}), the temperature $T = 3$ is sufficient for a significant division of agents for $\langle\mathcal C\rangle\ge 16$ clusters with a size exceeding the threshold $\theta$. The trend continues for simulations with available $K = 3$ different opinions (\Cref{sfig:clust_th12a6K3}). However, for high temperatures and a large number of possible opinions ($K = 4$, $T = 4$ and $K = 5$, $T = 3, 4$), the average number of clusters $\langle\mathcal C\rangle$ with a size $S$ greater than the threshold $\theta$ begins to decline due to too much fragmentation --- the system becomes an irregular set of many very small clusters (\Cref{sfig:a4K4T4-1,sfig:a4K4T4-2,sfig:a4K5T4-1,sfig:a4K5T4-2}), and none of the opinions can get a noticeable advantage. For $T=0$, the average number of clusters in the system remains very high and reaches $\langle\mathcal C\rangle= 32$.

For increased threshold $\theta=25$ (\Cref{fig:th25_clust}) noticeable differences appear for $K = 4, 5$ and $\alpha=3$ and the highest of the social temperatures studied $T = 4$ (\Cref{sfig:clust_th25a3K4,sfig:clust_th25a3K5}), where fewer clusters were recorded that met the condition $S> \theta = 25$. 

For the simulations with $K = 5$ and $T = 4$, where at least one cluster of an appropriate size has been preserved, it was so rare that the average number of clusters was a fraction ($\langle\mathcal C\rangle\approx  0.15$).
This value well reflects the division of agents who share the same opinion into small, randomly arranged clusters.

A further increase in the threshold $\theta$ (up to 50, \Cref{fig:th50_clust}) results in disappearing clusters of sizes $S$ larger than $\theta$ for $\alpha \ge 4$ and $K \ge 4$ (\Cref{sfig:clust_th50a4K4,sfig:clust_th50a4K5,sfig:clust_th50a6K4,sfig:clust_th50a6K5}).

\subsection{\label{disc:Smax}The sizes of the largest clusters}

We would like to recall the ambivalent role observed of the information noise level $T$ in shaping the largest cluster size $\mathcal S_\text{max}$ mentioned in Reference~\onlinecite[p. 14]{2002.05451}:
``[\ldots] the average size of the maximum cluster
$\mathcal S_\text{max}$ decreases with $\alpha$ for fixed $T$ values. The appearance of noise in the system ($T = 1$)
slightly organizes the system in relation to the noiseless situation with $T = 0$ (which is particularly visible for $\alpha > 2$ [\ldots]). Indeed, as in earlier studies \cite{PhysRevE.75.045101,Shirado2017}, small level of noise brought more order to the system. Furthermore, the introduction of noise ($T$) in the adoption of
opinions causes an increase in $\mathcal S_\text{max}$, and then its decrease, which is especially visible for
$\alpha > 2$ (this inflection point is nearly $T = 2$).''
and later: ``[\ldots] noise for certain values of $\alpha$ promotes unanimity. This situation occurs for $\alpha = 3$ (both for $K = 2$ and $K = 3$), when the frozen state system, with increasing noise $T$, achieves the consensus state for $T = 3$, before disordering for $T = 5$'' \cite[p. 18]{2002.05451}.

This nonmonotonous dependence $\mathcal S_\text{max}/L^2$ on the noise parameter $T$ is observed for any value of $\alpha$ but for larger values of $\alpha$ and larger values of the number $K$ of opinions available in the systems, this dependence becomes more and more spectacular.
For example, for $K=5$ (\Cref{sfig:biggest_cluster_K5}) we see a high peak of $\mathcal S_\text{max}/L^2\approx 50\%$ for $T=1$ and $\alpha=6$ deeply reduced to 2.4\% and 5.5\% for a larger ($T=2$) and lower ($T=0$) noise level. 
The similar behavior in $\mathcal S_\text{max}$ in dependence on $T$ is also observed for $\alpha=4$ with $\mathcal S_\text{max}/L^2\approx 60\%$ for $T=1, 2$ reduced to 1.2\% and 13.3\% for a higher ($T=3$) and lower ($T=0$) noise level.
The further increasing influence of more distance actors (decreasing $\alpha$) makes the $\mathcal S_\text{max}$ dependence more and more smoother, making it almost flat for $\alpha=2$ with only marginal deviation from $\mathcal S_\text{max}/L^2= 100\%$ at the edges of the range of values studied for the parameter $T$.

The picture presented above is also qualitatively reproduced for $K=4$ (see \Cref{sfig:biggest_cluster_K4}).

Independently of the number $K$ of opinions considered available in the system for fixed value of the noise parameter $T$ the average size of the largest cluster $\mathcal S_\text{max}$ decreases with increasing of $\alpha$, i.e., with limiting the influence of very long-range interactions. 

\subsection{\label{disc:Phi}Histograms of surviving opinions}
The histograms $f(\Phi;T)$ of the surviving opinions (\Cref{ssec:Phi}) presented in \Cref{fig:th12_sur,fig:th25_sur,fig:th50_sur} in \Cref{apx:num_sur} are almost untouched by the noise discrimination level $\theta$ for a high effective interaction range [$\alpha=2$, \Cref{sfig:th12a2K2,sfig:th12a2K3,sfig:th12a2K4,sfig:th12a2K5,sfig:th25a2K2,sfig:th25a2K3,sfig:th25a2K4,sfig:th25a2K5,sfig:th50a2K2,sfig:th50a2K3,sfig:th50a2K4,sfig:th50a2K5}] as well as for the lowest possible number of opinions available in the system [$K=2$, \Cref{sfig:th12a2K2,sfig:th12a3K2,sfig:th12a4K2,sfig:th12a6K2,sfig:th25a2K2,sfig:th25a3K2,sfig:th25a4K2,sfig:th25a6K2,sfig:th50a2K2,sfig:th50a3K2,sfig:th50a4K2,sfig:th50a6K2}].
This is a consequence of the appearance of consensus or system polarization and is consistent with the generally observed final system states presented earlier in \Cref{fig:opinions_a3K2,fig:opinions_a3K3,fig:opinions_a3K4,fig:opinions_a3K5,fig:opinions_a4K2,fig:opinions_a4K3,fig:opinions_a4K4,fig:opinions_a4K5}.

The most noticeable differences occur in \Cref{sfig:th12a3K4,sfig:th12a3K5,sfig:th12a4K4,sfig:th12a4K5,sfig:th12a6K4,sfig:th12a6K5,sfig:th25a3K4,sfig:th25a3K5,sfig:th25a4K4,sfig:th25a4K5,sfig:th25a6K4,sfig:th25a6K5,sfig:th50a3K4,sfig:th50a3K5,sfig:th50a4K4,sfig:th50a4K5,sfig:th50a6K4,sfig:th50a6K5}, that is, for $\alpha\ge 3$ and $K\ge 4$. 
For a high noise level ($T=4$) in this parameter regime, the frequency $f(\Phi=0)$ dominates the system (absence of sizes $S$ greater than $\theta$) except for the lowest assumed threshold $\theta=12$, allowing observation up to $\Phi=3$ surviving opinions, but of small cluster sizes.

\subsection{\label{dics:Phi-star}The most probable number of surviving opinions} 

We finalize the discussion of the results obtained with an analysis of the heat maps (\Cref{fig:most}) of the most probable final number $\Phi^*$ (\Cref{ssec:Phi-star}, \Cref{fig:most}) of the remaining opinions for various numbers $K$ of opinions available in the system and various noise discrimination numbers $\theta$.
These maps are constructed in the $(\alpha,T)$ plane.
With the assumed scanning accuracy of the parameters $\alpha$ i $T$ parameters, the shape of the obtained maps differs qualitatively from those reported in Figures \href{https://doi.org/10.1371/journal.pone.0235313.g006}{6} and \href{https://doi.org/10.1371/journal.pone.0235313.g007}{7} in Reference \cite{2002.05451}, particularly with well-visible juts for higher values of $\Phi^*$ for intermediate values of the level of information noise $2\le T\le 3$ and high values of $\alpha\approx 6$ (that is, for a long effective range of interaction between actors).

\section{\label{sec:conclusion}Conclusion}

In Reference \onlinecite{1902.03454} the model of opinion formation was introduced based on the Latan\'e theory of social impact with many available opinions. 
In computer simulations based on the Szamrej--Nowak--Latan\'e model, it was shown that increasing the number of opinions decreases the critical noise level separating ordered and disordered phases.
The observed results were followed by further studies \cite{2002.05451} in which both the noise level $T$ and the interaction range $\alpha$ were considered.
It was shown that the noise level has an ambiguous role: its lower value helps in system ordering (spatial clustering of opinions), while its higher value destroys any spatial correlations among actors and their opinions.
This useful role for the small noise level was also reported in References \onlinecite{PhysRevE.79.046108,Biondo2013,000414818100049,Shirado2017}.

In this paper, we follow the path indicated in the References \onlinecite{1902.03454,2002.05451} and with a computerized version of the social impact theory (\Cref{sec:model}) we simulate the formation of opinions in an artificial society. 
Images obtained from spatial opinion distributions (\Cref{sec:opinionmaps}) were analyzed in terms of the grouping of opinions and the characteristics of these opinion clusters (\Cref{sec:clustering}).
Based on the simulation results, we show how the number $\Phi^*$ of observed opinions (understood as spatial clusters of at least $\theta$ actors sharing the same opinion) depends on the model control parameters (effective range of interaction $\alpha$ and noise level $T$).
In contrast to the Reference \onlinecite{2002.05451}---were number of (arbitrarily recognized as small or large) cluster sizes were investigated---here we introduce the noise discrimination level $\theta$ allowing the finest analysis of histograms of cluster sizes.

As a square lattice is not best suited for modeling the social interaction, also checking another network topology seems to be a promising way for further studies.
On the other hand, the square lattice naturally produces a regular ego-centered network of actors \cite{Dunbar_2010,Sutcliffe_2012,Arnaboldi_2016}, where nodes in subsequent coordination zones may be equated with subsequent ``circles'' (in the ego-centered network theory terminology) containing the support clique (sites from the first and second coordination zones, \Cref{fig:neighbour_n09}), sympathy group (sites from the third to fifth coordination zones, the outermost ``ring'' in \Cref{fig:neighbour_n25}), affinity group (sites from the sixth to the ninth coordination zones, the outermost ``ring'' in \Cref{fig:neighbour_n49}) and active network (sites from the 10-th to 14-th coordination zones, not marked in \Cref{fig:neighbourhoods}).
Keeping the terminology of Reference~\onlinecite{Arnaboldi_2016}, a ``red'' actor presented in \Cref{fig:neighbour_n01} plays the role of ``ego'' while actors in subsequent coordination zones are his/her ``alters''.
Our results (\Cref{tab:In-to-I}) show that---independently of the number $K$ of opinions available in the system---from 57\% (low values of $\alpha$ in \Cref{eq:fg}) to 99\% (high values of $\alpha$ in \Cref{eq:fg}) social impact on ``ego'' comes from these five circles.
We note that this effect is purely geometrical and should be recognized in any other topology of the underlying network of social contacts.
 
The maps shown in \Cref{fig:opinions_a3K4,fig:opinions_a3K5} indicate the tendency of the system to ultimately dominate only one opinion for $T>1$. With the available opinions $K>3$, by introducing a higher temperature $T$ in the system, the share of dominant opinion in the entire system is reduced due to more spatially separated actors with different opinions. For the number of opinions $K=5$ and the social temperature $T=4$, this effect is magnified to such an extent that larger clusters in the system disappear, leading to an ever-changing random system state in which none of the available opinions prevail above the noise discrimination level $\theta$.

High social temperature (observed, e.g. before elections) can be identified with high-mood liability, where many often consecutive events cause constant changes in individual opinions. A large part of voters do not know who to vote for, they have just started to think about it, their opinions are poorly established, and the final opinion is determined by random events.

As the exponent $\alpha$ increases in the distance scaling function, the system tends to form more and more clusters. On the other hand, increasing the social temperature $T$ destroys the stability of the smaller clusters that exist in the system, which disappear in favor of the dominant clusters. However, as both values increase---especially for the large number of $K$ opinions available in the system---agents' opinions become highly dispersed and believers of the same opinion are unable to form large clusters. For high values of $K$, $\alpha$ and $T$ the system is fragmented, and the state of the system is represented by dynamically changing and randomly distributed clusters on the grid, and each opinion has a similar number of agents believing in it.

Increasing the discrimination coefficient decreases the importance of small---spatially separated---groups of agents sharing a given opinion in the measurement of opinions. This may contribute to the impression of strong polarization in the system, giving a vision of the presence of well-established divisions in society.
This, in turn, may promote the image of a deep conflict between members of society, for example, between the voters of the two main political forces, creating the impression of a high electoral threshold. This effect is clearly visible in \Cref{fig:most}, where the successive increase in $\theta$ leads to the systematic impression that the opinions of minorities (or at least their spatial dispersion) successively decrease the measured number $\Phi^*$ of the remaining opinions. This effect is best visible in the last row of \Cref{fig:most} (\Cref{fig:probable_th12K5,fig:probable_th25K5,fig:probable_th50K5}), that is, for a large number of available options ($K=5$), where for the threshold $\theta=50$ (\Cref{fig:probable_th50K5}) regardless of the influence of the effective interaction range $\alpha$ or the social temperature $T$, we do not observe a group of followers of the fifth opinion, and followers of the fourth opinion appear only marginally with only one of the examined sets of parameters ($\alpha=3$ and $T=0$).
On the one hand, this can be a hint for manipulators of public opinion, and on the other hand, it can suggest how to effectively oppose such manipulation.

We emphasize that the concept of multiple opinions ($K\ge 3$) seems to be essential for the possibility of speaking about system polarization (which term is probably often overused in binary models of opinion formation). 
Based on the results collected in \Cref{tab:In-to-I} we conclude that the larger $\alpha$ the more influential the nearest neighbors are (see \Cref{sec:beta}).
The level of noise discrimination $\theta$ (allowing for detailed studies of the number $\Phi^*$ of surviving opinions) may be a useful tool for the analysis of social systems not only in models of opinion dynamics. 

The further direction of investigating this model may include checking the computational complexity, that is, the time to reach the equilibrium of the system as dependent on the size of the system or checking the influence of setting $s_i$ and $p_i$ in a way other than proposed here (i.e., taking them from normal instead of uniform distribution, or setting all of them to the same arbitrarily chosen values and reducing their space into only two parameters: $\forall i: s_i=s, p_i=p$).

\begin{acknowledgments}
We thank Krzysztof Ku{\l}akowski for a fruitful discussion and Jacek Tarasiuk for providing a nice random number generator.
We thank anonymous Reviewers for pointing out the deficiency of the original manuscript, possible directions of further research and References \onlinecite{cond-mat_0111494,Laguna_2004} and particularly Reference~\onlinecite{Arnaboldi_2016} as well as for encouraging us to make the source code available online \cite{app_Dworak}.
\end{acknowledgments}

%

\appendix

\section{\label{apx:example_opinions}Examples of final spatial opinion distribution}

\begin{figure}[!hp]
\subcaptionbox{\label{sfig:a3K2T0-1}$T=0$}{\includegraphics[width=.21\textwidth]{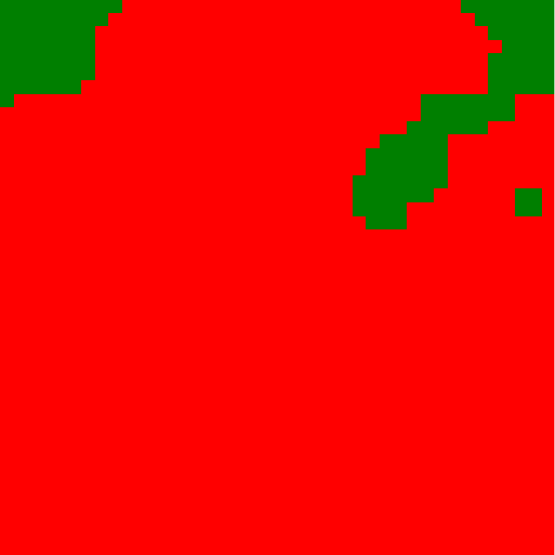}}\hfill 
\subcaptionbox{\label{sfig:a3K2T0-2}$T=0$}{\includegraphics[width=.21\textwidth]{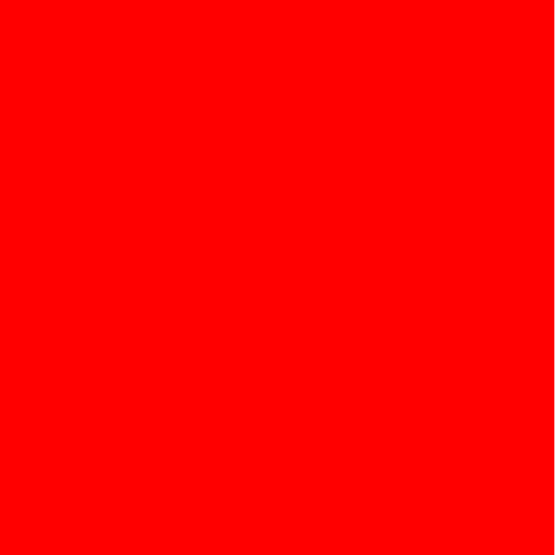}}\\  
\subcaptionbox{\label{sfig:a3K2T1-1}$T=1$}{\includegraphics[width=.21\textwidth]{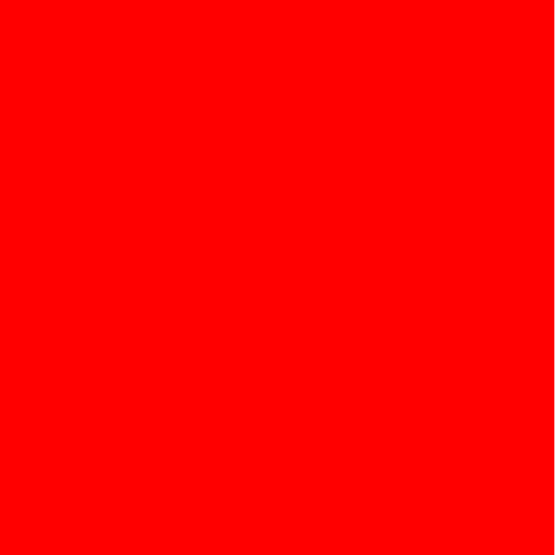}}\hfill 
\subcaptionbox{\label{sfig:a3K2T1-2}$T=1$}{\includegraphics[width=.21\textwidth]{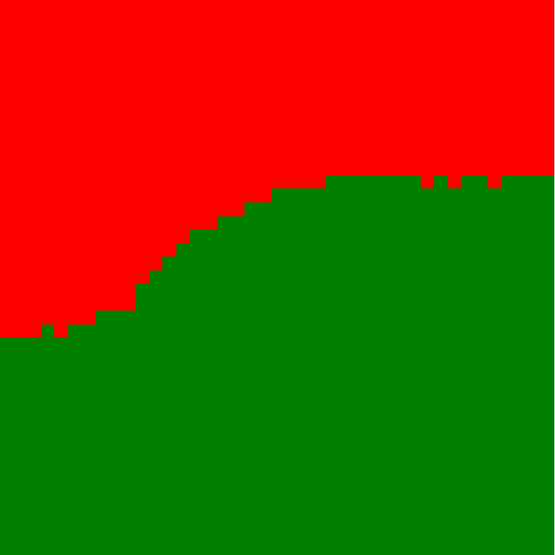}}\\ 
\subcaptionbox{\label{sfig:a3K2T2-1}$T=2$}{\includegraphics[width=.21\textwidth]{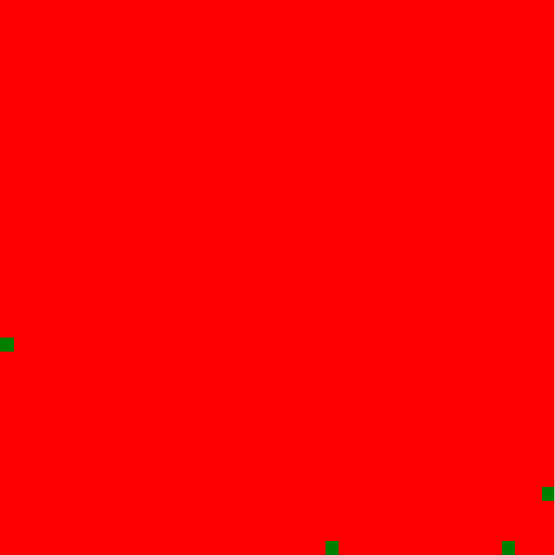}}\hfill 
\subcaptionbox{\label{sfig:a3K2T2-2}$T=2$}{\includegraphics[width=.21\textwidth]{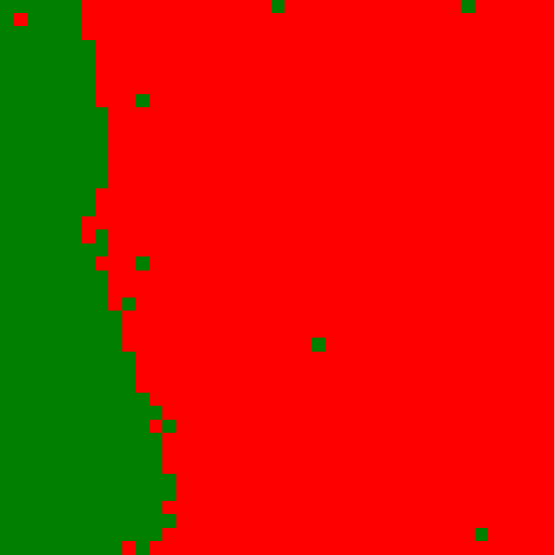}}\\ 
\subcaptionbox{\label{sfig:a3K2T3}  $T=3$}{\includegraphics[width=.21\textwidth]{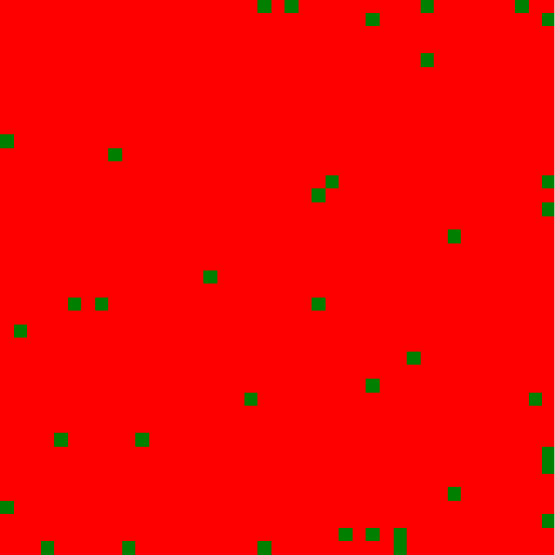}}\\ 
\subcaptionbox{\label{sfig:a3K2T4}  $T=4$}{\includegraphics[width=.21\textwidth]{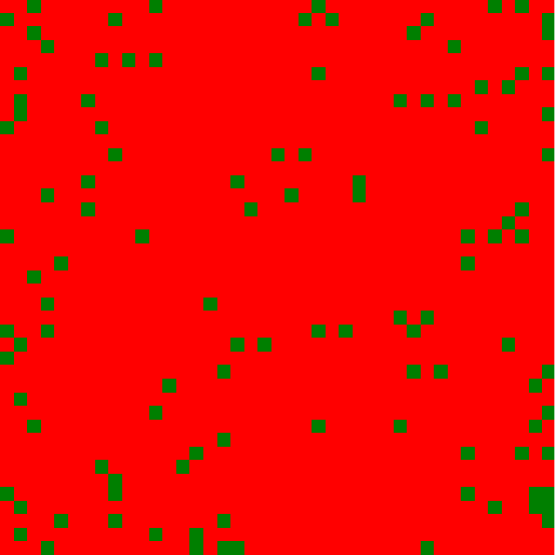}}
\caption{\label{fig:opinions_a3K2}Examples of two most probable spatial distributions of the final opinion after $10^3$ time steps. $L=41$, $\alpha=3$, $K=2$ and various levels of noise $T$.}
\end{figure}

\begin{figure}[!hp]
\subcaptionbox{\label{sfig:a3K3T0}  $T=0$}{\includegraphics[width=.21\textwidth]{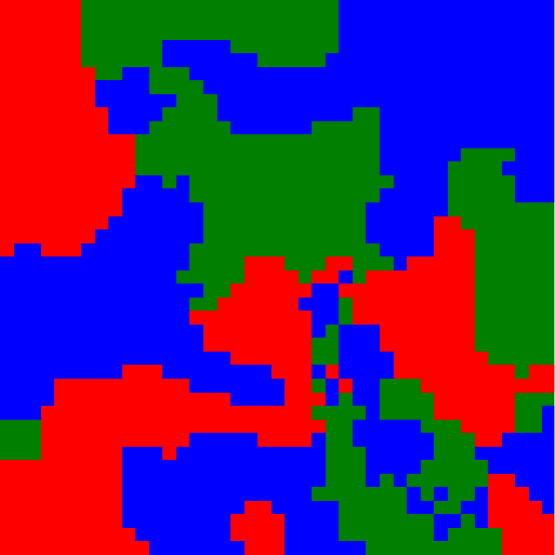}}\\
\subcaptionbox{\label{sfig:a3K3T1-1}$T=1$}{\includegraphics[width=.21\textwidth]{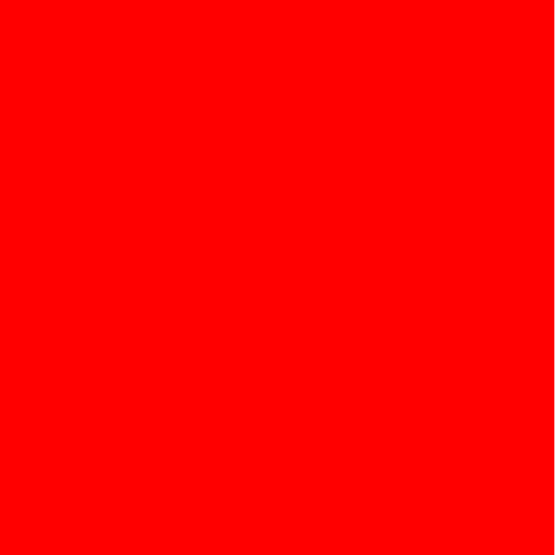}}\hfill
\subcaptionbox{\label{sfig:a3K3T1-2}$T=1$}{\includegraphics[width=.21\textwidth]{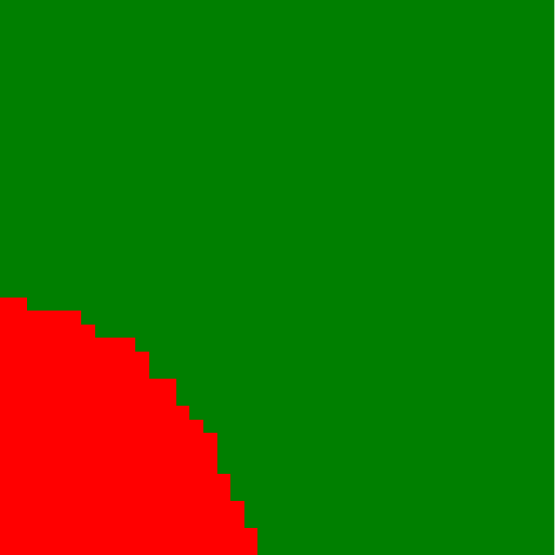}}\\
\subcaptionbox{\label{sfig:a3K3T2-1}$T=2$}{\includegraphics[width=.21\textwidth]{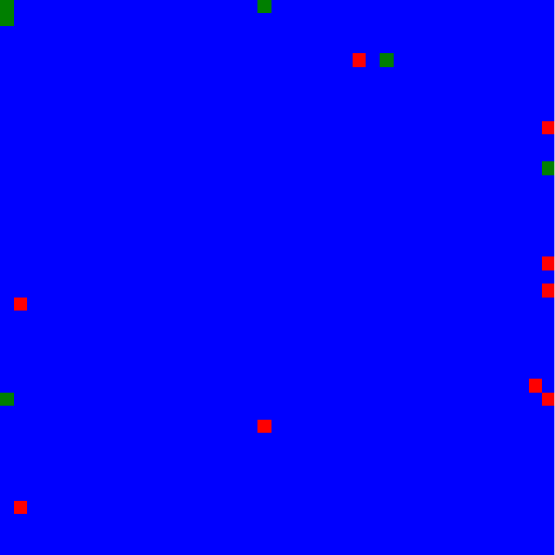}}\hfill
\subcaptionbox{\label{sfig:a3K3T2-2}$T=2$}{\includegraphics[width=.21\textwidth]{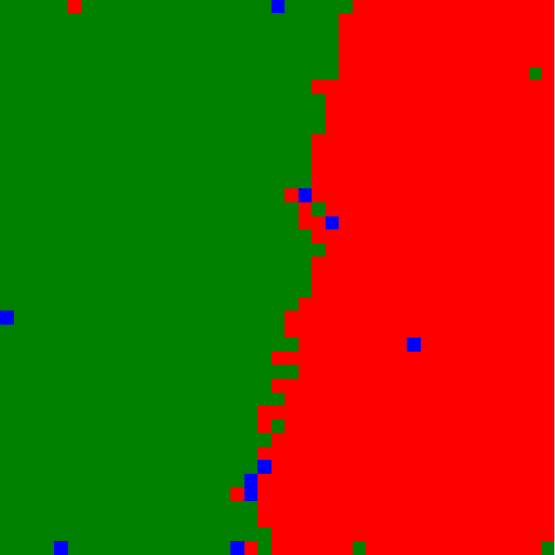}}\\
\subcaptionbox{\label{sfig:a3K3T3-1}$T=3$}{\includegraphics[width=.21\textwidth]{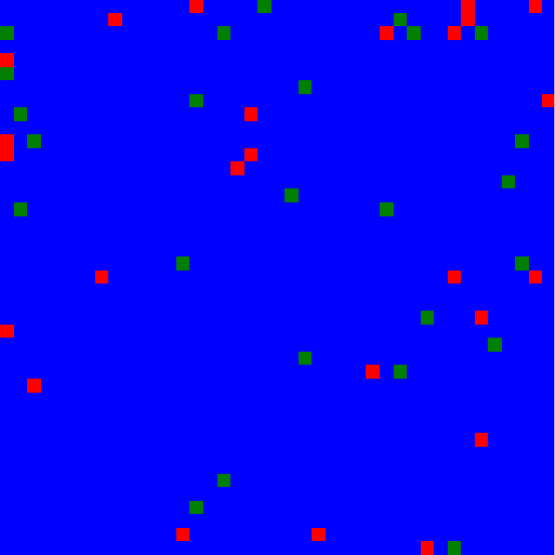}}\hfill
\subcaptionbox{\label{sfig:a3K3T3-2}$T=3$}{\includegraphics[width=.21\textwidth]{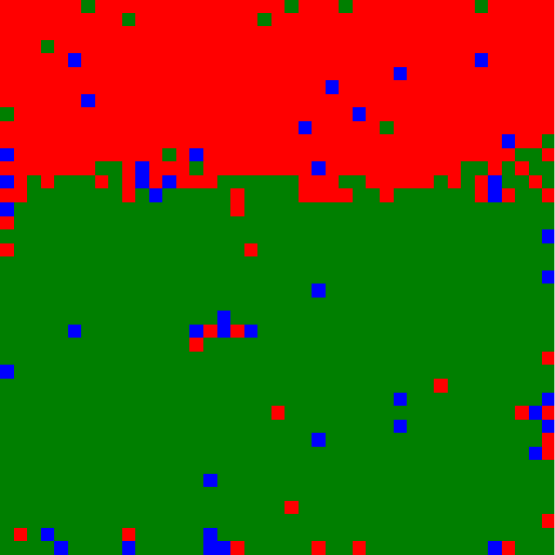}}\\
\subcaptionbox{\label{sfig:a3K3T4-1}$T=4$}{\includegraphics[width=.21\textwidth]{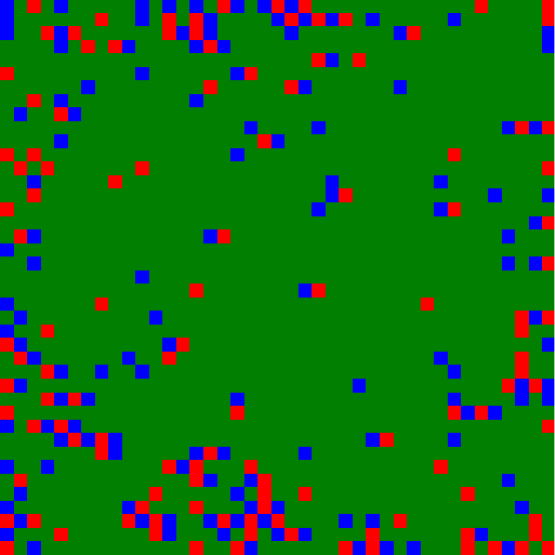}}\hfill
\subcaptionbox{\label{sfig:a3K3T4-2}$T=4$}{\includegraphics[width=.21\textwidth]{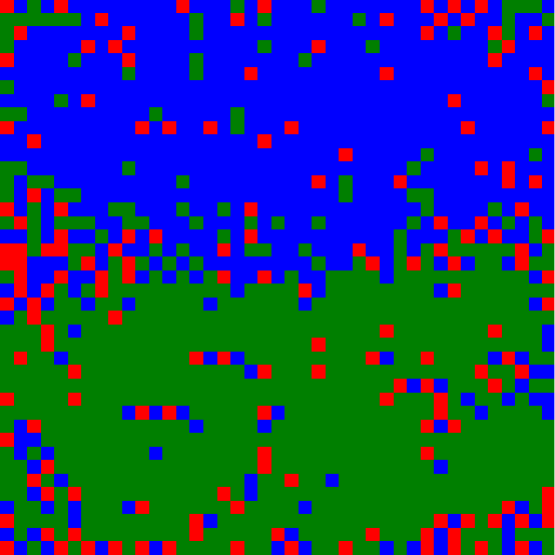}}
\caption{\label{fig:opinions_a3K3}Examples of two most probable spatial distributions of the final opinion after $10^3$ time steps. $L=41$, $\alpha=3$, $K=3$ and various levels of noise $T$.}
\end{figure}

\begin{figure}[!hp]
\subcaptionbox{\label{sfig:a3K5T0-1}$T=0$}{\includegraphics[width=.21\textwidth]{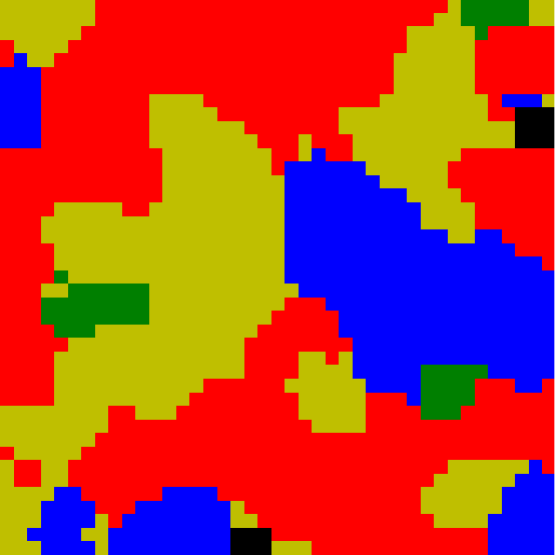}}\hfill
\subcaptionbox{\label{sfig:a3K5T0-2}$T=0$}{\includegraphics[width=.21\textwidth]{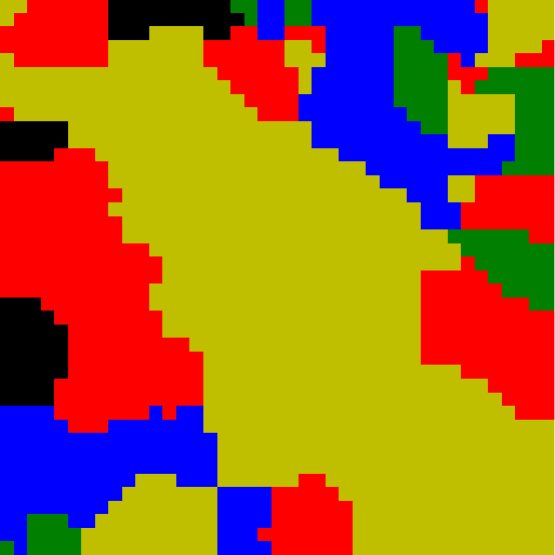}}\\
\subcaptionbox{\label{sfig:a3K5T1-1}$T=1$}{\includegraphics[width=.21\textwidth]{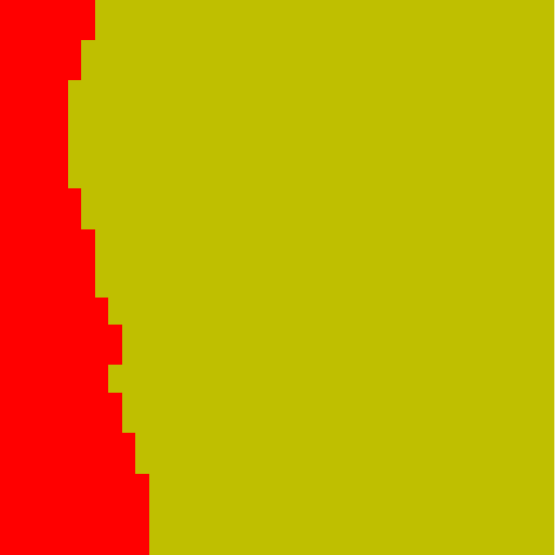}}\hfill
\subcaptionbox{\label{sfig:a3K5T1-2}$T=1$}{\includegraphics[width=.21\textwidth]{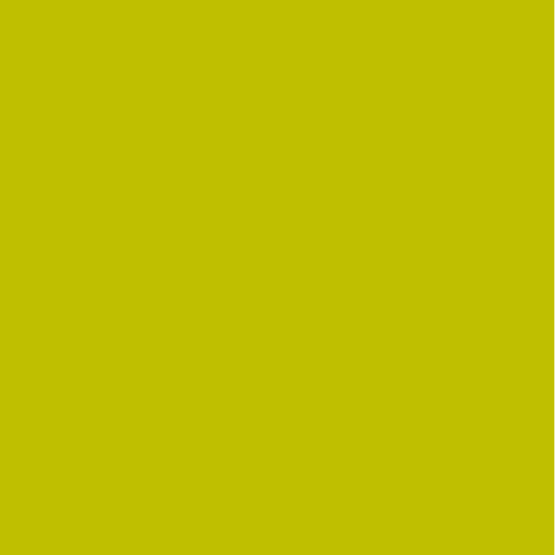}}\\
\subcaptionbox{\label{sfig:a3K5T2-1}$T=2$}{\includegraphics[width=.21\textwidth]{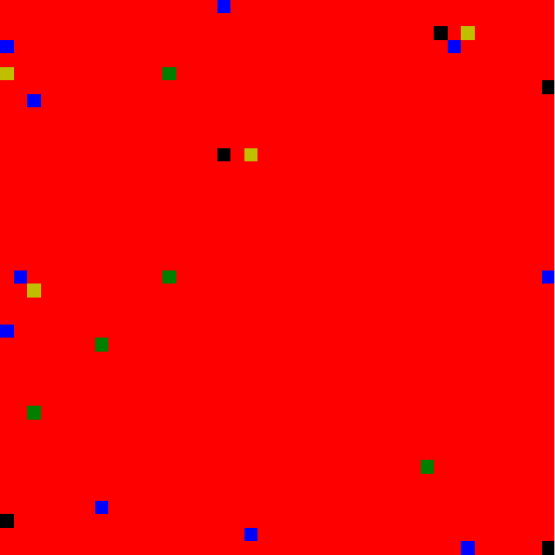}}\hfill
\subcaptionbox{\label{sfig:a3K5T2-2}$T=2$}{\includegraphics[width=.21\textwidth]{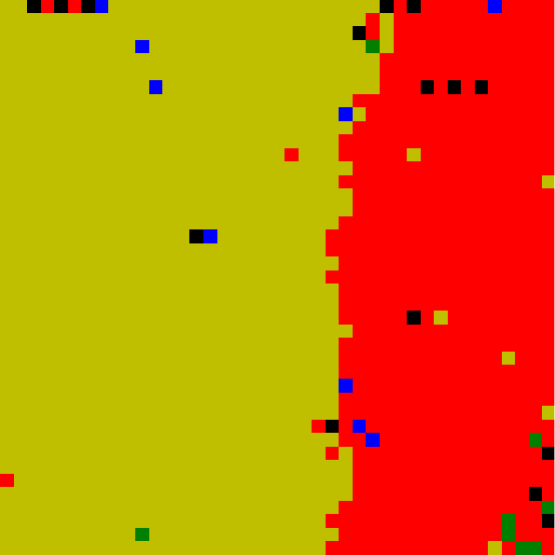}}\\
\subcaptionbox{\label{sfig:a3K5T3-1}$T=3$}{\includegraphics[width=.21\textwidth]{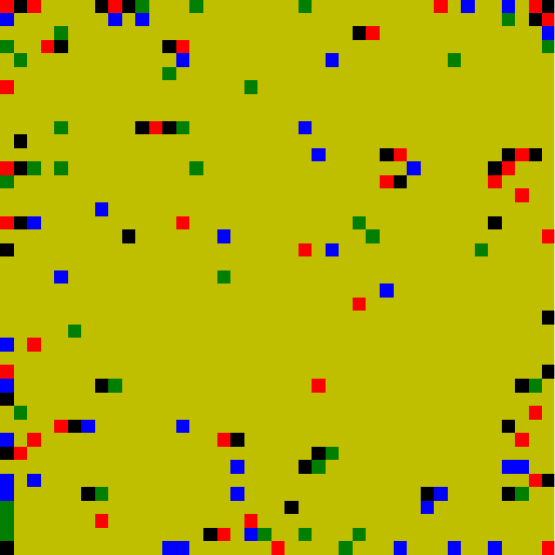}}\hfill
\subcaptionbox{\label{sfig:a3K5T3-2}$T=3$}{\includegraphics[width=.21\textwidth]{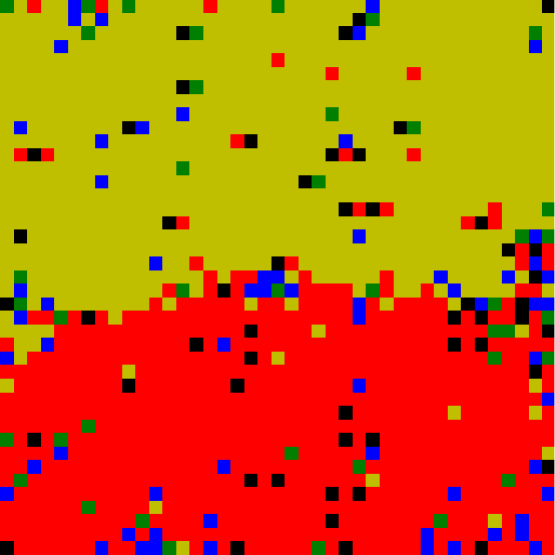}}\\
\subcaptionbox{\label{sfig:a3K5T4-1}$T=4$}{\includegraphics[width=.21\textwidth]{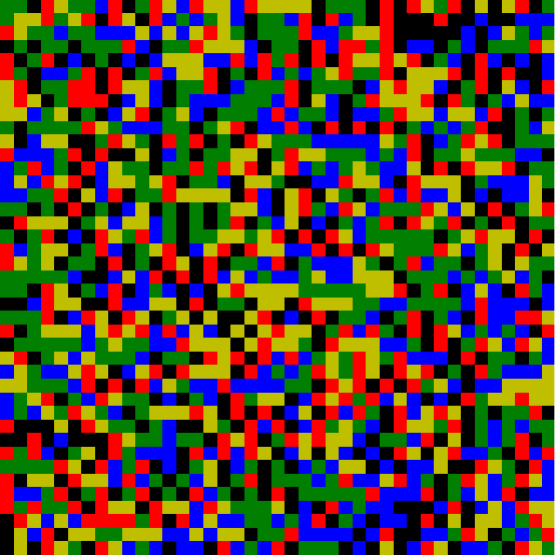}}\hfill
\subcaptionbox{\label{sfig:a3K5T4-2}$T=4$}{\includegraphics[width=.21\textwidth]{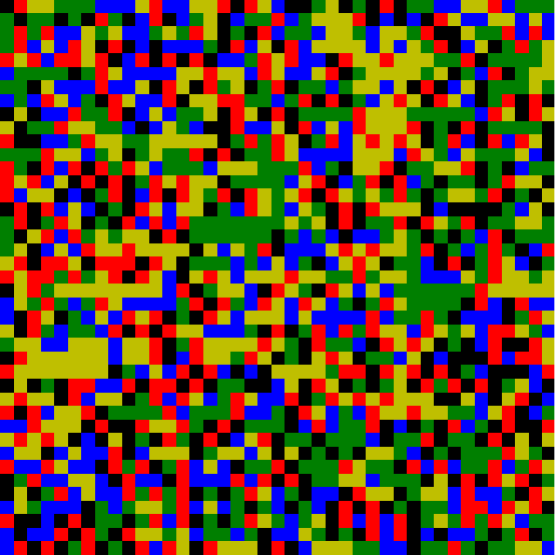}}
\caption{\label{fig:opinions_a3K5}Examples of two most probable spatial distributions of the final opinion after $10^3$ time steps. $L=41$, $\alpha=3$, $K=5$ and various levels of noise $T$.}
\end{figure}

\begin{figure}[!hp]
\subcaptionbox{\label{sfig:a4K2T0}  $T=0$}{\includegraphics[width=.21\textwidth]{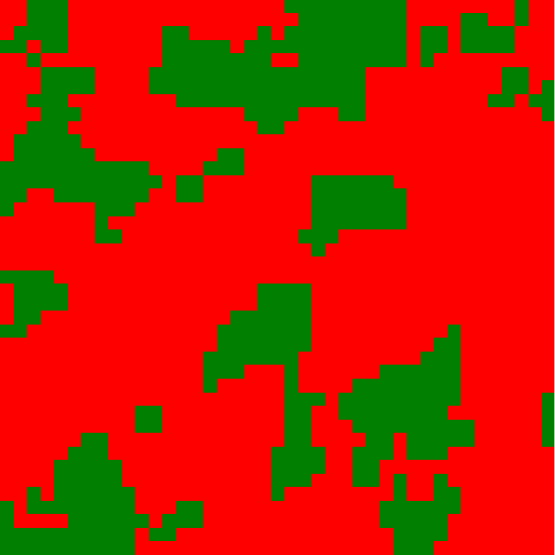}}\\
\subcaptionbox{\label{sfig:a4K2T1-1}$T=1$}{\includegraphics[width=.21\textwidth]{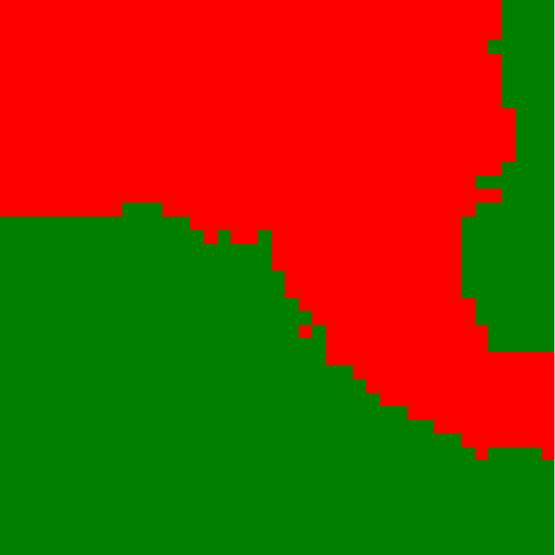}}\hfill
\subcaptionbox{\label{sfig:a4K2T1-2}$T=1$}{\includegraphics[width=.21\textwidth]{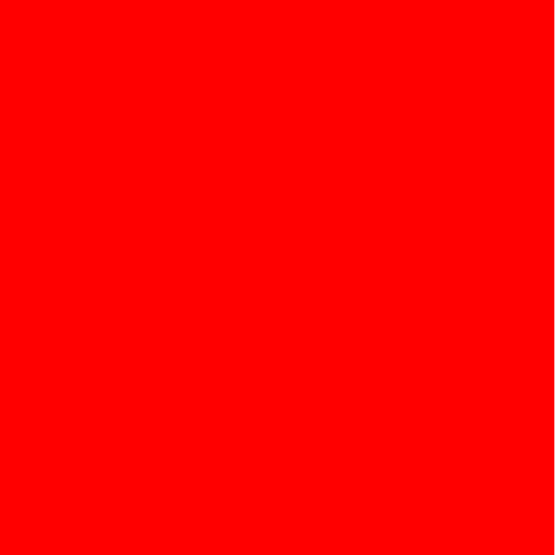}}\\
\subcaptionbox{\label{sfig:a4K2T2-1}$T=2$}{\includegraphics[width=.21\textwidth]{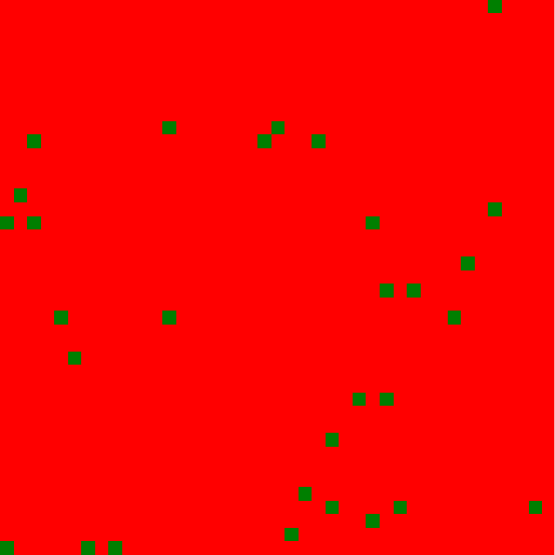}}\hfill
\subcaptionbox{\label{sfig:a4K2T2-2}$T=2$}{\includegraphics[width=.21\textwidth]{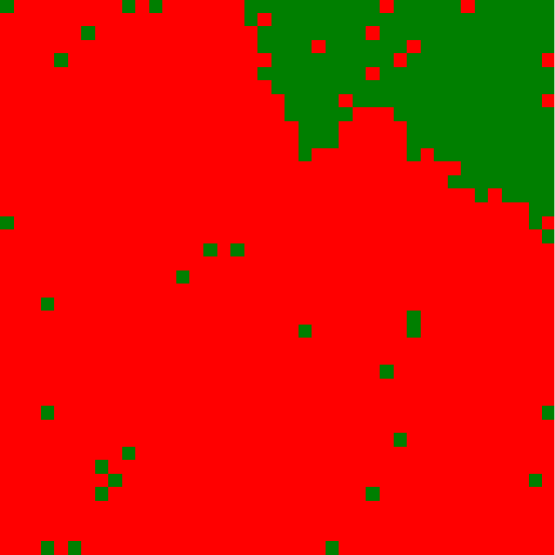}}\\
\subcaptionbox{\label{sfig:a4K2T3-1}$T=3$}{\includegraphics[width=.21\textwidth]{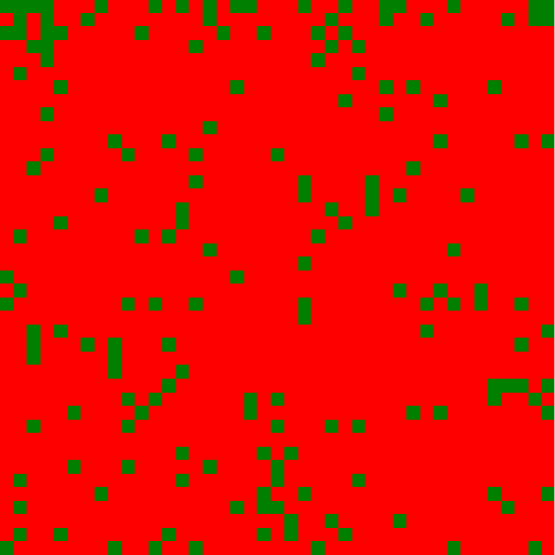}}\hfill
\subcaptionbox{\label{sfig:a4K2T3-2}$T=3$}{\includegraphics[width=.21\textwidth]{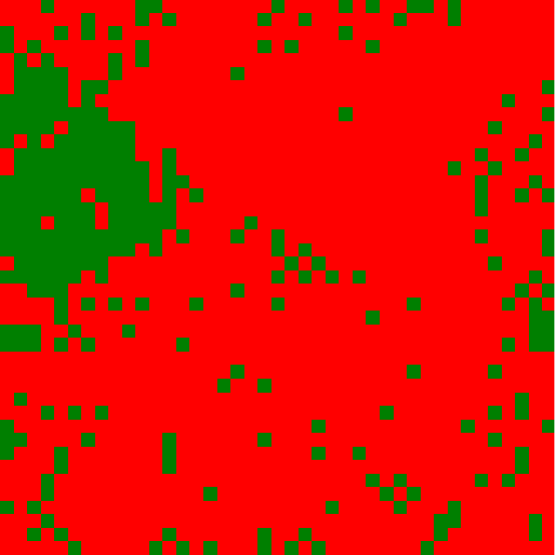}}\\ 
\subcaptionbox{\label{sfig:a4K2T4-1}$T=4$}{\includegraphics[width=.21\textwidth]{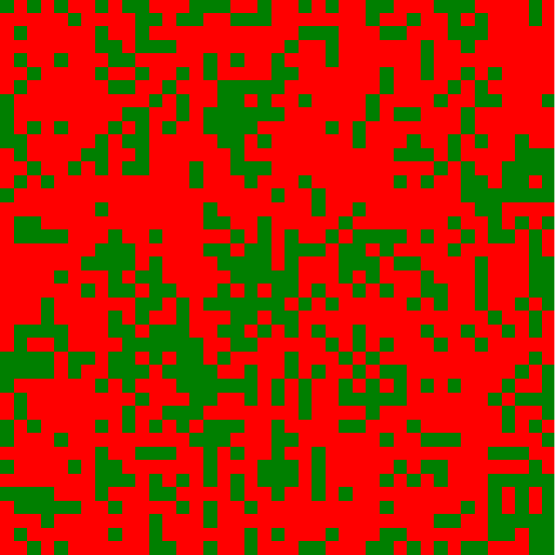}}\hfill
\subcaptionbox{\label{sfig:a4K2T4-2}$T=4$}{\includegraphics[width=.21\textwidth]{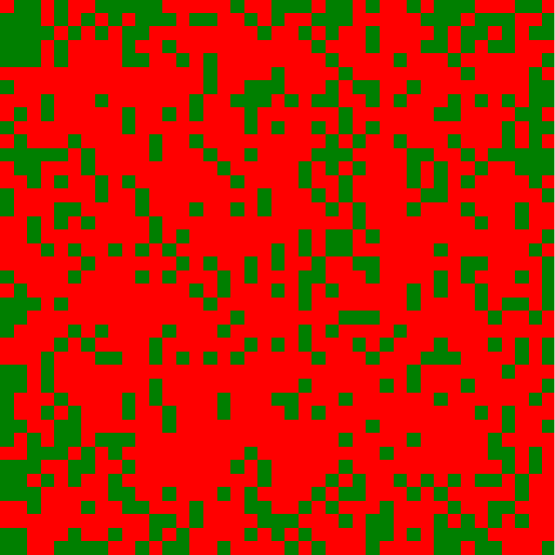}}
\caption{\label{fig:opinions_a4K2}Examples of two most probable spatial distributions of the final opinion after $10^3$ time steps. $L=41$, $\alpha=4$, $K=2$ and various levels of noise $T$.}
\end{figure}

\begin{figure}[!hp]
\subcaptionbox{\label{sfig:a4K3T0}  $T=0$}{\includegraphics[width=.21\textwidth]{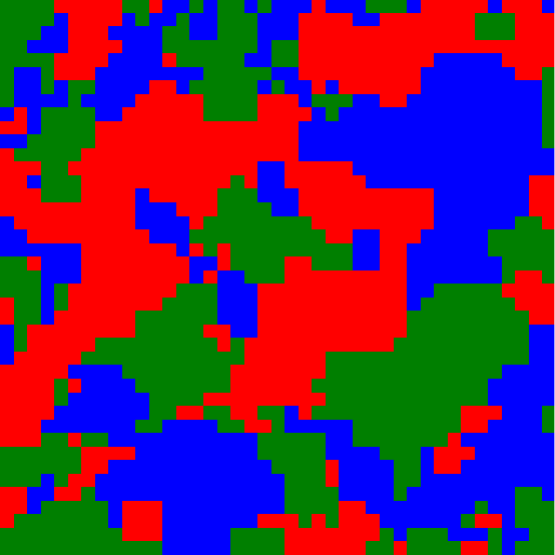}}\\
\subcaptionbox{\label{sfig:a4K3T1-1}$T=1$}{\includegraphics[width=.21\textwidth]{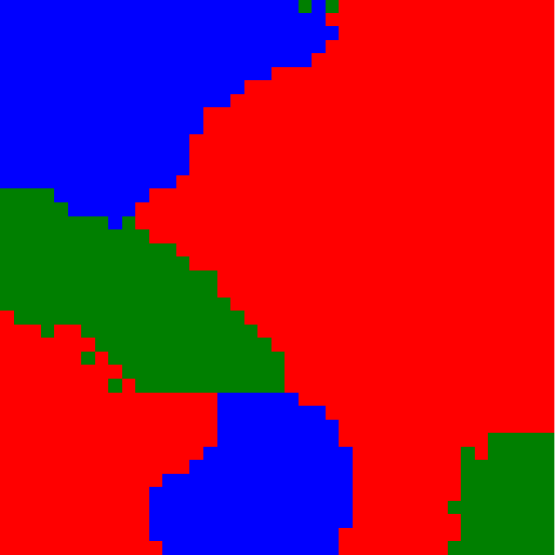}}\hfill
\subcaptionbox{\label{sfig:a4K3T1-2}$T=1$}{\includegraphics[width=.21\textwidth]{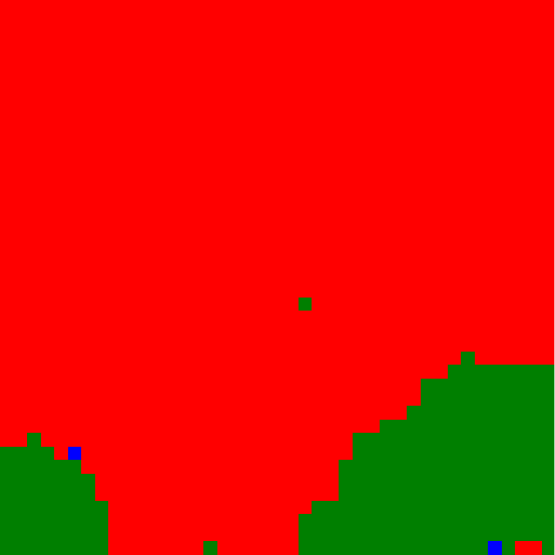}}\\
\subcaptionbox{\label{sfig:a4K3T2-1}$T=2$}{\includegraphics[width=.21\textwidth]{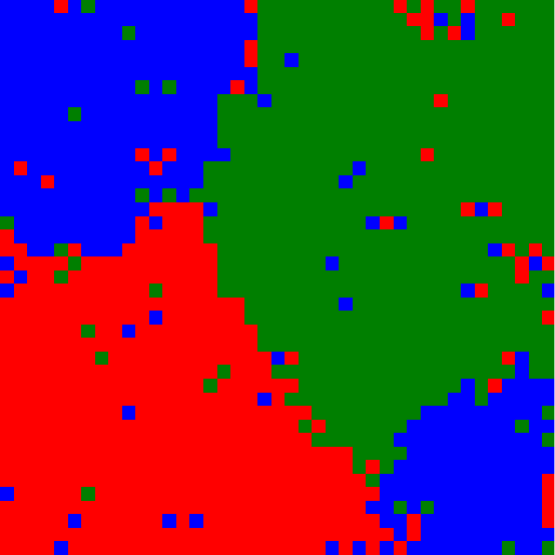}}\hfill
\subcaptionbox{\label{sfig:a4K3T2-2}$T=2$}{\includegraphics[width=.21\textwidth]{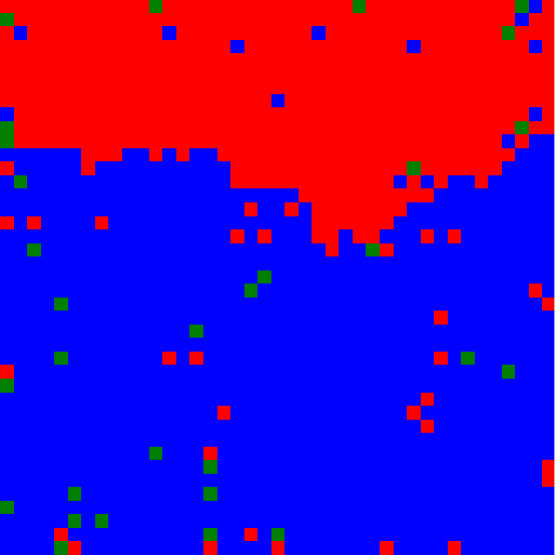}}\\
\subcaptionbox{\label{sfig:a4K3T3-1}$T=3$}{\includegraphics[width=.21\textwidth]{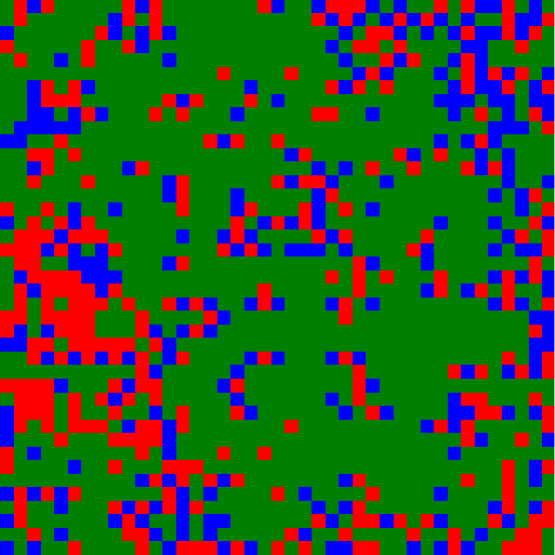}}\hfill
\subcaptionbox{\label{sfig:a4K3T3-2}$T=3$}{\includegraphics[width=.21\textwidth]{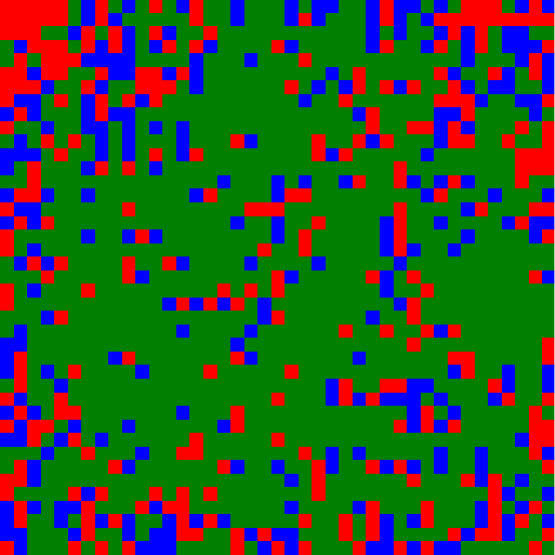}}\\
\subcaptionbox{\label{sfig:a4K3T4-1}$T=4$}{\includegraphics[width=.21\textwidth]{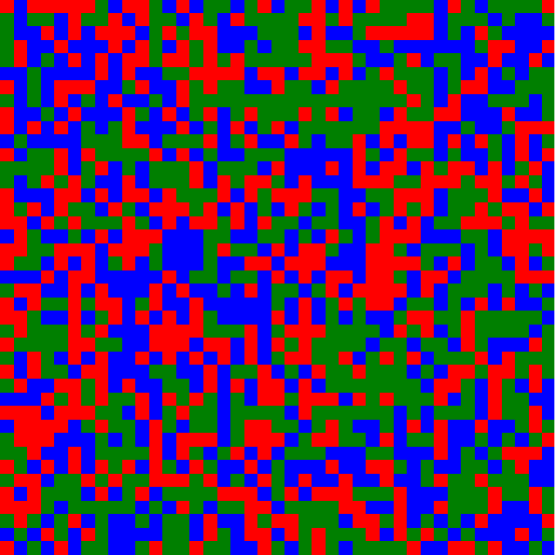}}\hfill
\subcaptionbox{\label{sfig:a4K3T4-2}$T=4$}{\includegraphics[width=.21\textwidth]{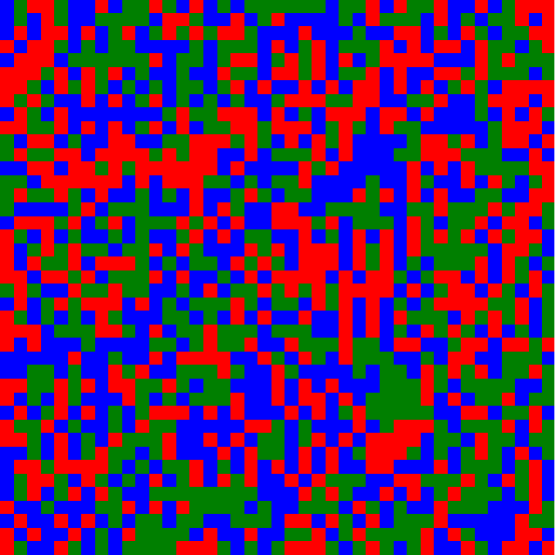}}
\caption{\label{fig:opinions_a4K3}Examples of two most probable spatial distributions of the final opinion after $10^3$ time steps. $L=41$, $\alpha=4$, $K=3$ and various levels of noise $T$.}
\end{figure}

\begin{figure}[htbp]
\subcaptionbox{\label{sfig:a4K4T0-1}$T=0$}{\includegraphics[width=.21\textwidth]{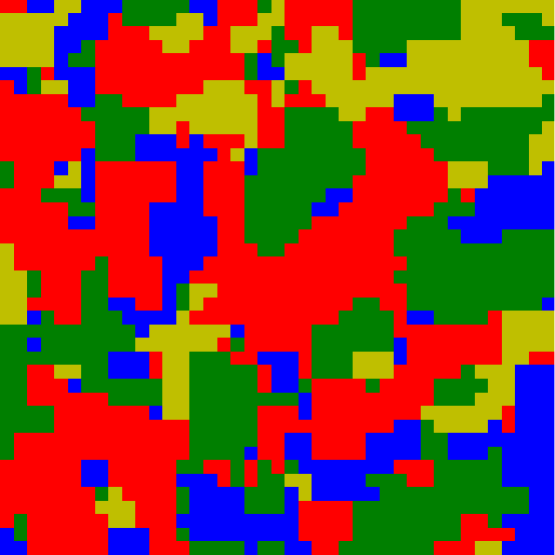}}\hfill
\subcaptionbox{\label{sfig:a4K4T0-2}$T=0$}{\includegraphics[width=.21\textwidth]{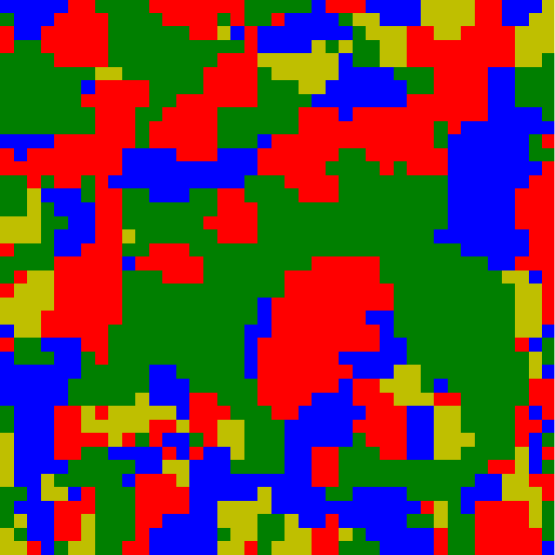}}\\
\subcaptionbox{\label{sfig:a4K4T1-1}$T=1$}{\includegraphics[width=.21\textwidth]{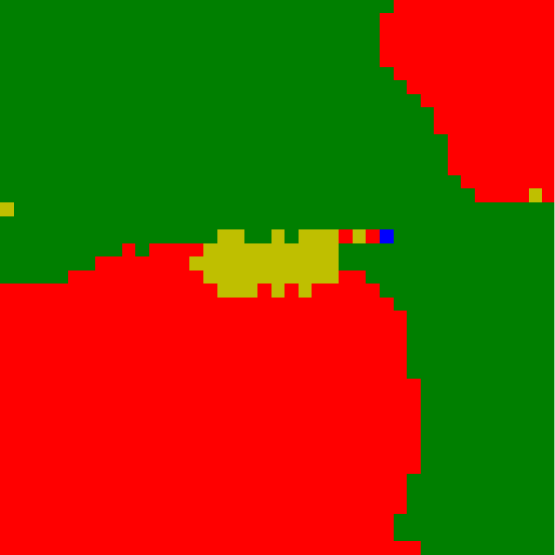}}\hfill
\subcaptionbox{\label{sfig:a4K4T1-2}$T=1$}{\includegraphics[width=.21\textwidth]{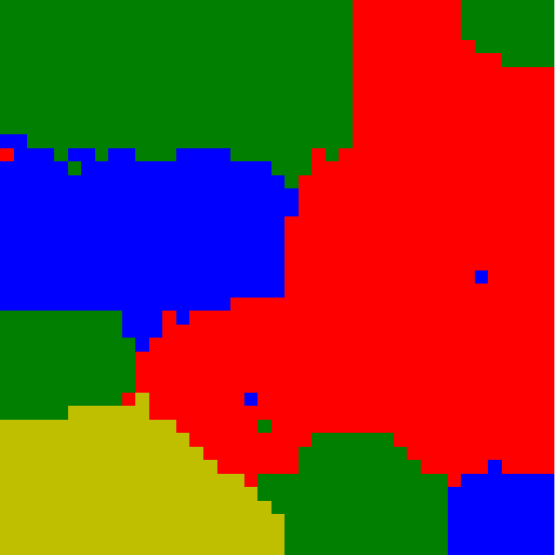}}\\
\subcaptionbox{\label{sfig:a4K4T2-1}$T=2$}{\includegraphics[width=.21\textwidth]{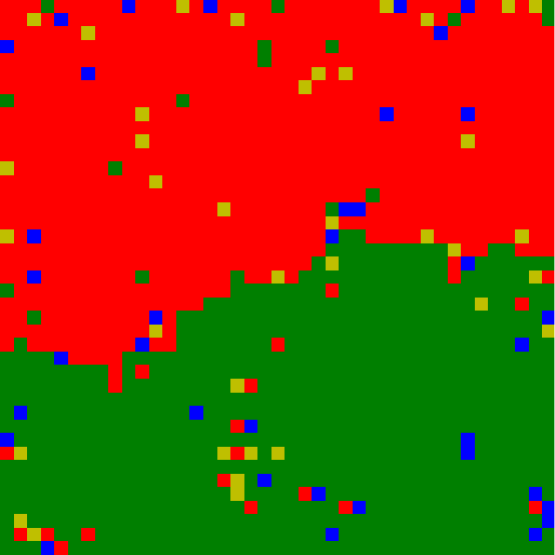}}\hfill
\subcaptionbox{\label{sfig:a4K4T2-2}$T=2$}{\includegraphics[width=.21\textwidth]{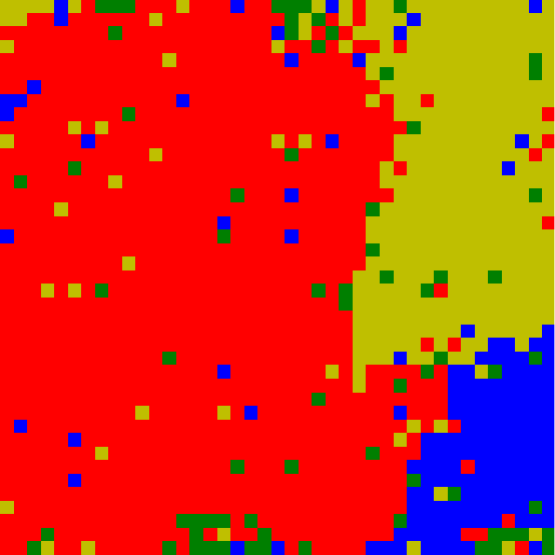}}\\
\subcaptionbox{\label{sfig:a4K4T3-1}$T=3$}{\includegraphics[width=.21\textwidth]{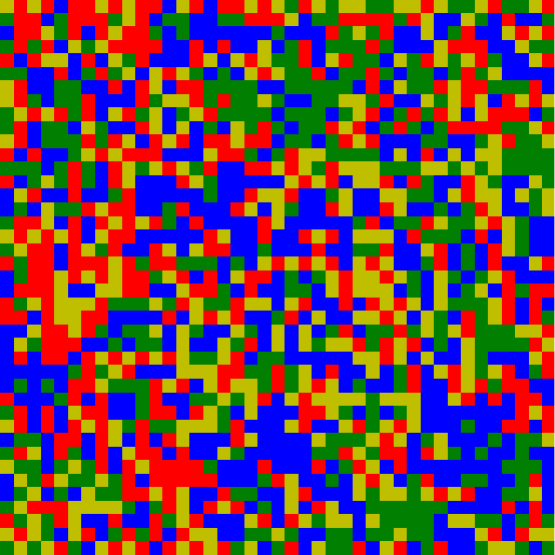}}\hfill
\subcaptionbox{\label{sfig:a4K4T3-2}$T=3$}{\includegraphics[width=.21\textwidth]{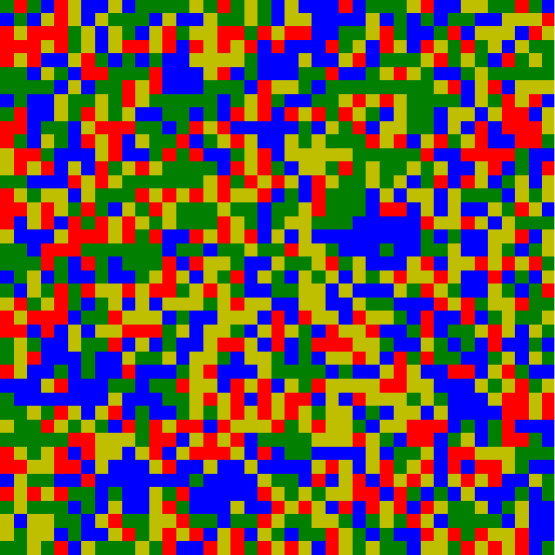}}\\
\subcaptionbox{\label{sfig:a4K4T4-1}$T=4$}{\includegraphics[width=.21\textwidth]{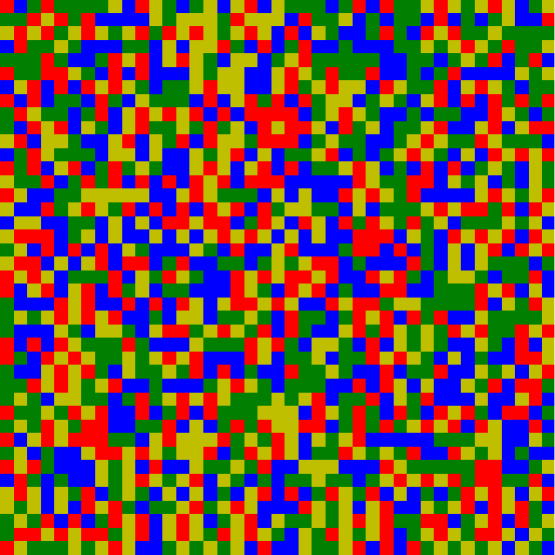}}\hfill
\subcaptionbox{\label{sfig:a4K4T4-2}$T=4$}{\includegraphics[width=.21\textwidth]{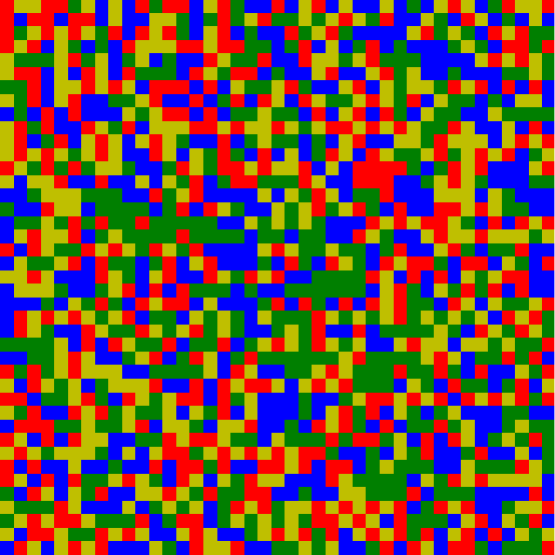}}\\
\caption{\label{fig:opinions_a4K4}Examples of two most probable spatial distributions of the final opinion after $10^3$ time steps. $L=41$, $\alpha=4$, $K=4$ and various levels of noise $T$.}
\end{figure}

\begin{figure}[!hp]
\subcaptionbox{\label{sfig:a4K5T0-1}$T=0$}{\includegraphics[width=.21\textwidth]{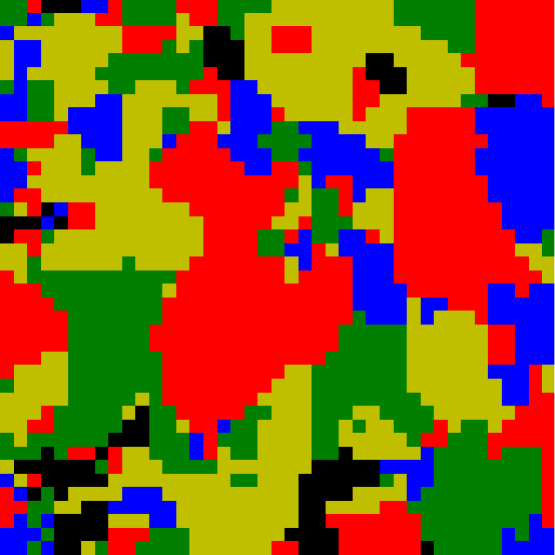}}\hfill
\subcaptionbox{\label{sfig:a4K5T0-2}$T=0$}{\includegraphics[width=.21\textwidth]{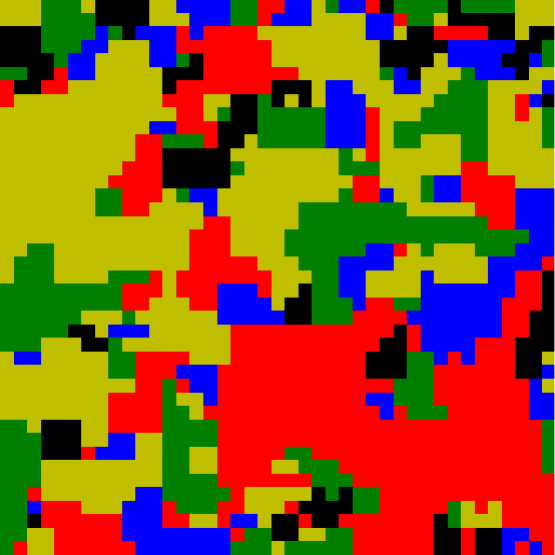}}\\
\subcaptionbox{\label{sfig:a4K5T1-1}$T=1$}{\includegraphics[width=.21\textwidth]{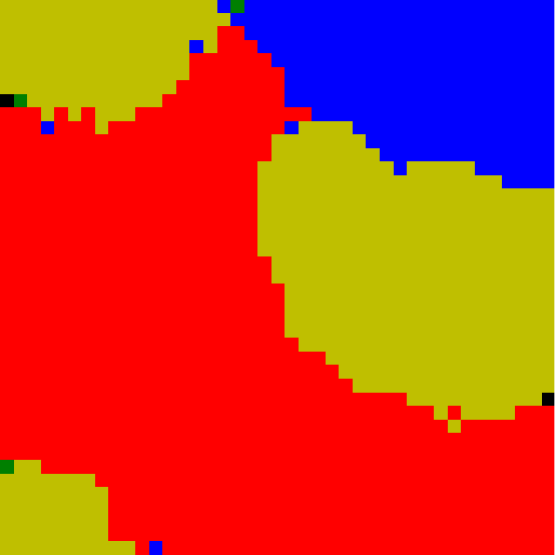}}\hfill
\subcaptionbox{\label{sfig:a4K5T1-2}$T=1$}{\includegraphics[width=.21\textwidth]{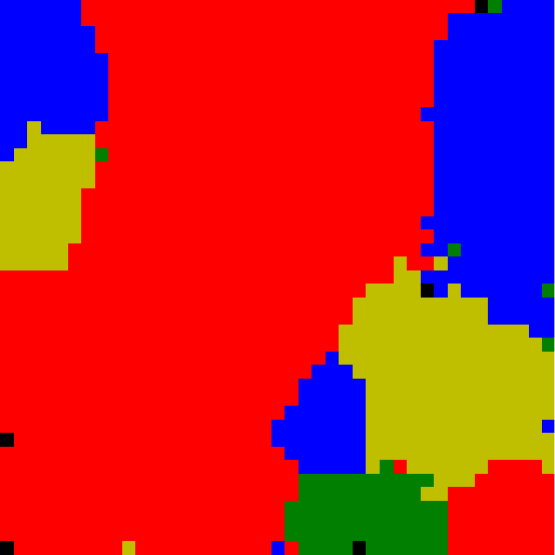}}\\
\subcaptionbox{\label{sfig:a4K5T2-1}$T=2$}{\includegraphics[width=.21\textwidth]{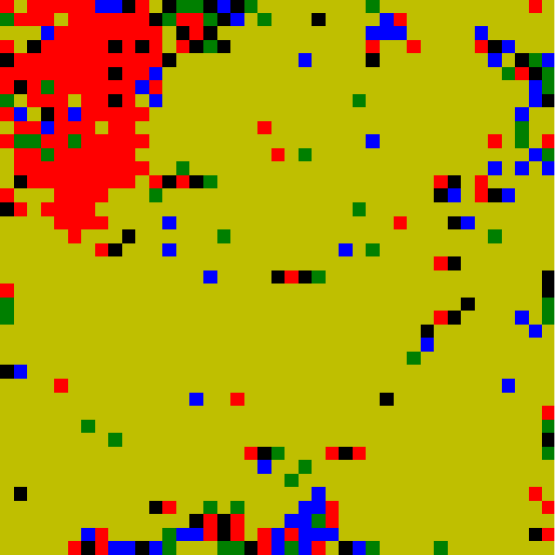}}\hfill
\subcaptionbox{\label{sfig:a4K5T2-2}$T=2$}{\includegraphics[width=.21\textwidth]{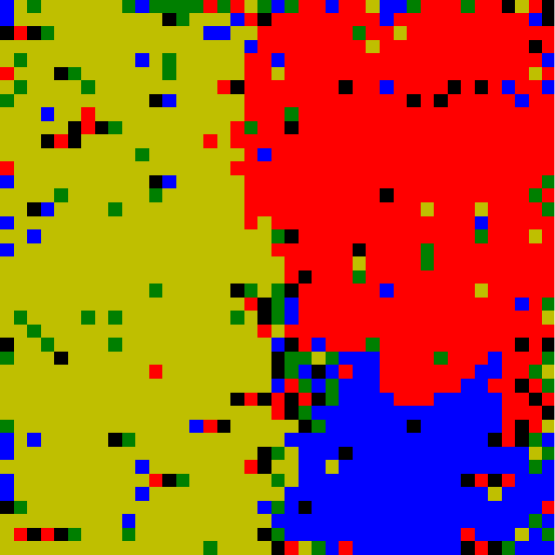}}\\
\subcaptionbox{\label{sfig:a4K5T3-1}$T=3$}{\includegraphics[width=.21\textwidth]{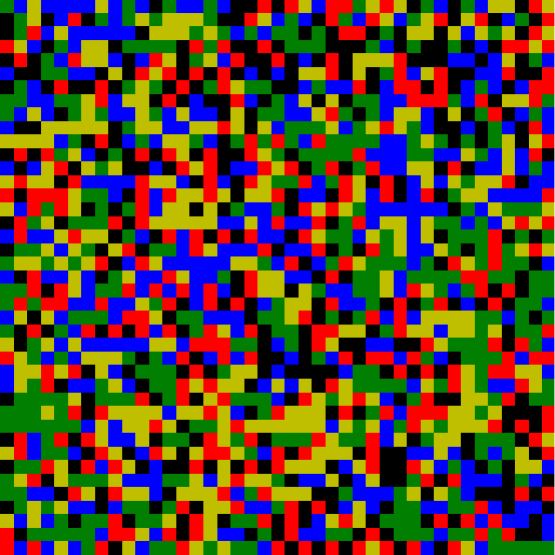}}\hfill
\subcaptionbox{\label{sfig:a4K5T3-2}$T=3$}{\includegraphics[width=.21\textwidth]{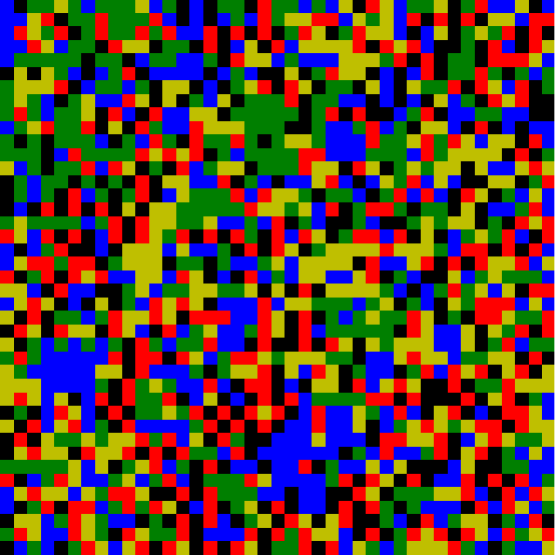}}\\
\subcaptionbox{\label{sfig:a4K5T4-1}$T=4$}{\includegraphics[width=.21\textwidth]{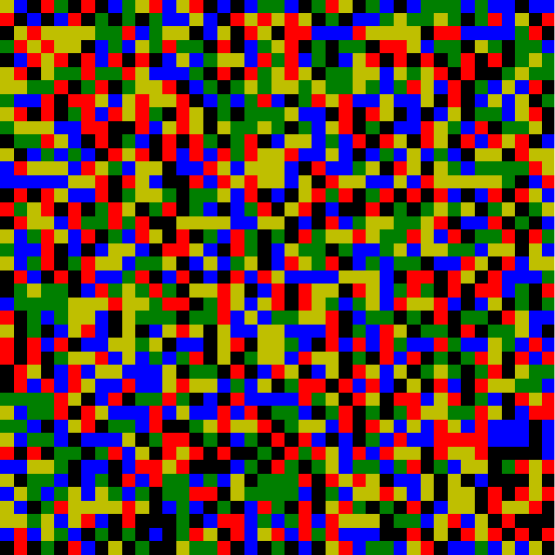}}\hfill
\subcaptionbox{\label{sfig:a4K5T4-2}$T=4$}{\includegraphics[width=.21\textwidth]{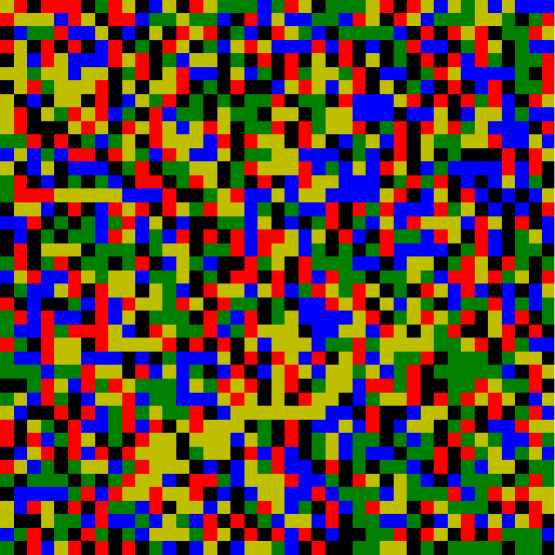}}\\
\caption{\label{fig:opinions_a4K5}Examples of two most probable spatial distributions of the final opinion after $10^3$ time steps. $L=41$, $\alpha=4$, $K=5$ and various levels of noise $T$.}
\end{figure}

Examples of the two most probable spatial distributions of the final opinion after $t=10^3$ time steps of the system evolution for various noise levels $T$.
The system contains $L^2=41^2$ actors.
The exponent of the distance scaling function $\alpha=3$ and the number of available opinions $K=2$ (\Cref{fig:opinions_a3K2}), 
$\alpha=3$ and $K=2$ (\Cref{fig:opinions_a3K3}),
$\alpha=3$ and $K=5$ (\Cref{fig:opinions_a3K5}),
$\alpha=4$ and $K=2$ (\Cref{fig:opinions_a4K2}),
$\alpha=4$ and $K=3$ (\Cref{fig:opinions_a4K3}),
$\alpha=4$ and $K=4$ (\Cref{fig:opinions_a4K4}),
$\alpha=4$ and $K=5$ (\Cref{fig:opinions_a4K5}).

\begin{figure*}[htbp]
\subcaptionbox{$\alpha = 2$, $K = 2$\label{sfig:clust_th12a2K2}}{\includegraphics[width=.24\textwidth, ]{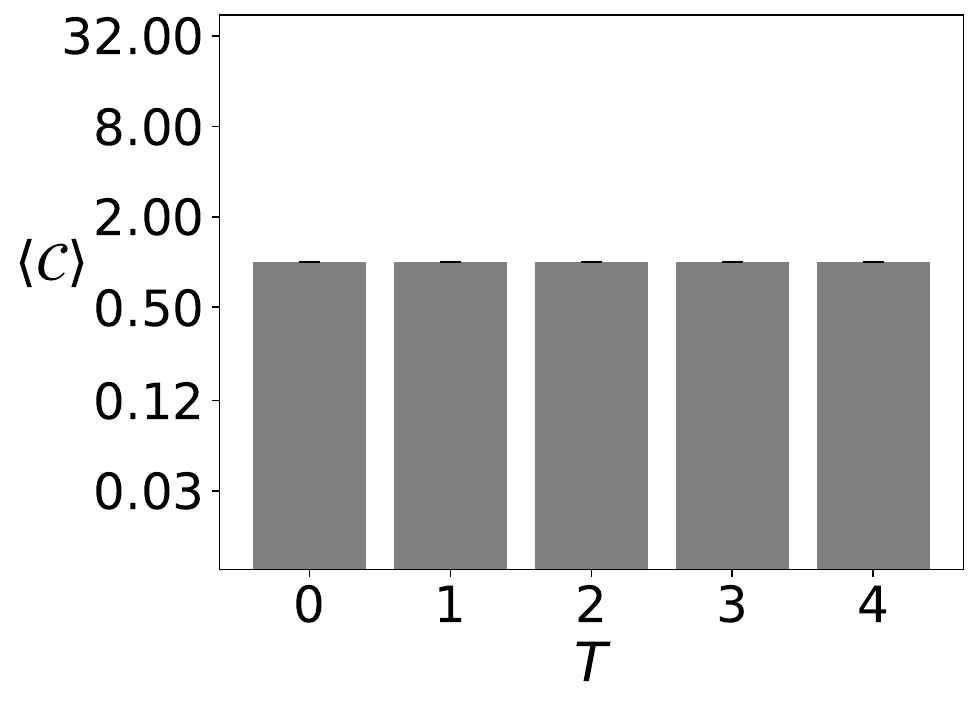}}\hfill%
\subcaptionbox{$\alpha = 2$, $K = 3$\label{sfig:clust_th12a2K3}}{\includegraphics[width=.24\textwidth, ]{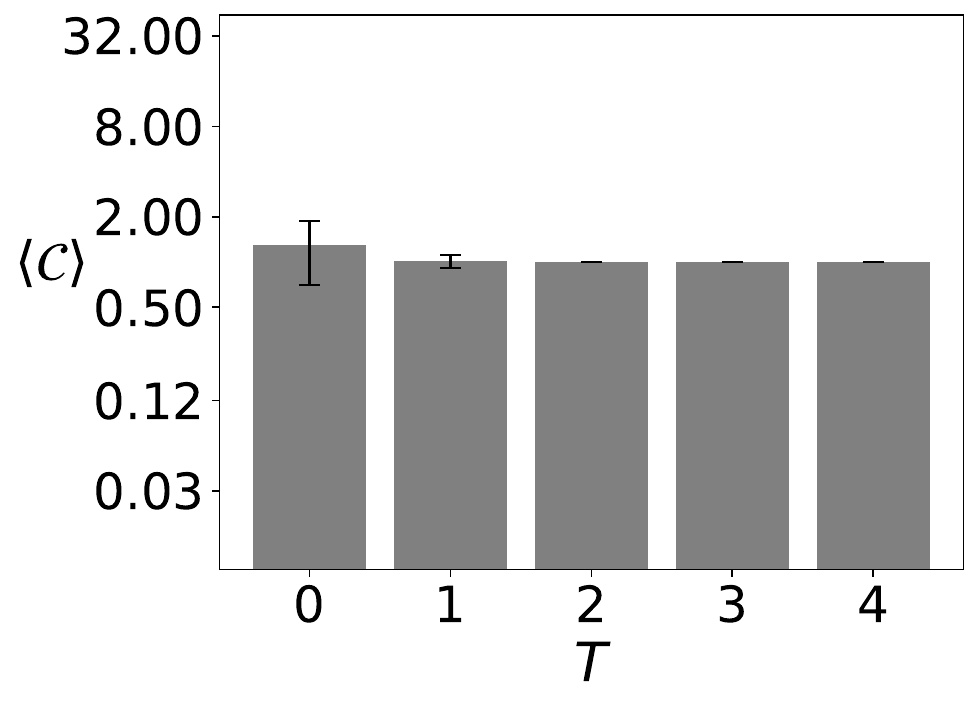}}\hfill%
\subcaptionbox{$\alpha = 2$, $K = 4$\label{sfig:clust_th12a2K4}}{\includegraphics[width=.24\textwidth, ]{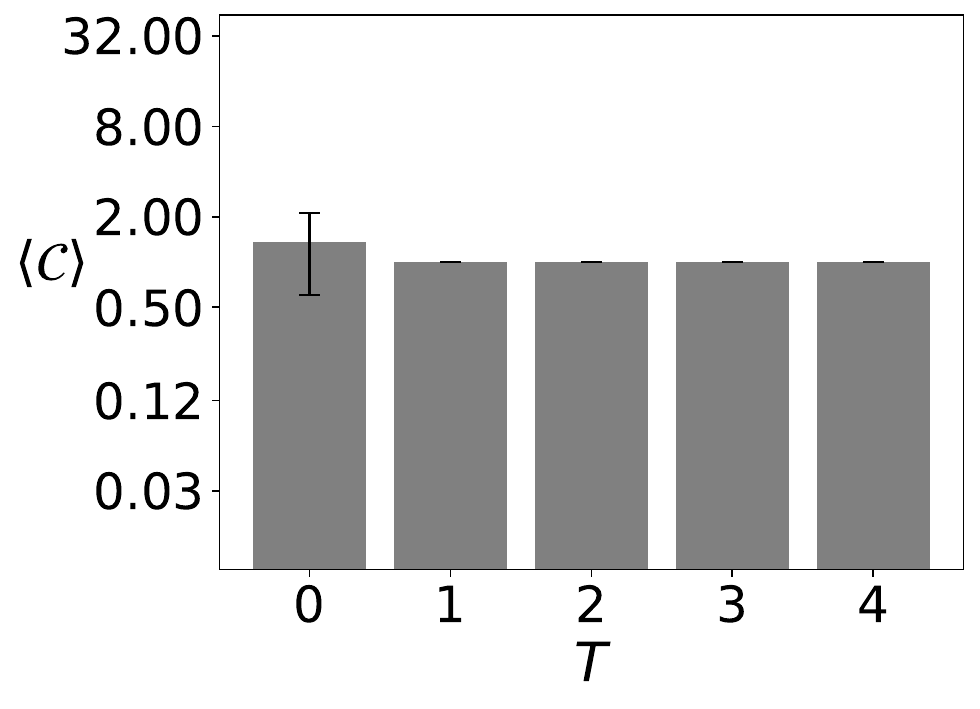}}\hfill%
\subcaptionbox{$\alpha = 2$, $K = 5$\label{sfig:clust_th12a2K5}}{\includegraphics[width=.24\textwidth, ]{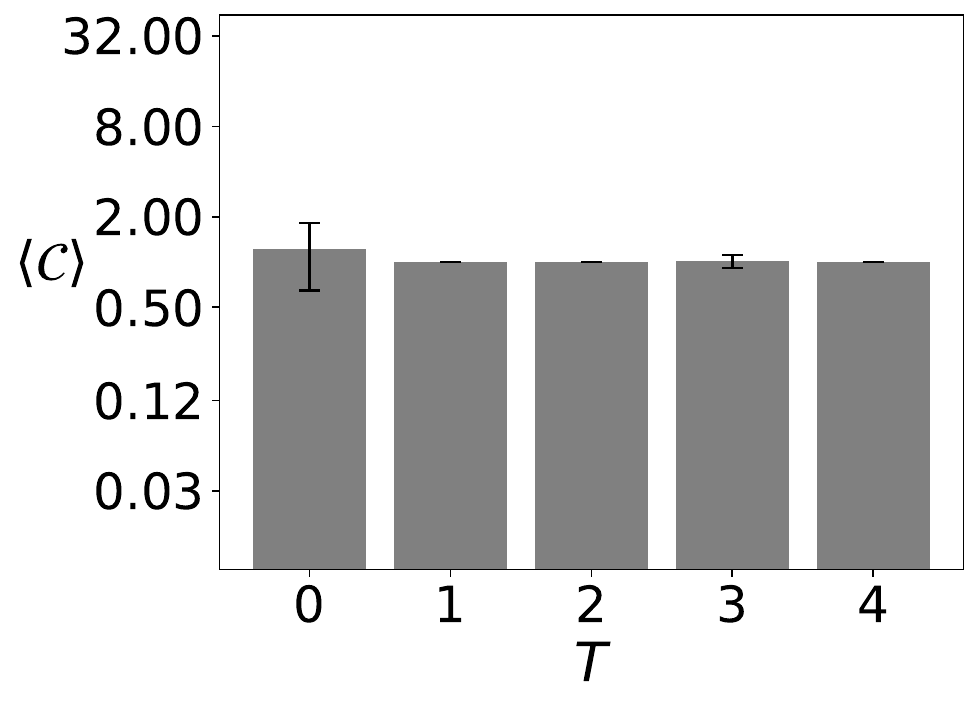}}\\
\subcaptionbox{$\alpha = 3$, $K = 2$\label{sfig:clust_th12a3K2}}{\includegraphics[width=.24\textwidth, ]{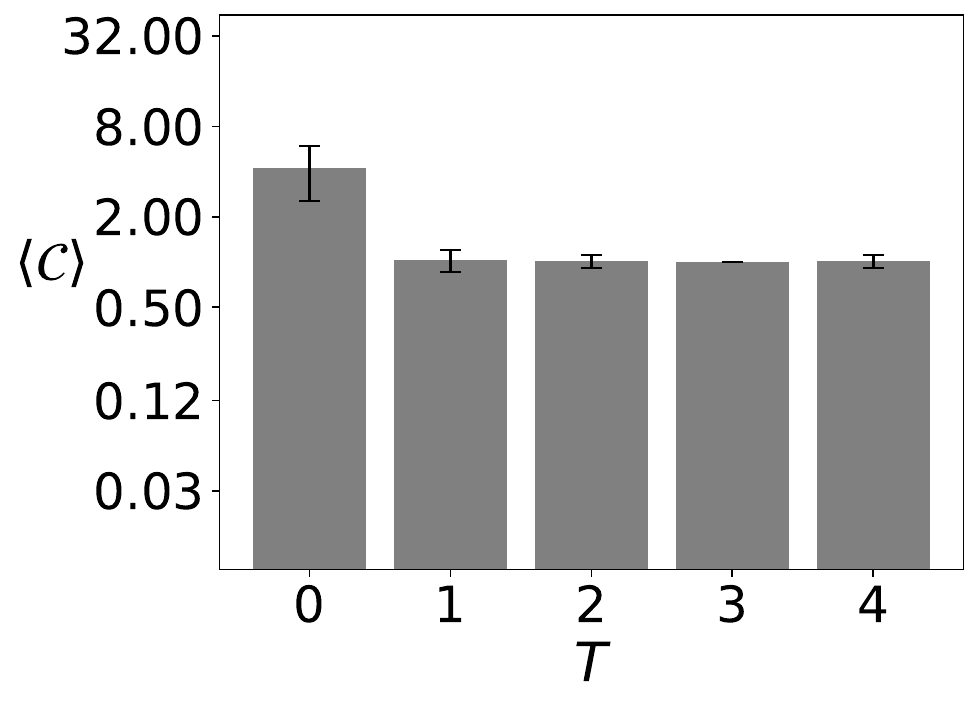}}\hfill%
\subcaptionbox{$\alpha = 3$, $K = 3$\label{sfig:clust_th12a3K3}}{\includegraphics[width=.24\textwidth, ]{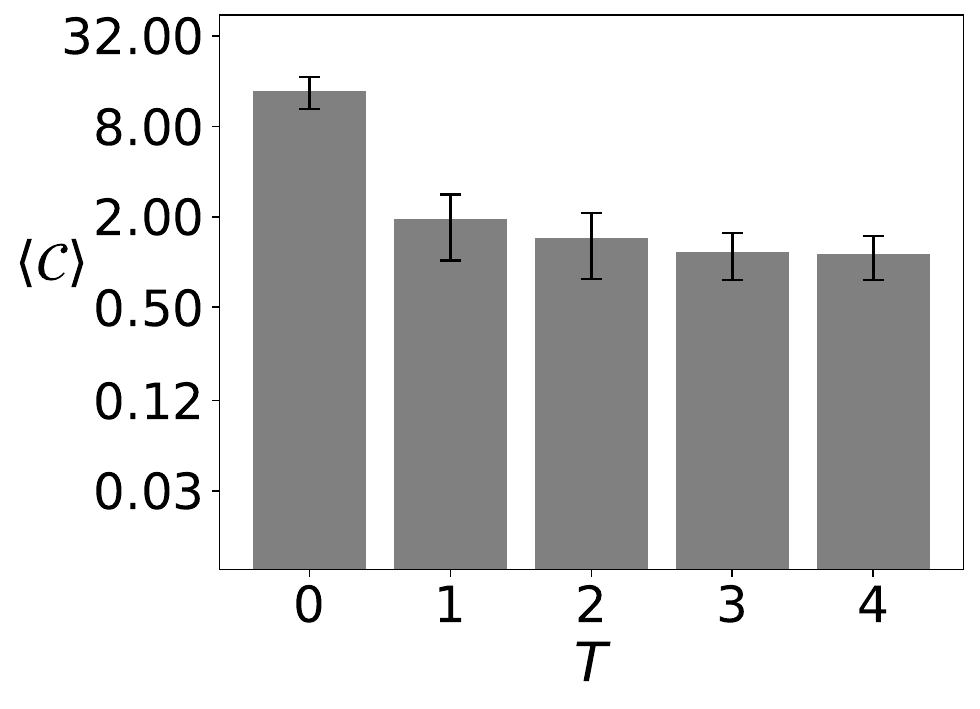}}\hfill%
\subcaptionbox{$\alpha = 3$, $K = 4$\label{sfig:clust_th12a3K4}}{\includegraphics[width=.24\textwidth, ]{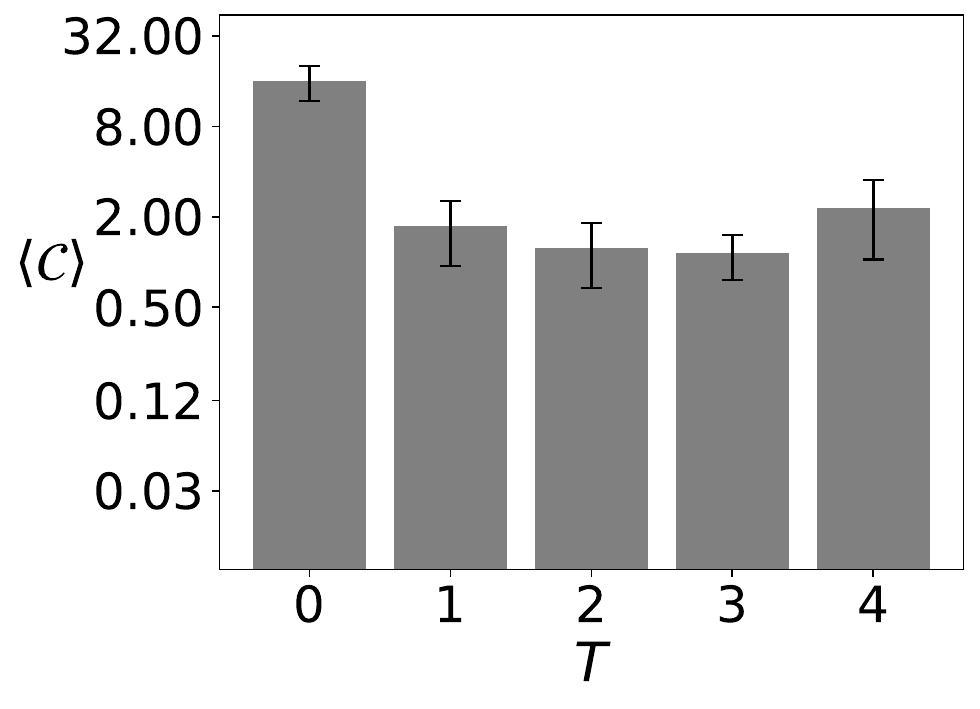}}\hfill%
\subcaptionbox{$\alpha = 3$, $K = 5$\label{sfig:clust_th12a3K5}}{\includegraphics[width=.24\textwidth, ]{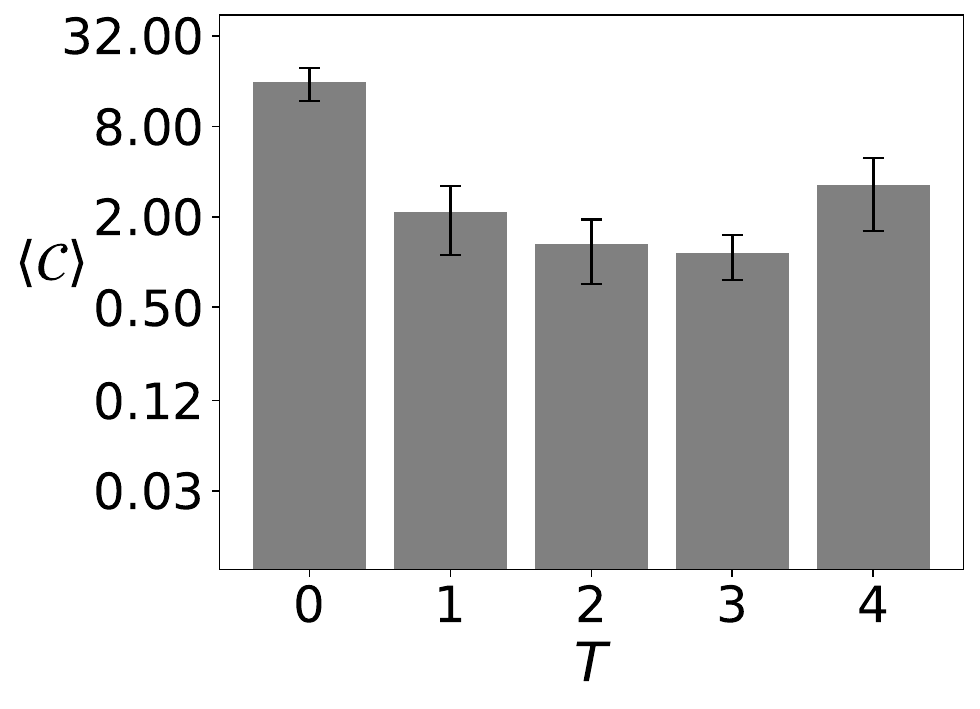}}\\
\subcaptionbox{$\alpha = 4$, $K = 2$\label{sfig:clust_th12a4K2}}{\includegraphics[width=.24\textwidth, ]{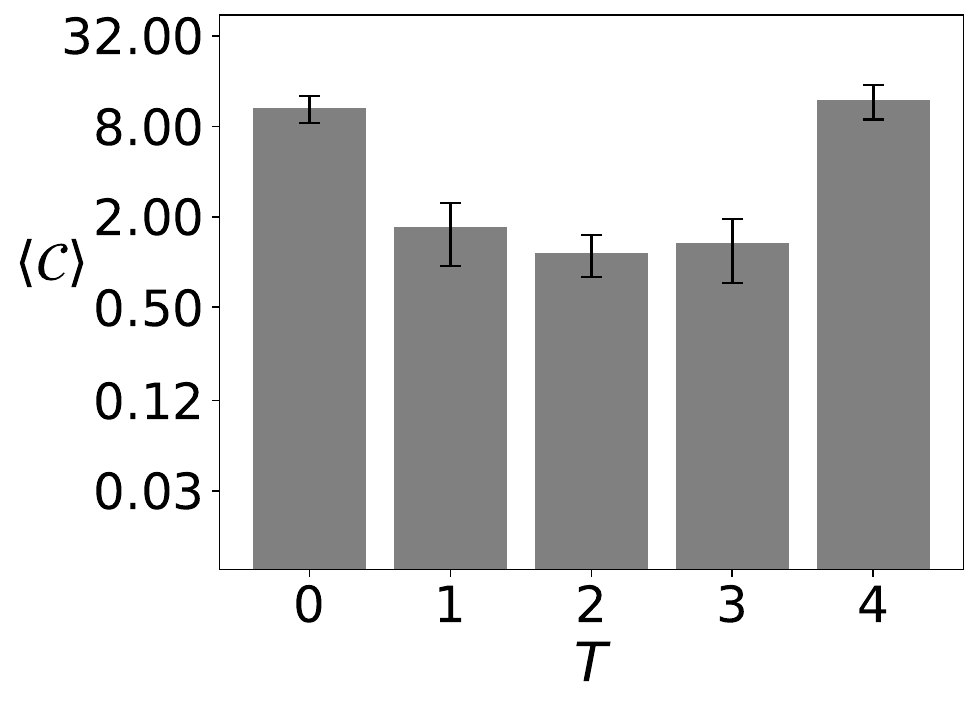}}\hfill%
\subcaptionbox{$\alpha = 4$, $K = 3$\label{sfig:clust_th12a4K3}}{\includegraphics[width=.24\textwidth, ]{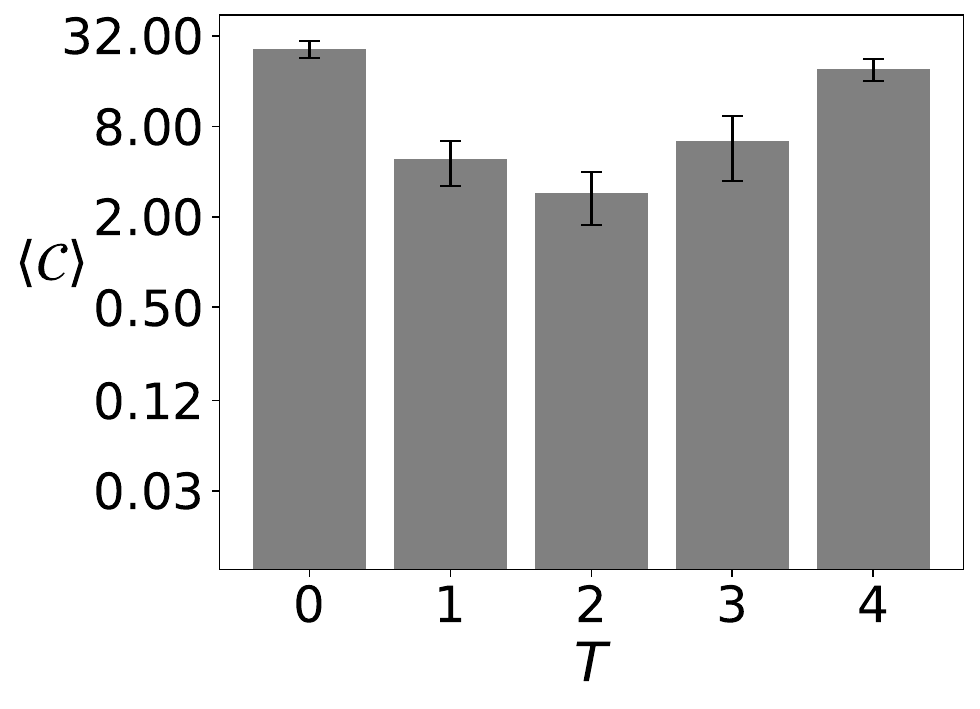}}\hfill%
\subcaptionbox{$\alpha = 4$, $K = 4$\label{sfig:clust_th12a4K4}}{\includegraphics[width=.24\textwidth, ]{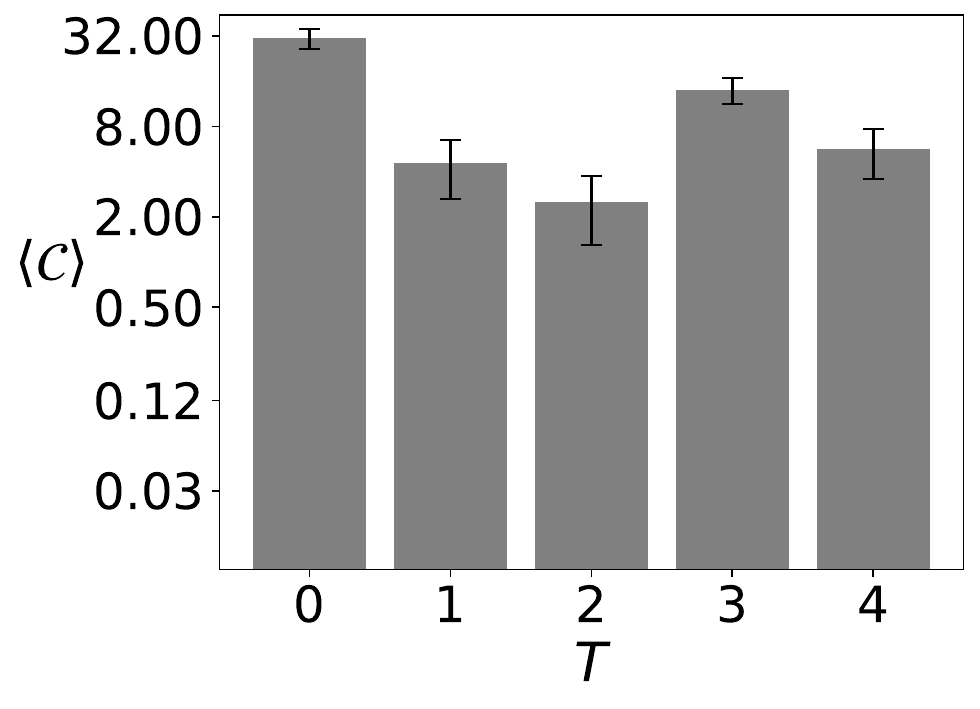}}\hfill%
\subcaptionbox{$\alpha = 4$, $K = 5$\label{sfig:clust_th12a4K5}}{\includegraphics[width=.24\textwidth, ]{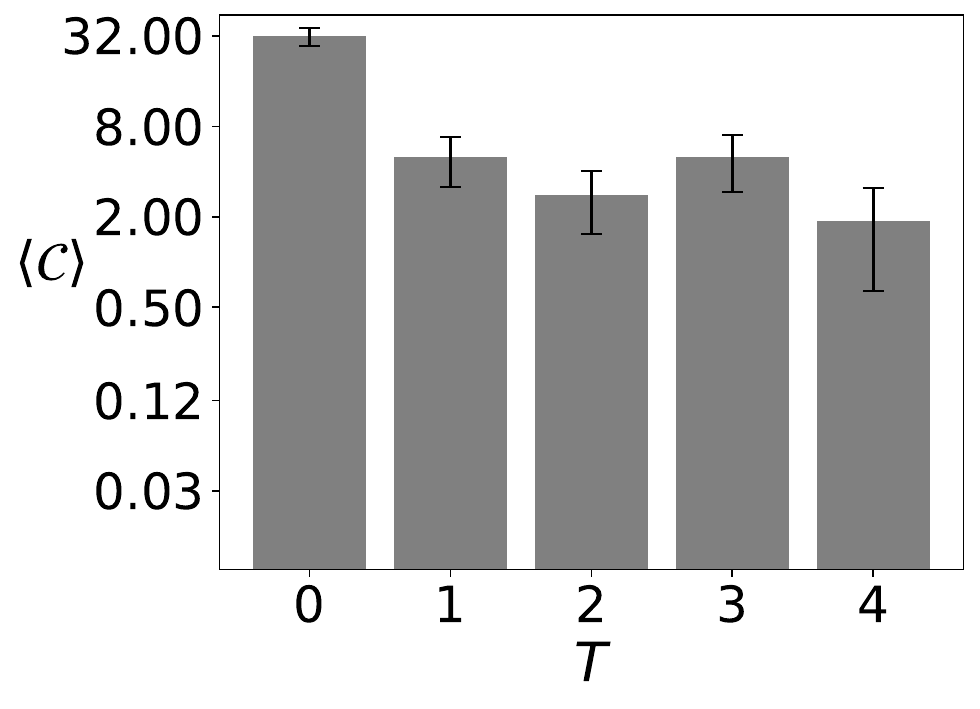}}\\
\subcaptionbox{$\alpha = 6$, $K = 2$\label{sfig:clust_th12a6K2}}{\includegraphics[width=.24\textwidth, ]{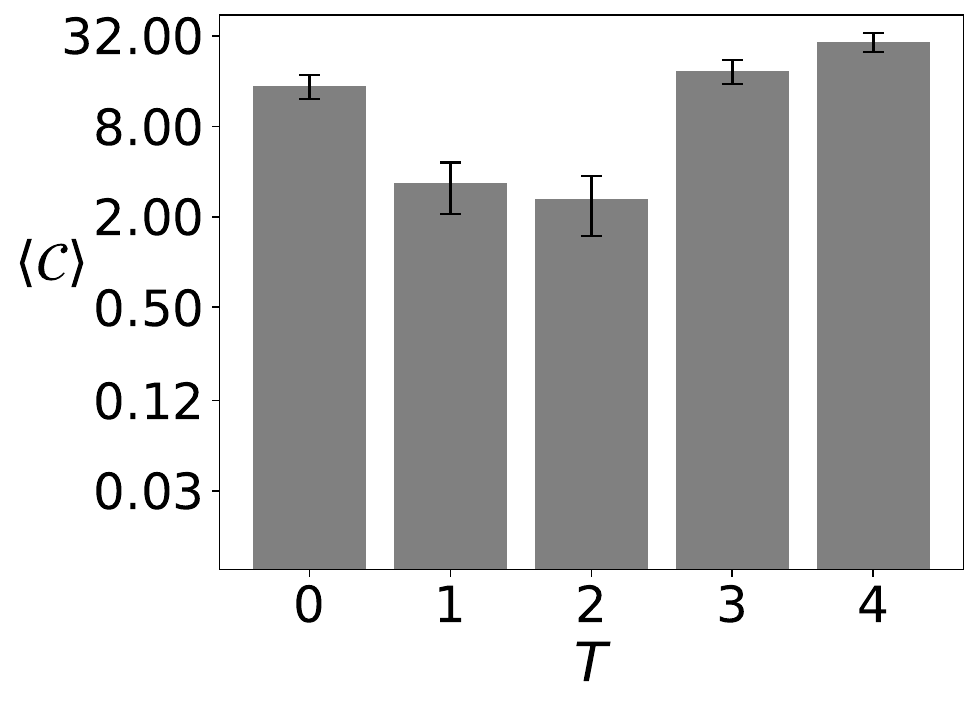}}\hfill%
\subcaptionbox{$\alpha = 6$, $K = 3$\label{sfig:clust_th12a6K3}}{\includegraphics[width=.24\textwidth, ]{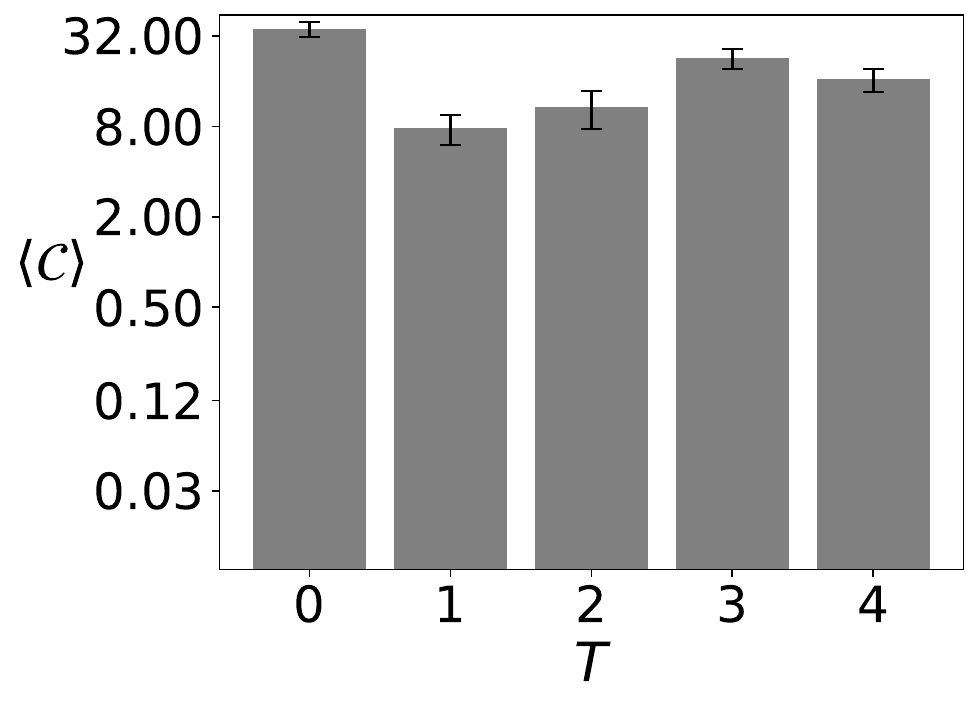}}\hfill%
\subcaptionbox{$\alpha = 6$, $K = 4$\label{sfig:clust_th12a6K4}}{\includegraphics[width=.24\textwidth, ]{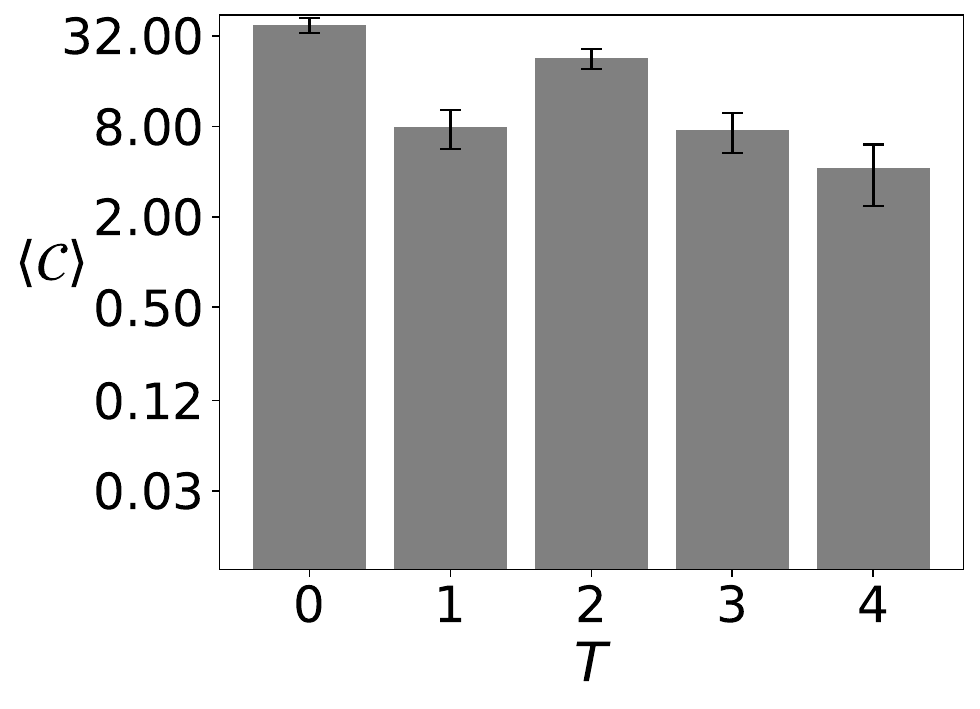}}\hfill%
\subcaptionbox{$\alpha = 6$, $K = 5$\label{sfig:clust_th12a6K5}}{\includegraphics[width=.24\textwidth, ]{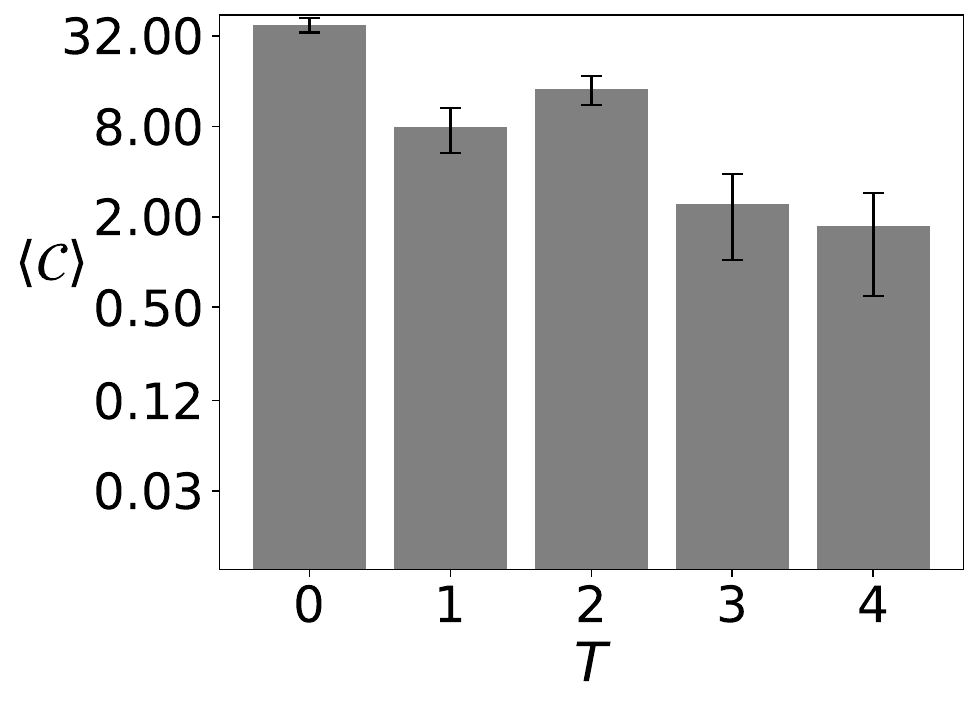}}\\
\caption{\label{fig:th12_clust}Average number $\langle\mathcal C\rangle$ of opinion clusters after $t=10^3$ time steps for various exponents of the distance scaling function $\alpha$ and various numbers of available opinions $K$ in the system. Noise discrimination threshold $\theta=12$. The system contains $L^2=41^2$ actors. The results are averaged over $R=100$ independent system realizations.}
\end{figure*}

\begin{figure*}[htbp]
\subcaptionbox{$\alpha = 2$, $K = 2$\label{sfig:clust_th25a2K2}}{\includegraphics[width=.24\textwidth, ]{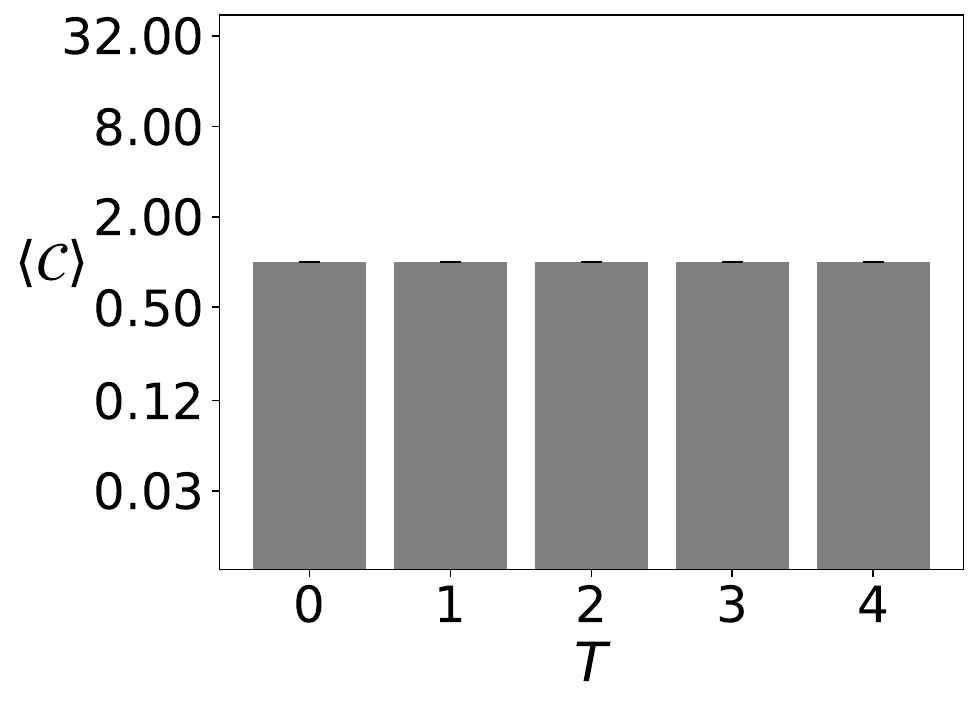}}\hfill%
\subcaptionbox{$\alpha = 2$, $K = 3$\label{sfig:clust_th25a2K3}}{\includegraphics[width=.24\textwidth, ]{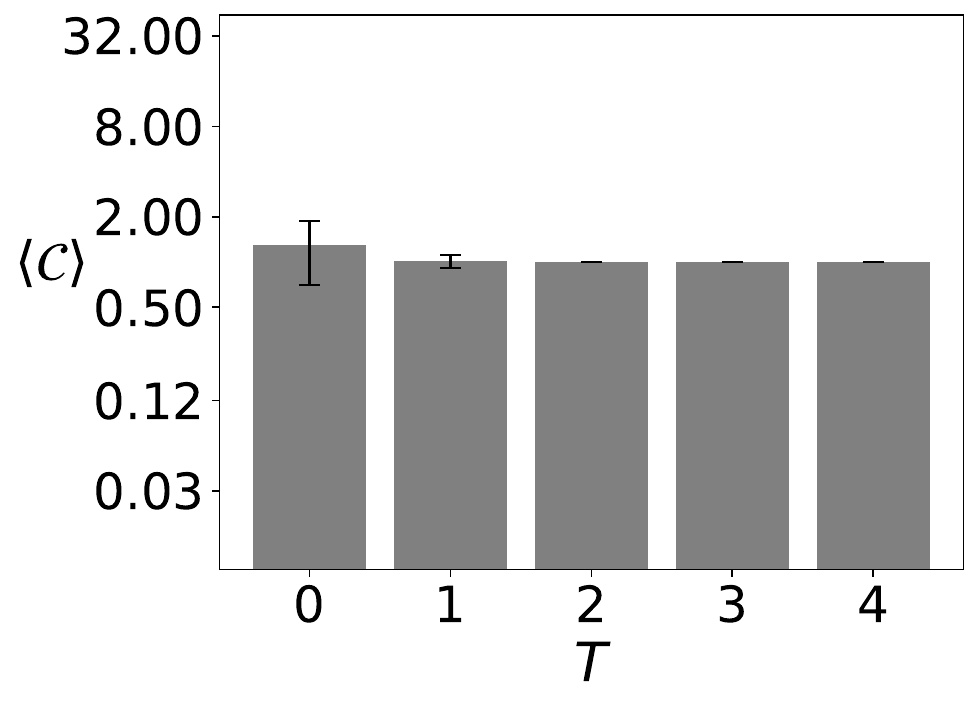}}\hfill%
\subcaptionbox{$\alpha = 2$, $K = 4$\label{sfig:clust_th25a2K4}}{\includegraphics[width=.24\textwidth, ]{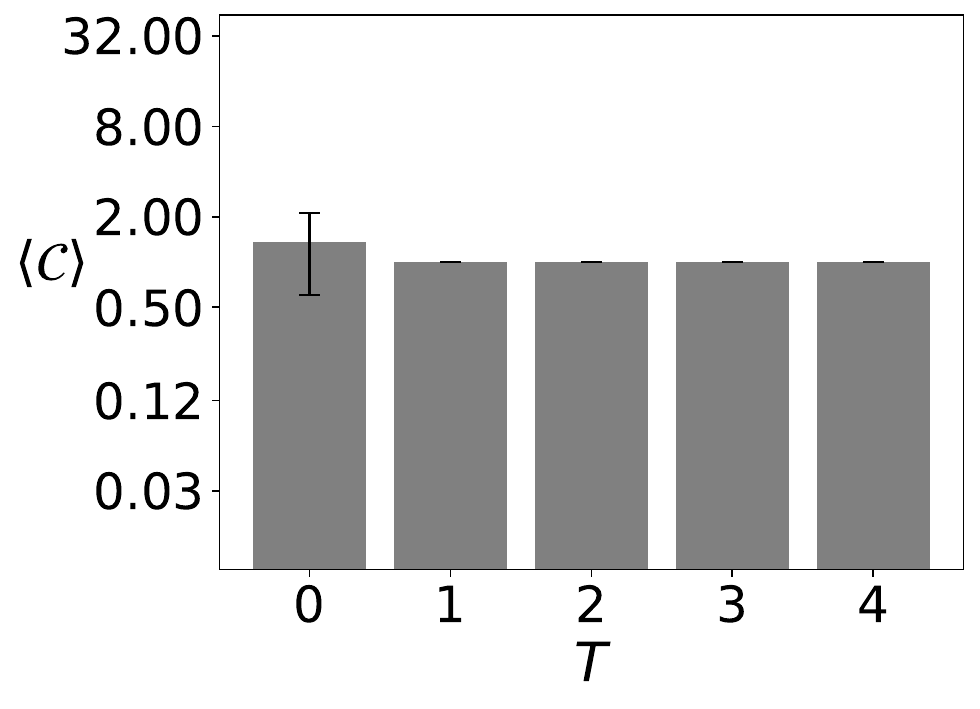}}\hfill%
\subcaptionbox{$\alpha = 2$, $K = 5$\label{sfig:clust_th25a2K5}}{\includegraphics[width=.24\textwidth, ]{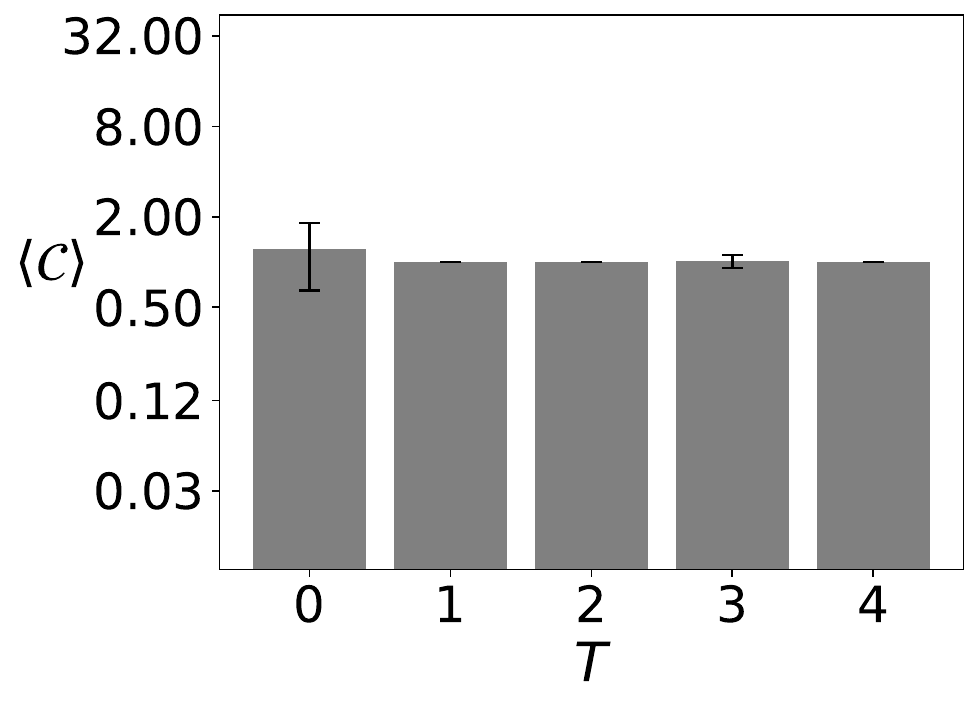}}\\
\subcaptionbox{$\alpha = 3$, $K = 2$\label{sfig:clust_th25a3K2}}{\includegraphics[width=.24\textwidth, ]{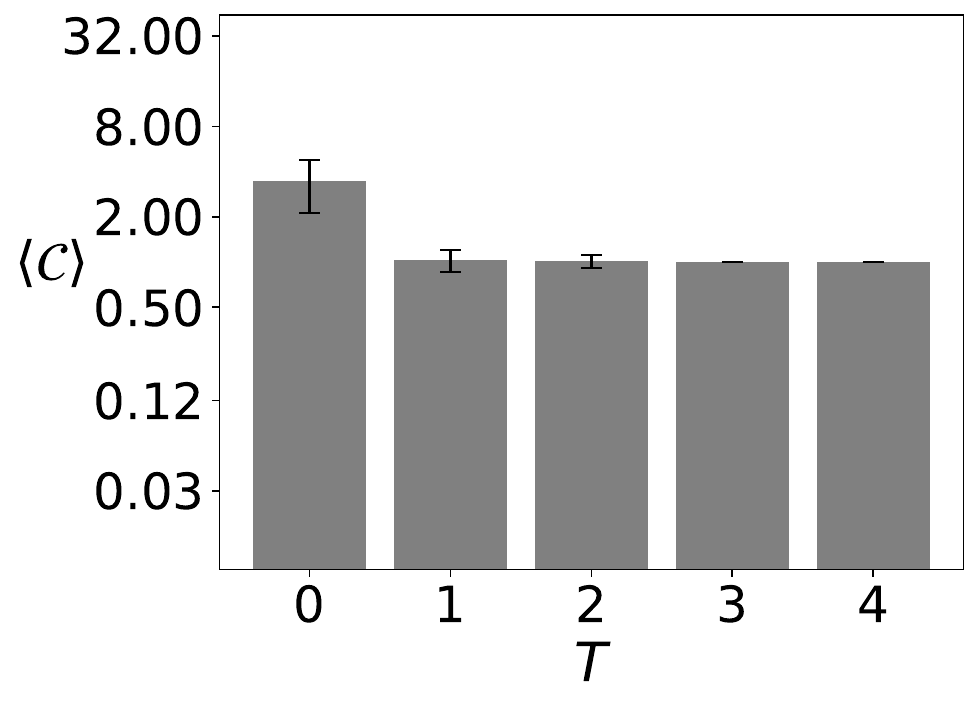}}\hfill%
\subcaptionbox{$\alpha = 3$, $K = 3$\label{sfig:clust_th25a3K3}}{\includegraphics[width=.24\textwidth, ]{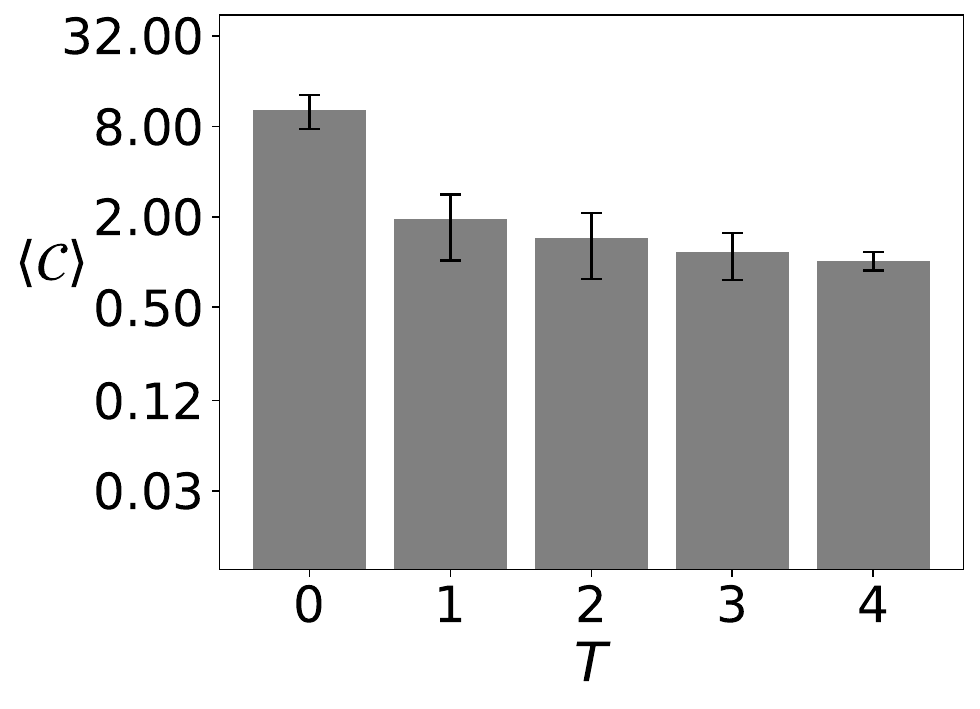}}\hfill%
\subcaptionbox{$\alpha = 3$, $K = 4$\label{sfig:clust_th25a3K4}}{\includegraphics[width=.24\textwidth, ]{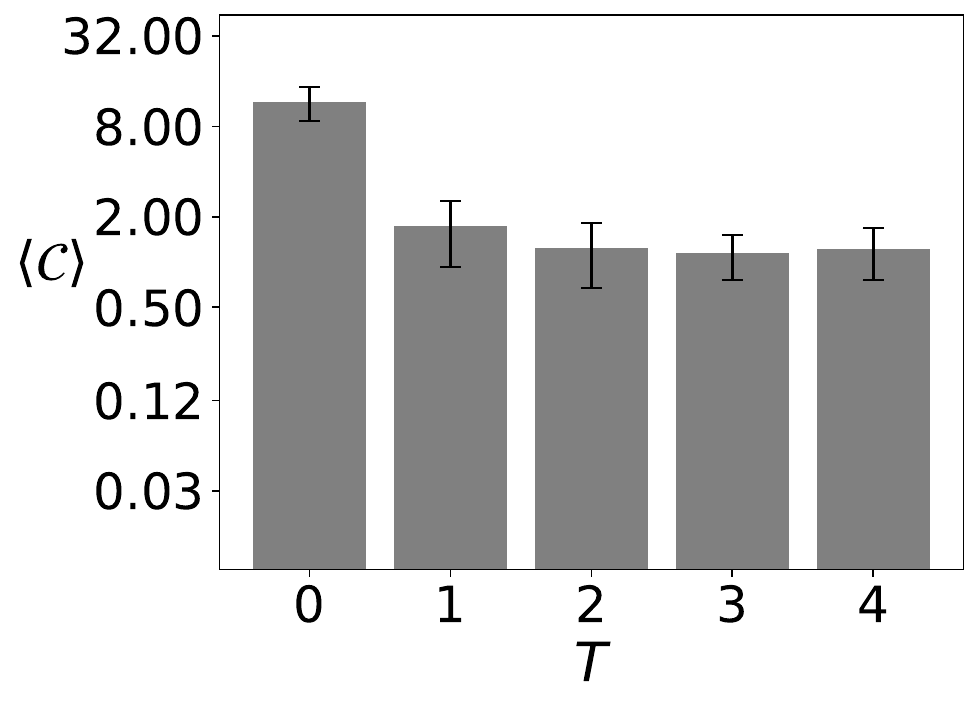}}\hfill%
\subcaptionbox{$\alpha = 3$, $K = 5$\label{sfig:clust_th25a3K5}}{\includegraphics[width=.24\textwidth, ]{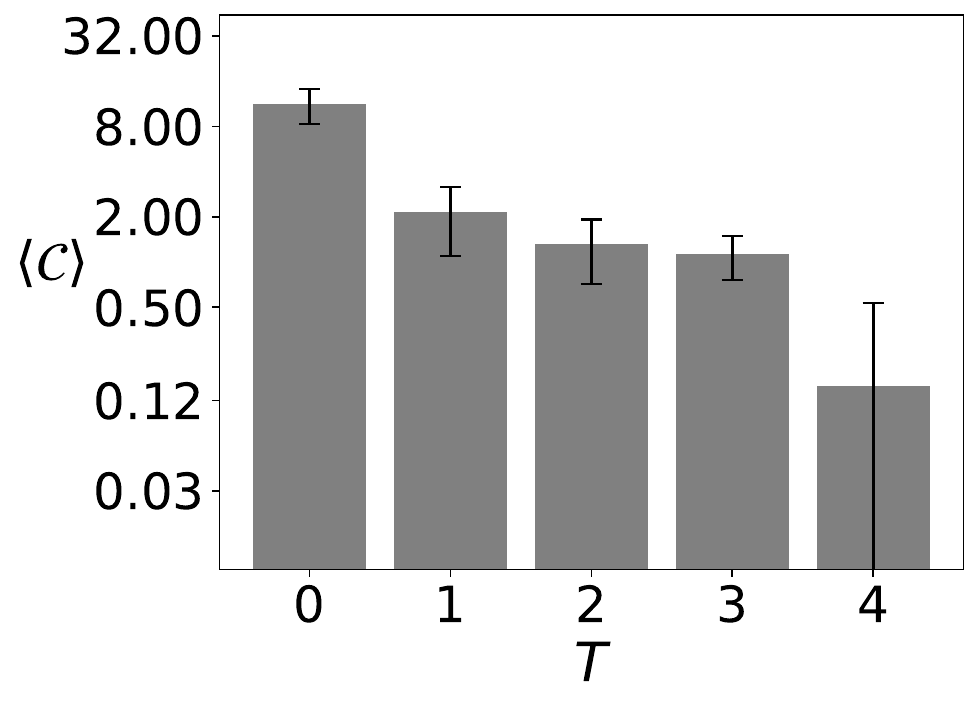}}\\
\subcaptionbox{$\alpha = 4$, $K = 2$\label{sfig:clust_th25a4K2}}{\includegraphics[width=.24\textwidth, ]{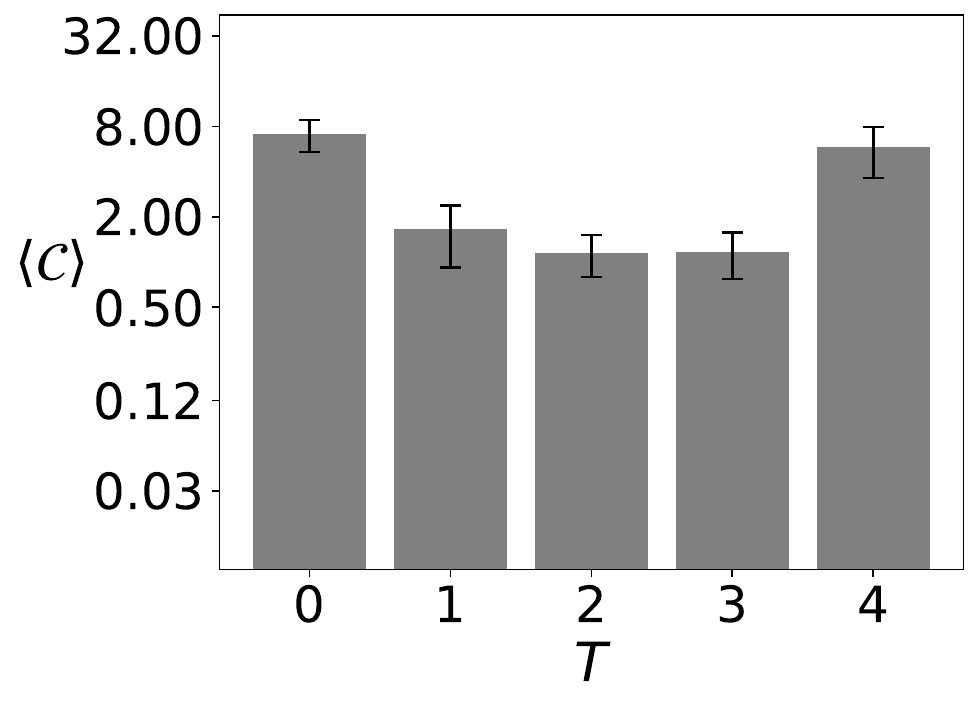}}\hfill%
\subcaptionbox{$\alpha = 4$, $K = 3$\label{sfig:clust_th25a4K3}}{\includegraphics[width=.24\textwidth, ]{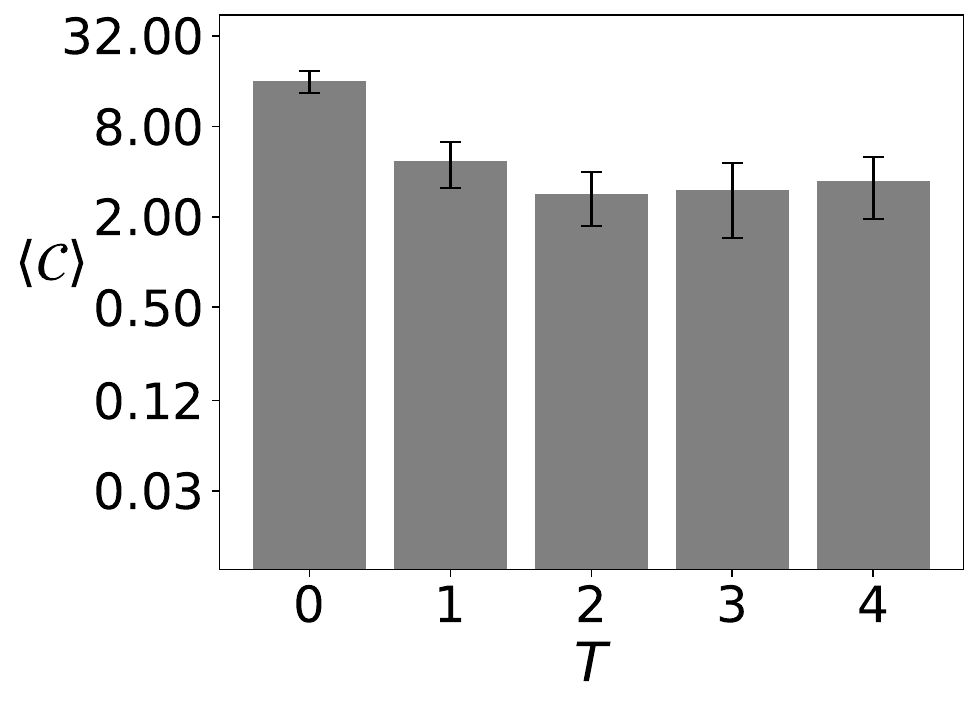}}\hfill%
\subcaptionbox{$\alpha = 4$, $K = 4$\label{sfig:clust_th25a4K4}}{\includegraphics[width=.24\textwidth, ]{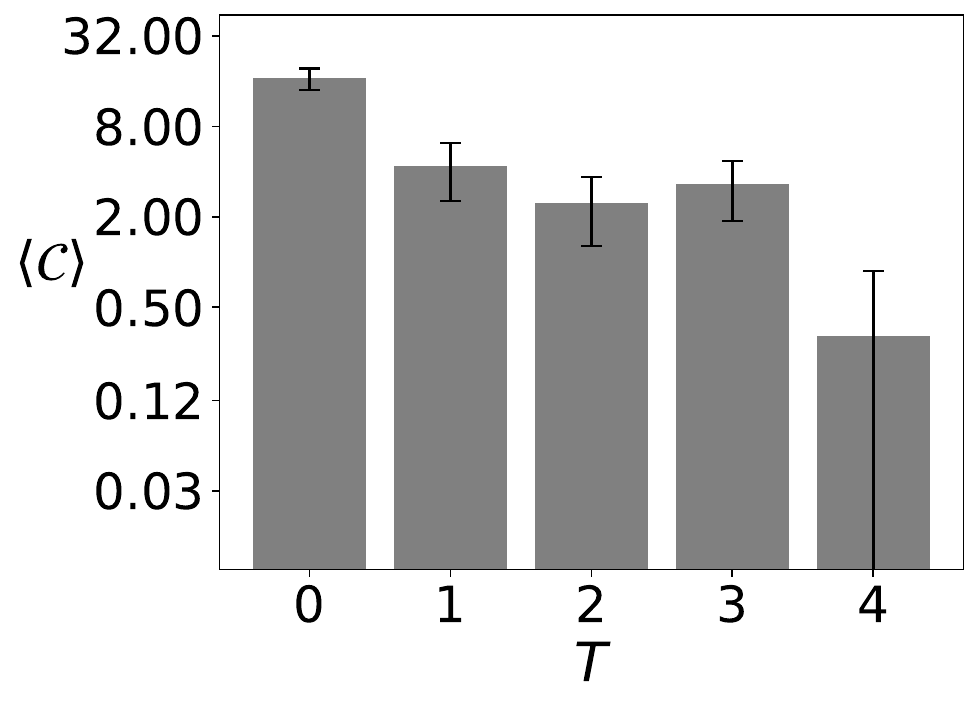}}\hfill%
\subcaptionbox{$\alpha = 4$, $K = 5$\label{sfig:clust_th25a4K5}}{\includegraphics[width=.24\textwidth, ]{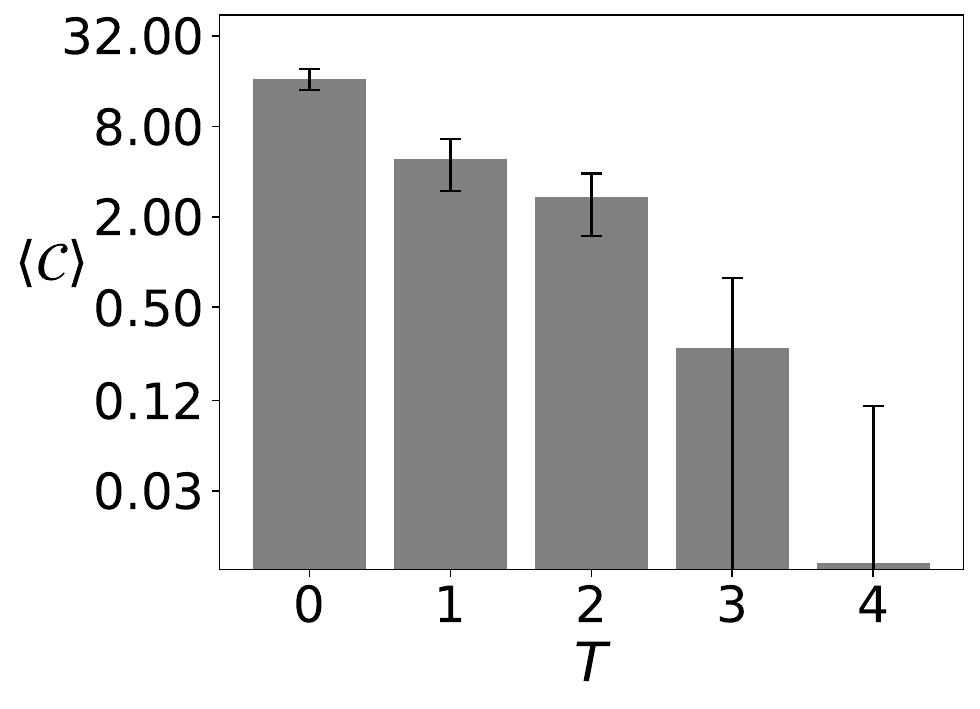}}\\
\subcaptionbox{$\alpha = 6$, $K = 2$\label{sfig:clust_th25a6K2}}{\includegraphics[width=.24\textwidth, ]{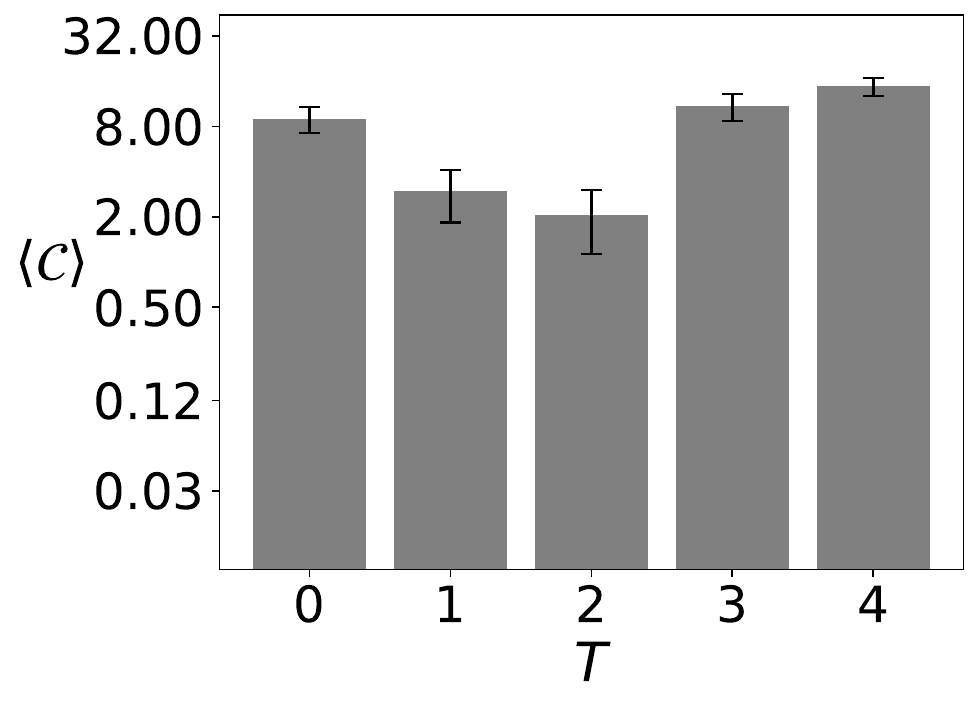}}\hfill%
\subcaptionbox{$\alpha = 6$, $K = 3$\label{sfig:clust_th25a6K3}}{\includegraphics[width=.24\textwidth, ]{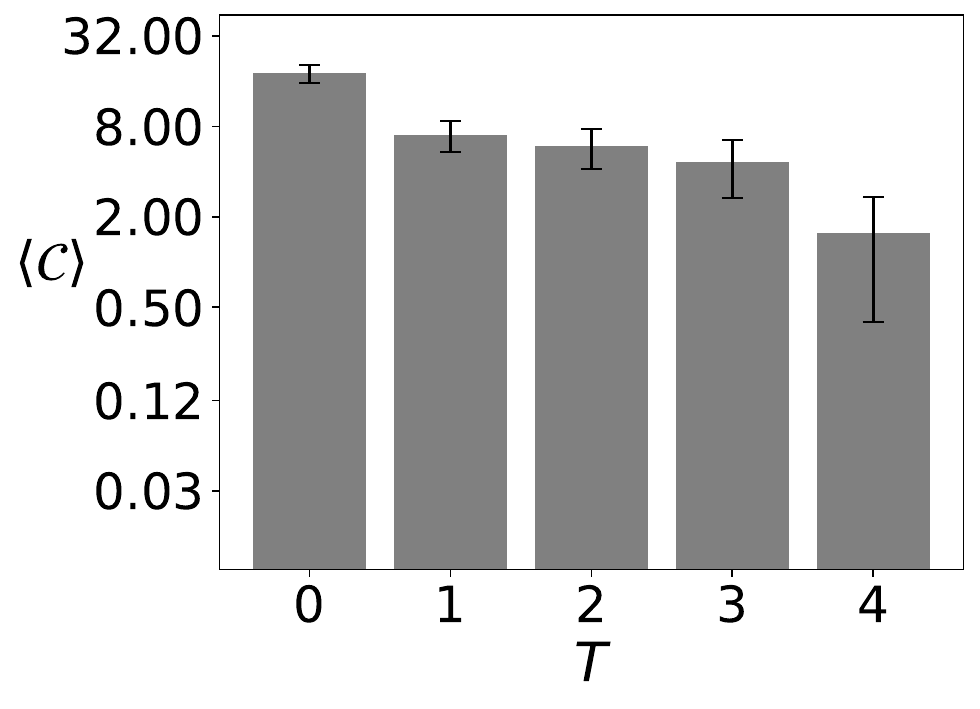}}\hfill%
\subcaptionbox{$\alpha = 6$, $K = 4$\label{sfig:clust_th25a6K4}}{\includegraphics[width=.24\textwidth, ]{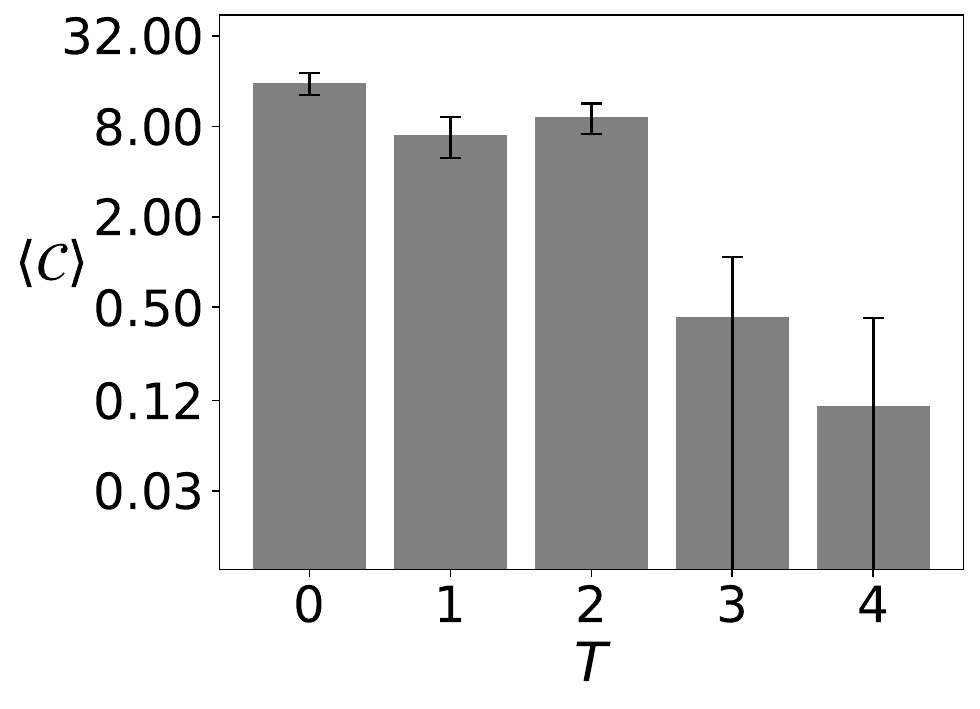}}\hfill%
\subcaptionbox{$\alpha = 6$, $K = 5$\label{sfig:clust_th25a6K5}}{\includegraphics[width=.24\textwidth, ]{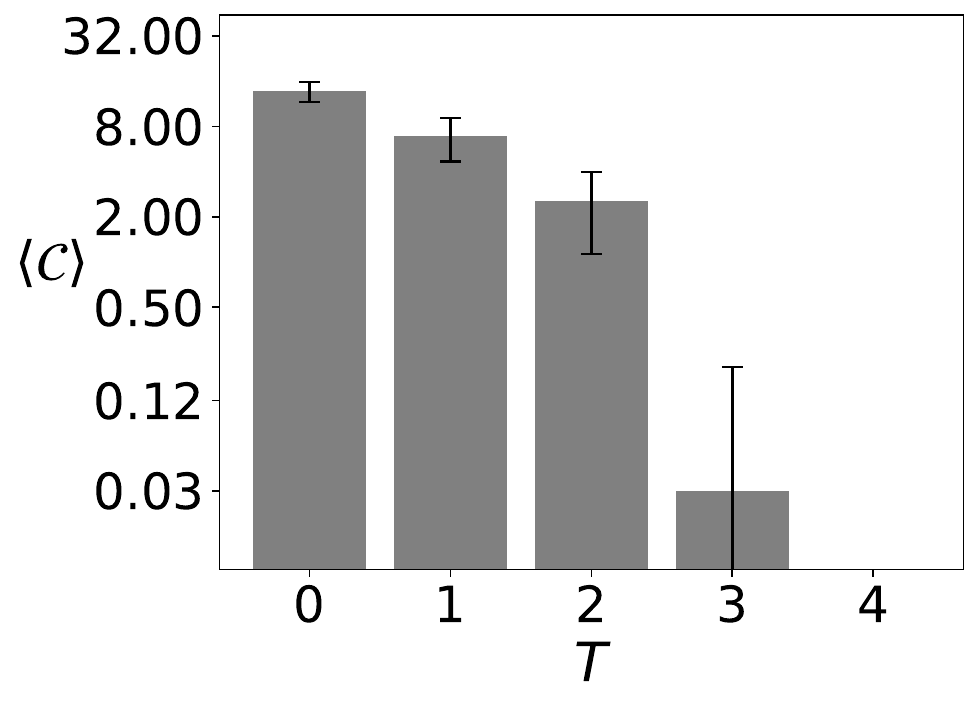}}\\
\caption{\label{fig:th25_clust}Average number $\langle\mathcal C\rangle$ of opinion clusters after $t=10^3$ time steps for various exponents of the distance scaling function $\alpha$ and various numbers of available opinions $K$ in the system. The noise discrimination threshold $\theta=25$. The system contains $L^2=41^2$ actors. The results are averaged over $R=100$ independent system realizations.}
\end{figure*}

\begin{figure*}[htbp]
\subcaptionbox{$\alpha = 2$, $K = 2$\label{sfig:clust_th50a2K2}}{\includegraphics[width=.24\textwidth, ]{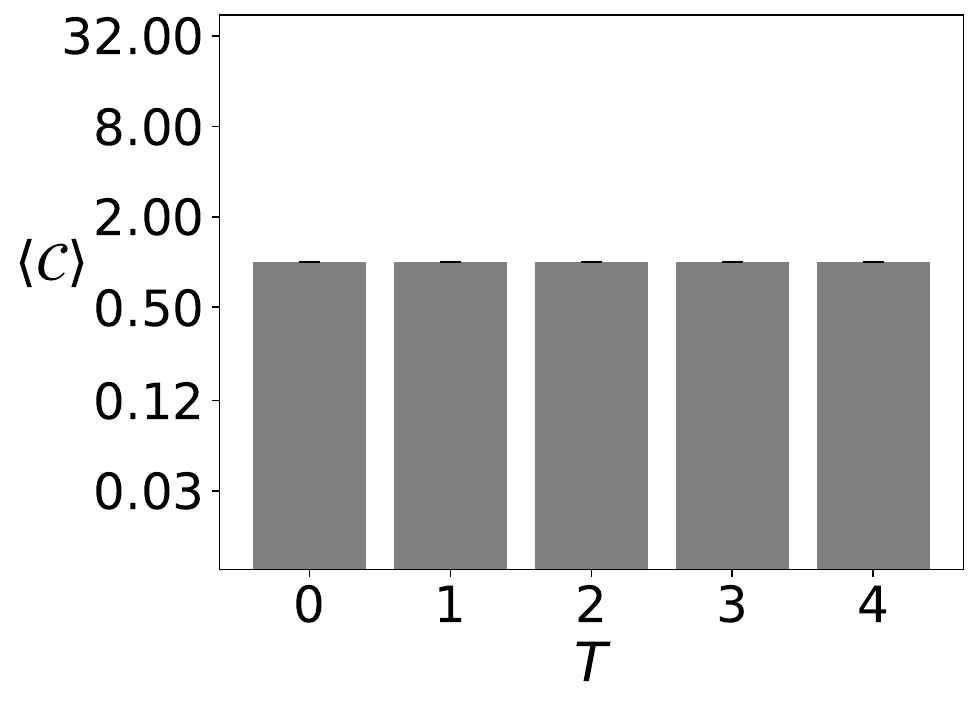}}\hfill%
\subcaptionbox{$\alpha = 2$, $K = 3$\label{sfig:clust_th50a2K3}}{\includegraphics[width=.24\textwidth, ]{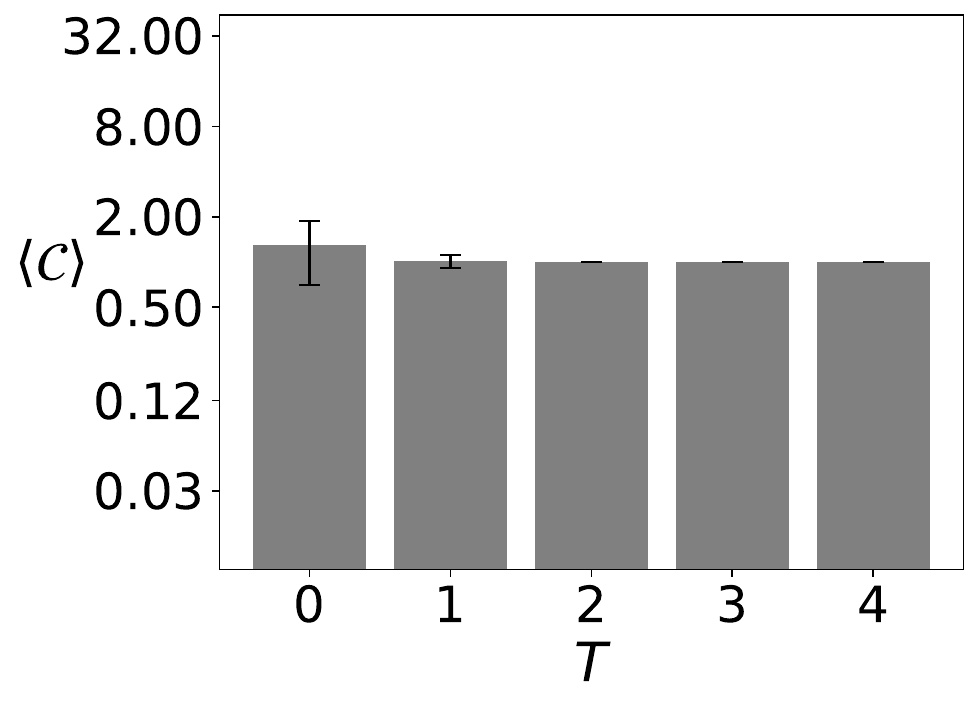}}\hfill%
\subcaptionbox{$\alpha = 2$, $K = 4$\label{sfig:clust_th50a2K4}}{\includegraphics[width=.24\textwidth, ]{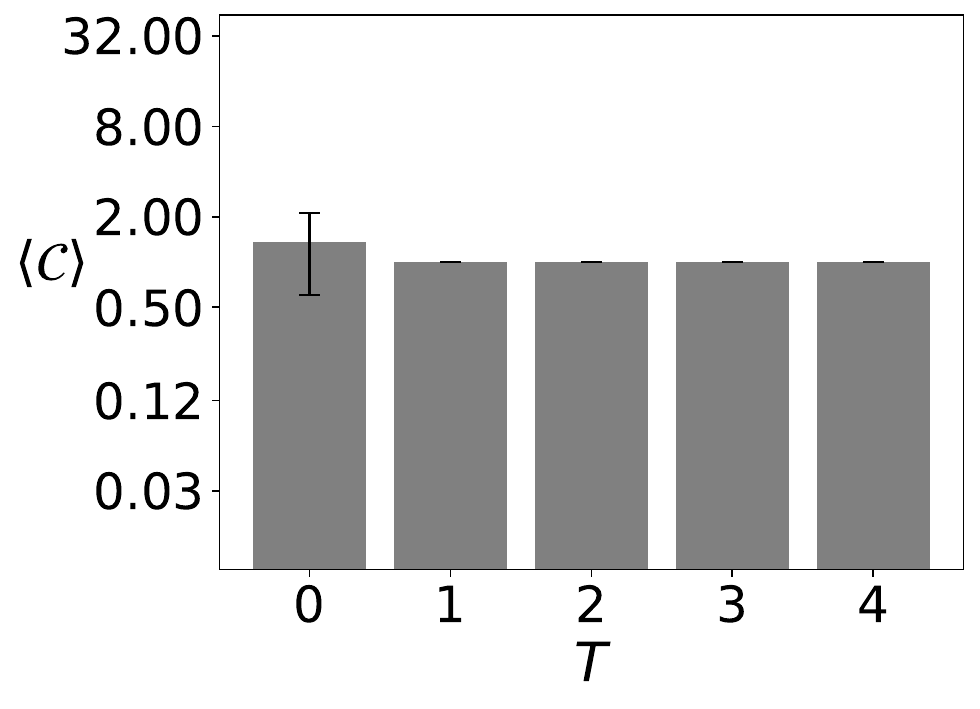}}\hfill%
\subcaptionbox{$\alpha = 2$, $K = 5$\label{sfig:clust_th50a2K5}}{\includegraphics[width=.24\textwidth, ]{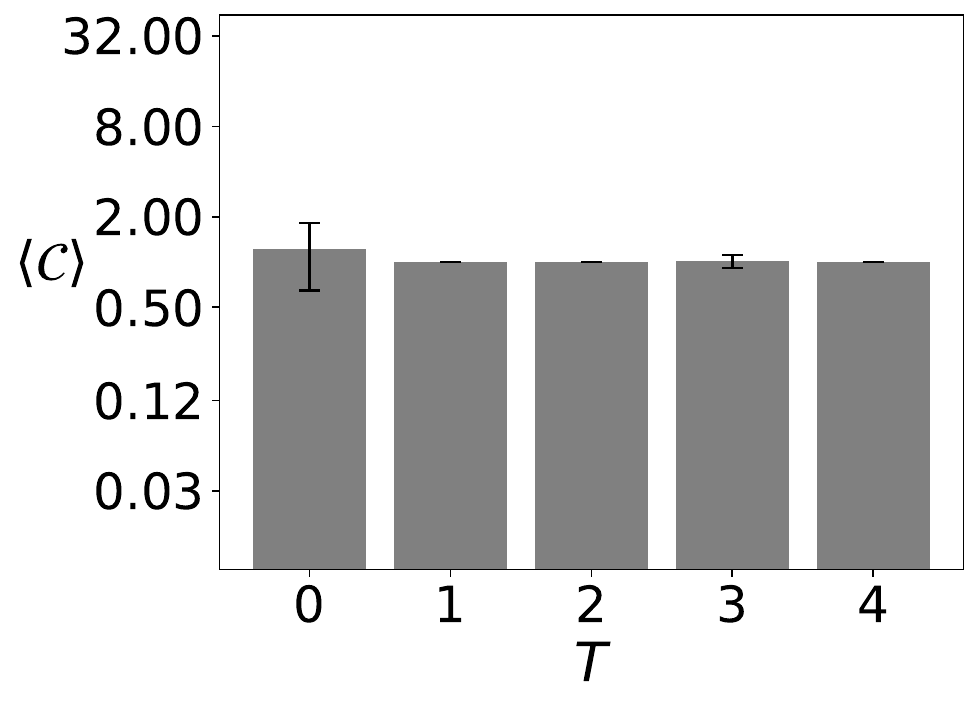}}\\
\subcaptionbox{$\alpha = 3$, $K = 2$\label{sfig:clust_th50a3K2}}{\includegraphics[width=.24\textwidth, ]{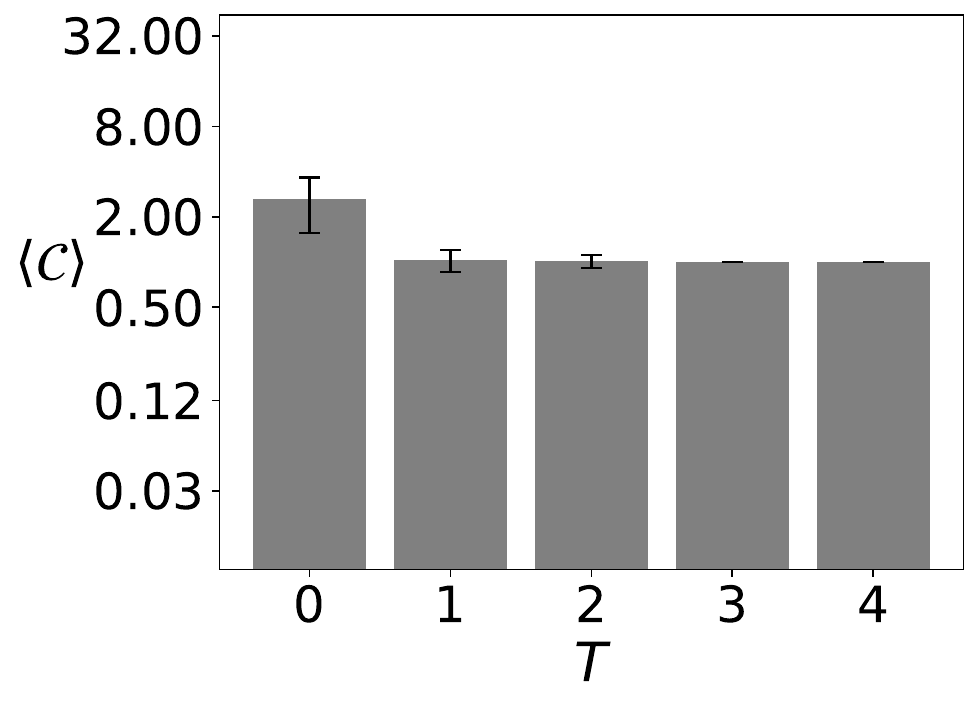}}\hfill%
\subcaptionbox{$\alpha = 3$, $K = 3$\label{sfig:clust_th50a3K3}}{\includegraphics[width=.24\textwidth, ]{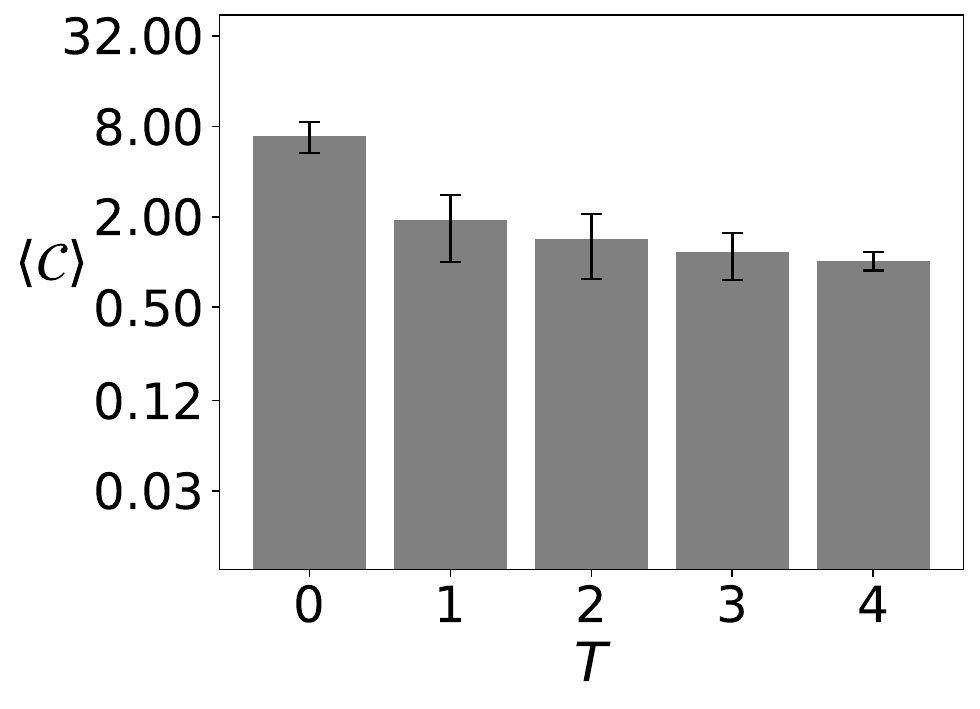}}\hfill%
\subcaptionbox{$\alpha = 3$, $K = 4$\label{sfig:clust_th50a3K4}}{\includegraphics[width=.24\textwidth, ]{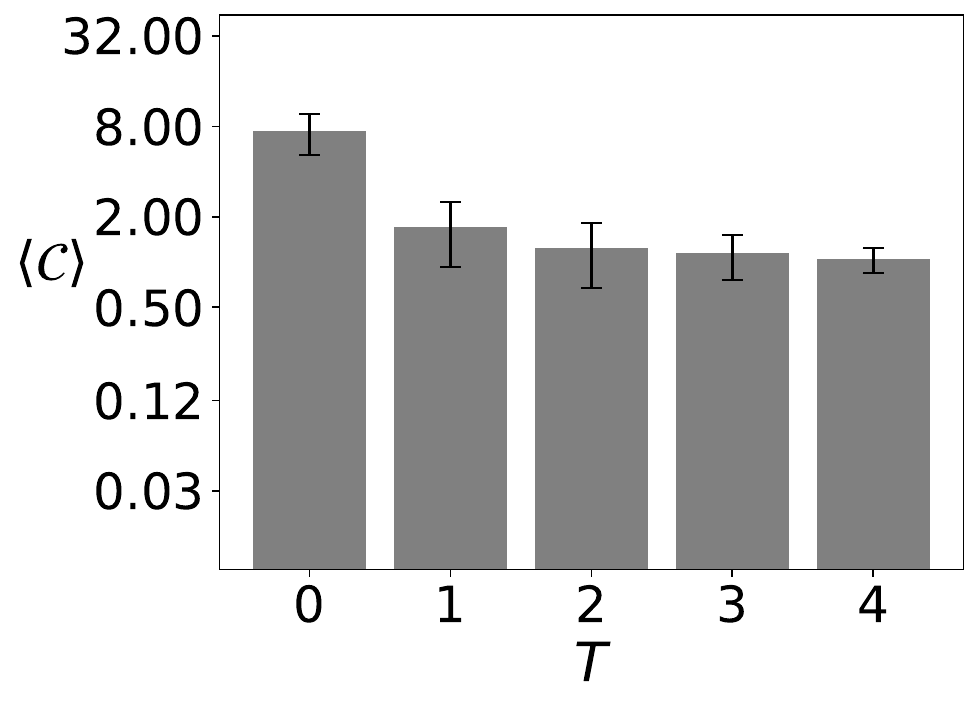}}\hfill%
\subcaptionbox{$\alpha = 3$, $K = 5$\label{sfig:clust_th50a3K5}}{\includegraphics[width=.24\textwidth, ]{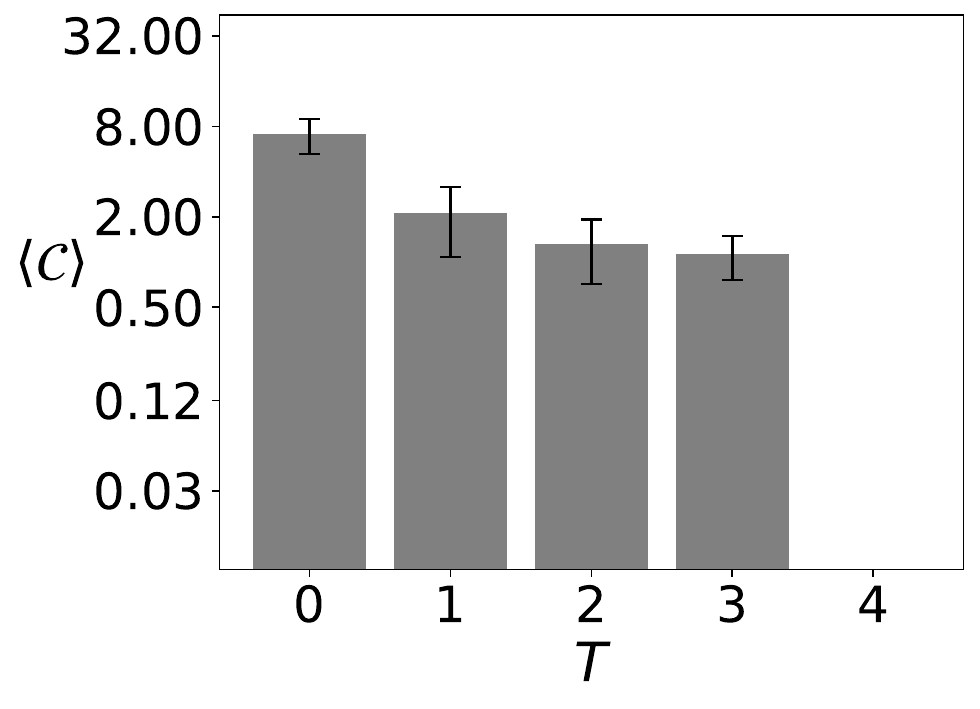}}\\
\subcaptionbox{$\alpha = 4$, $K = 2$\label{sfig:clust_th50a4K2}}{\includegraphics[width=.24\textwidth, ]{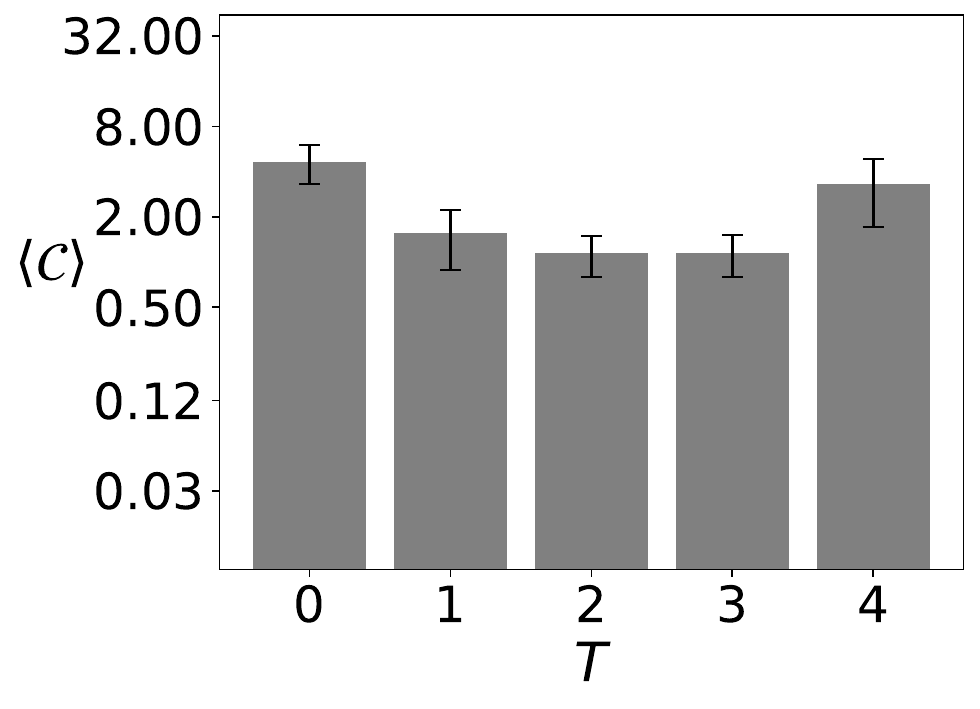}}\hfill%
\subcaptionbox{$\alpha = 4$, $K = 3$\label{sfig:clust_th50a4K3}}{\includegraphics[width=.24\textwidth, ]{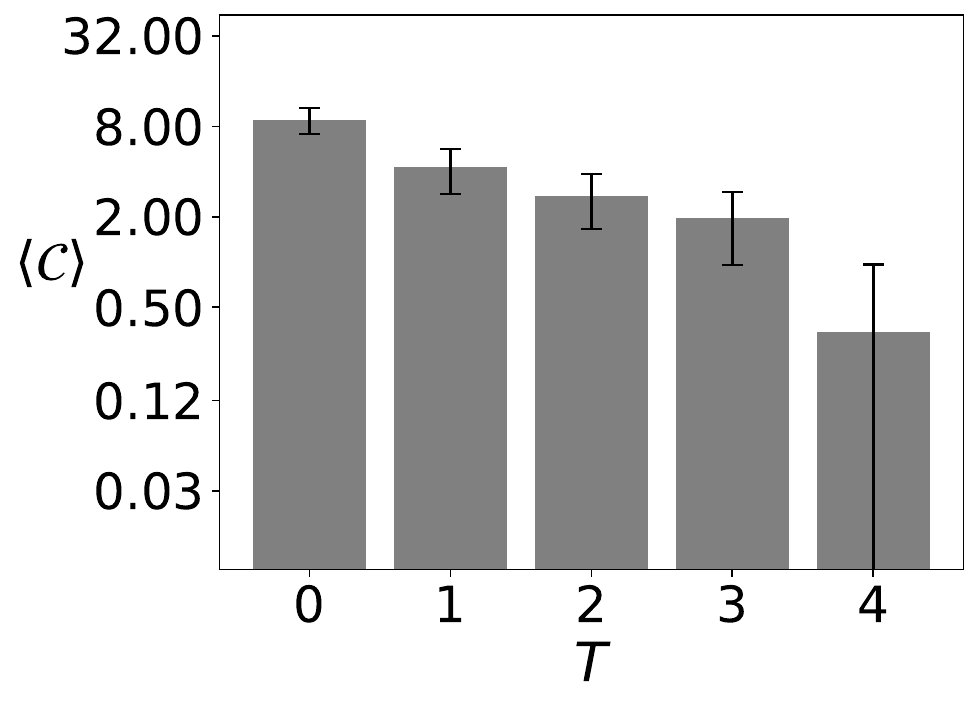}}\hfill%
\subcaptionbox{$\alpha = 4$, $K = 4$\label{sfig:clust_th50a4K4}}{\includegraphics[width=.24\textwidth, ]{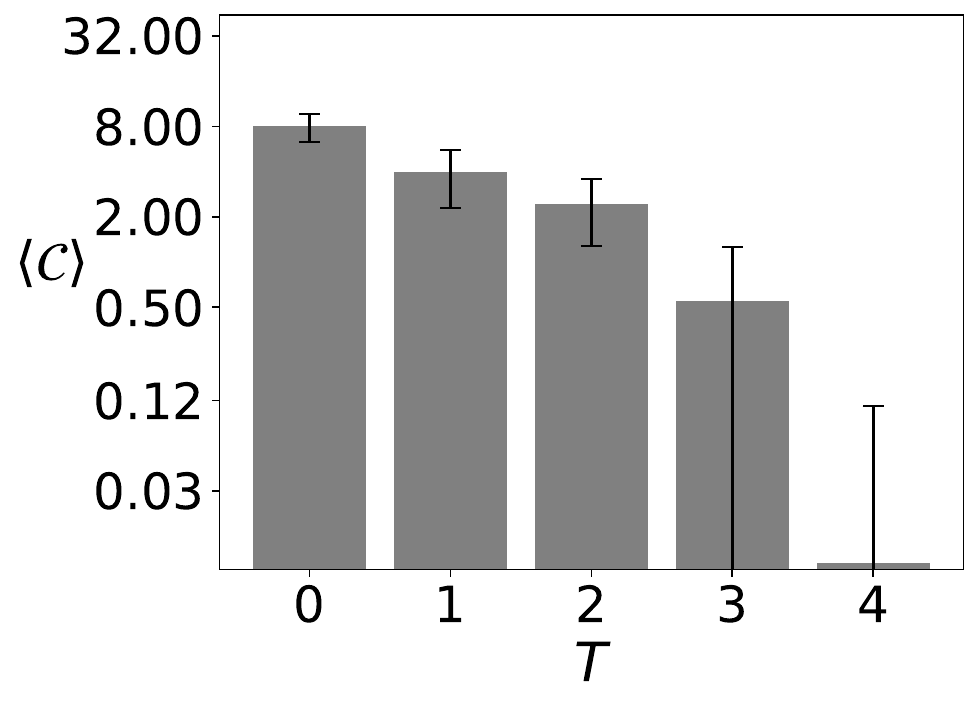}}\hfill%
\subcaptionbox{$\alpha = 4$, $K = 5$\label{sfig:clust_th50a4K5}}{\includegraphics[width=.24\textwidth, ]{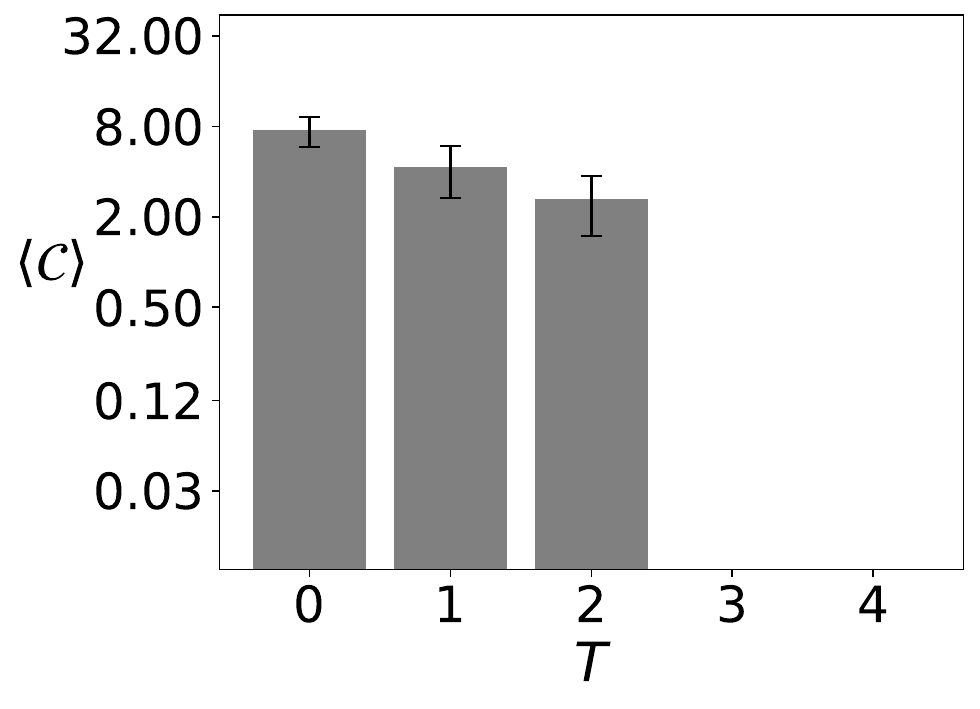}}\\
\subcaptionbox{$\alpha = 6$, $K = 2$\label{sfig:clust_th50a6K2}}{\includegraphics[width=.24\textwidth, ]{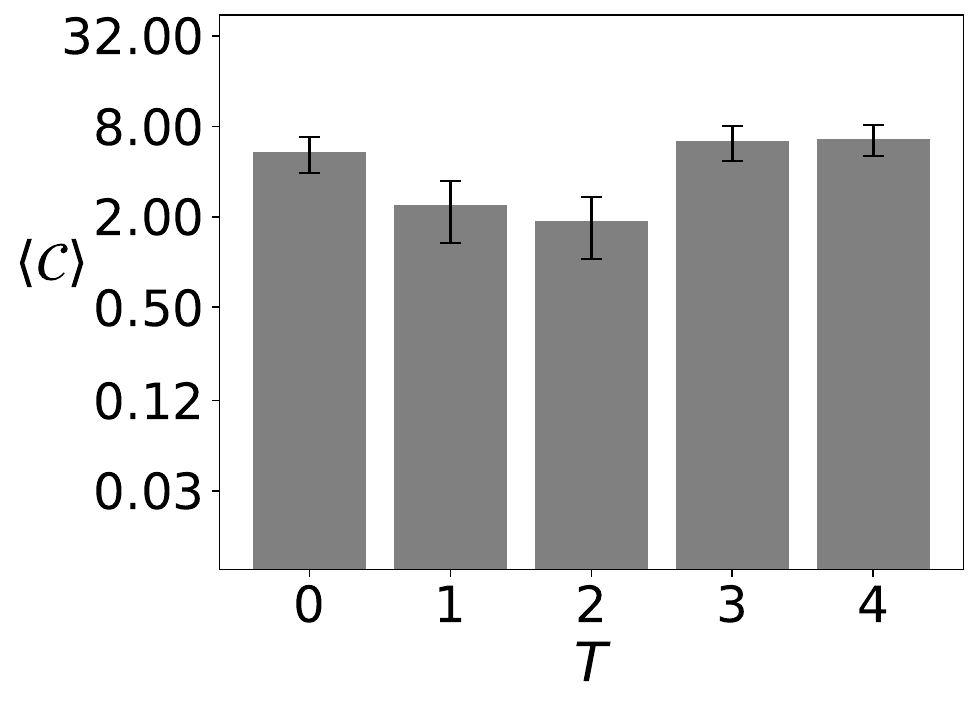}}\hfill%
\subcaptionbox{$\alpha = 6$, $K = 3$\label{sfig:clust_th50a6K3}}{\includegraphics[width=.24\textwidth, ]{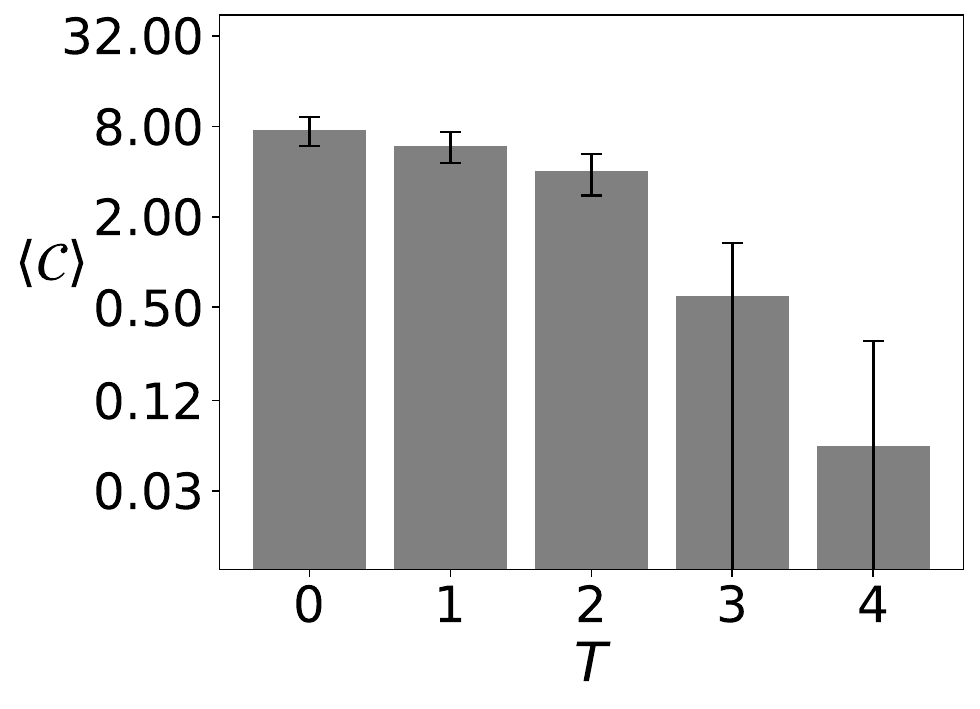}}\hfill%
\subcaptionbox{$\alpha = 6$, $K = 4$\label{sfig:clust_th50a6K4}}{\includegraphics[width=.24\textwidth, ]{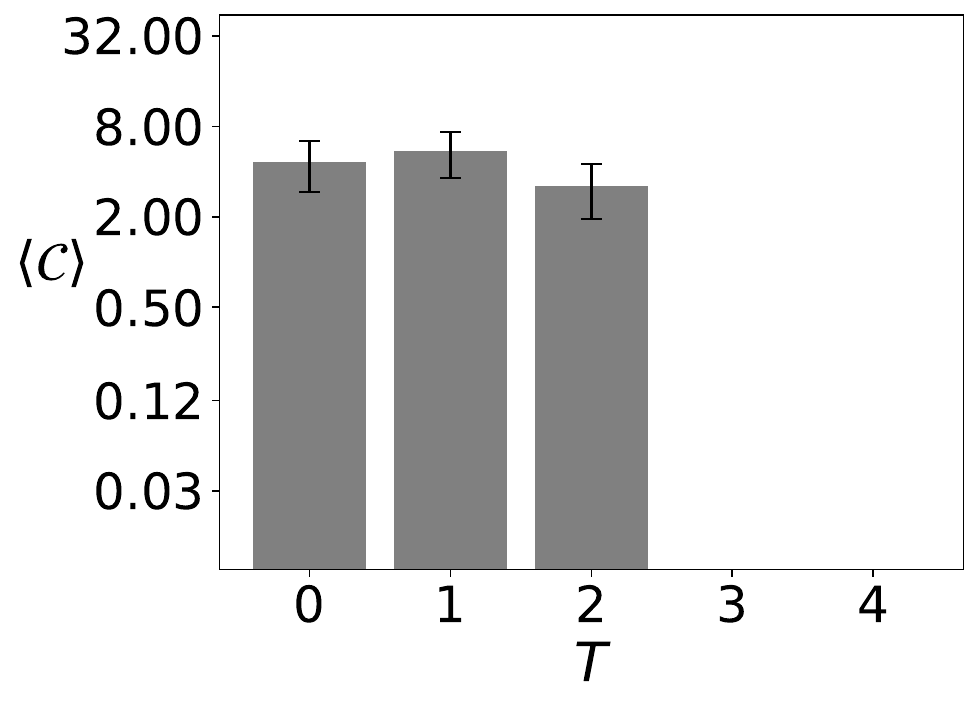}}\hfill%
\subcaptionbox{$\alpha = 6$, $K = 5$\label{sfig:clust_th50a6K5}}{\includegraphics[width=.24\textwidth, ]{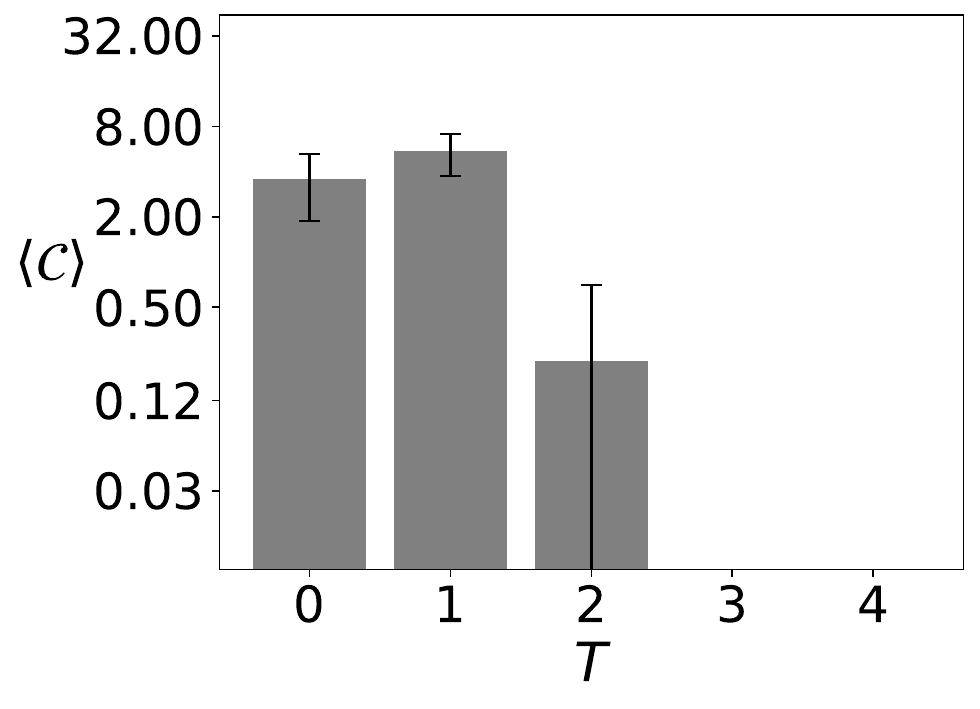}}\\
\caption{\label{fig:th50_clust}Average number $\langle\mathcal C\rangle$ of opinion clusters after $t=10^3$ time steps for various exponents of the distance scaling function $\alpha$ and various numbers of available opinions $K$ in the system. The noise discrimination threshold $\theta=50$. The system contains $L^2=41^2$ actors. The results are averaged over $R=100$ independent system realizations.}
\end{figure*}

\begin{figure*}[htbp]
\hfill\includegraphics[width=.12\textwidth]{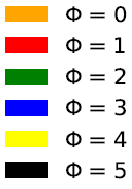}\\
\subcaptionbox{$\alpha = 2$, $K = 2$\label{sfig:th12a2K2}}{\includegraphics[width=.24\textwidth]{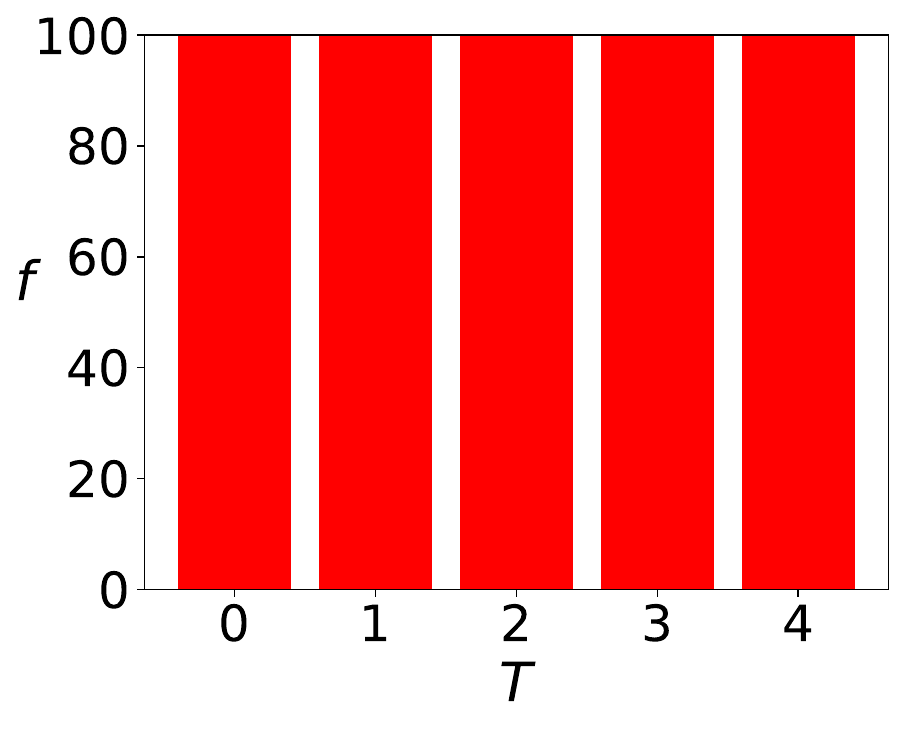}}\hfill
\subcaptionbox{$\alpha = 2$, $K = 3$\label{sfig:th12a2K3}}{\includegraphics[width=.24\textwidth]{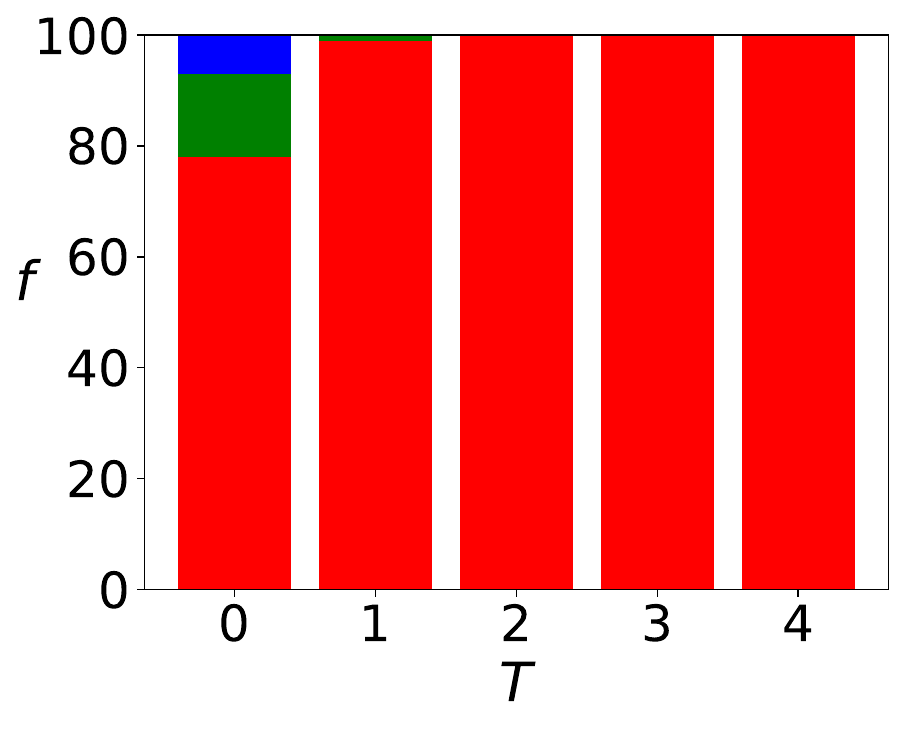}}\hfill
\subcaptionbox{$\alpha = 2$, $K = 4$\label{sfig:th12a2K4}}{\includegraphics[width=.24\textwidth]{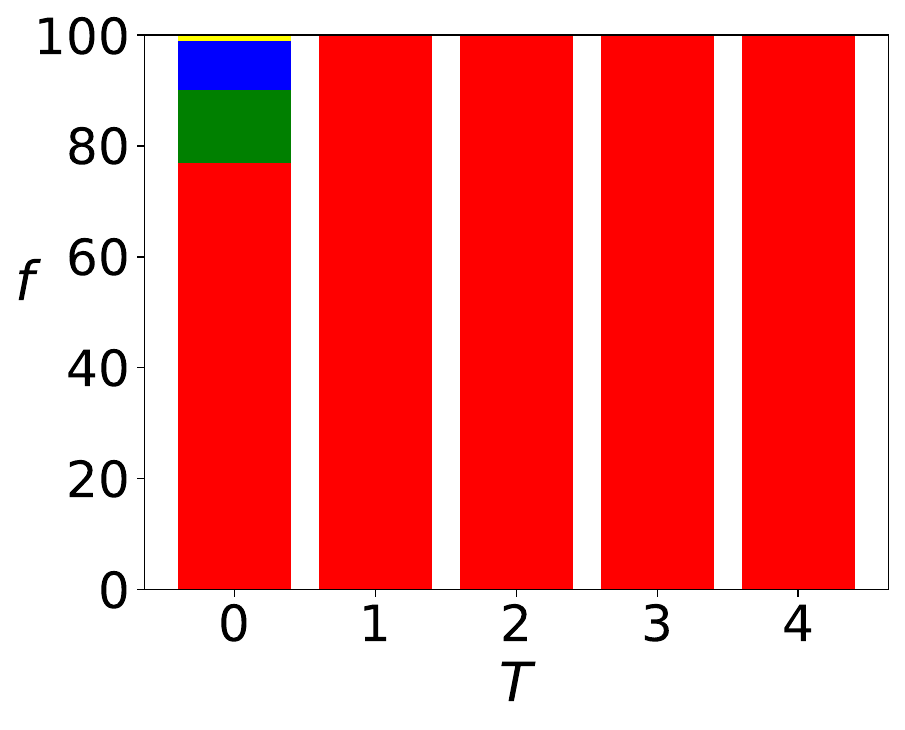}}\hfill
\subcaptionbox{$\alpha = 2$, $K = 5$\label{sfig:th12a2K5}}{\includegraphics[width=.24\textwidth]{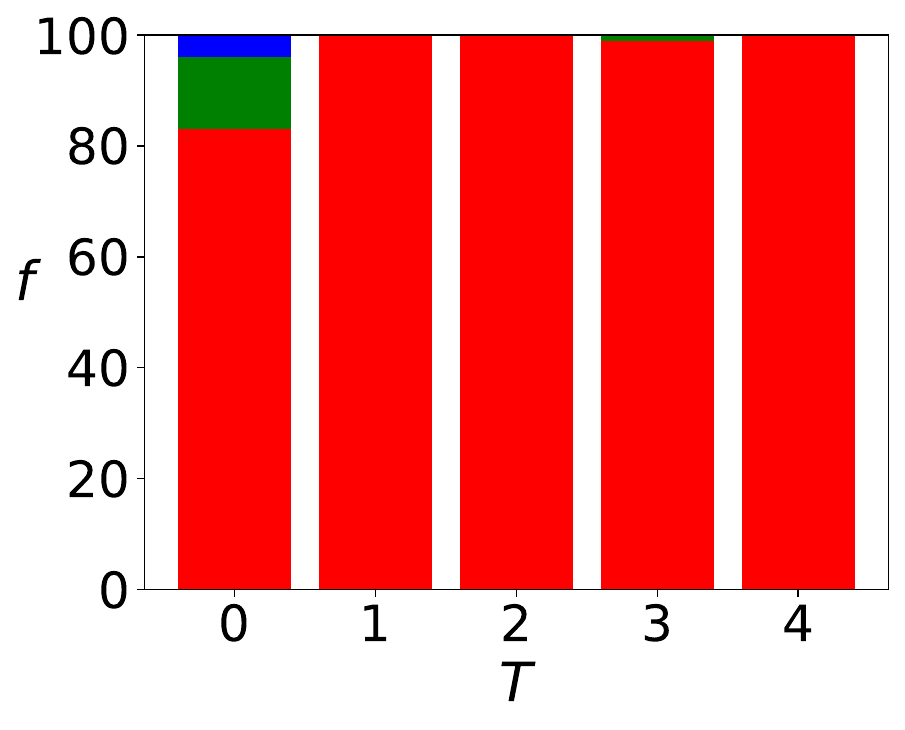}}\\
\subcaptionbox{$\alpha = 3$, $K = 2$\label{sfig:th12a3K2}}{\includegraphics[width=.24\textwidth]{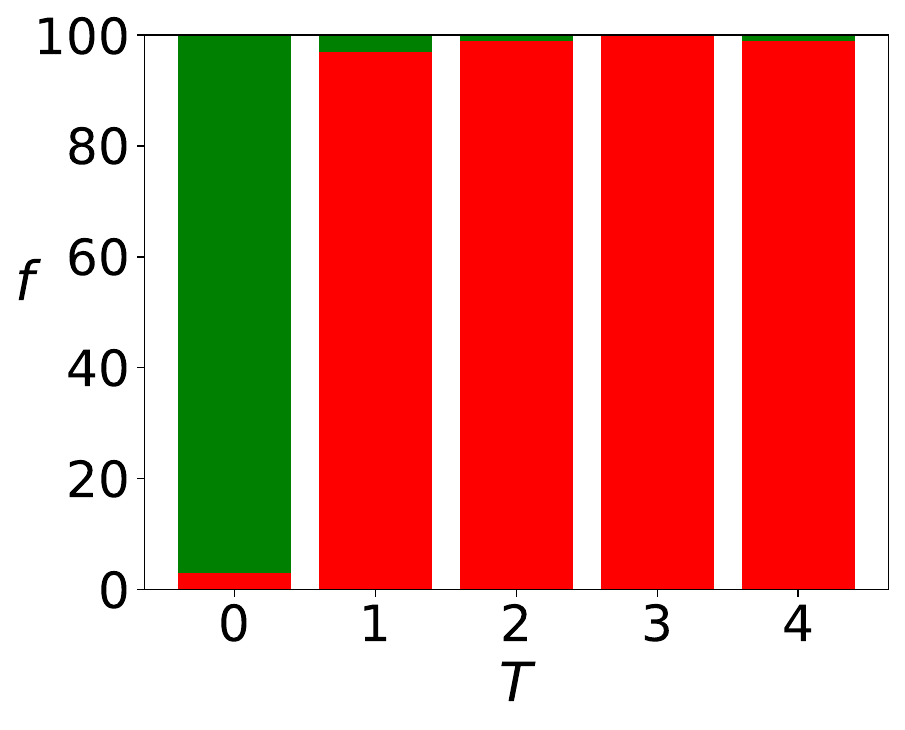}}\hfill
\subcaptionbox{$\alpha = 3$, $K = 3$\label{sfig:th12a3K3}}{\includegraphics[width=.24\textwidth]{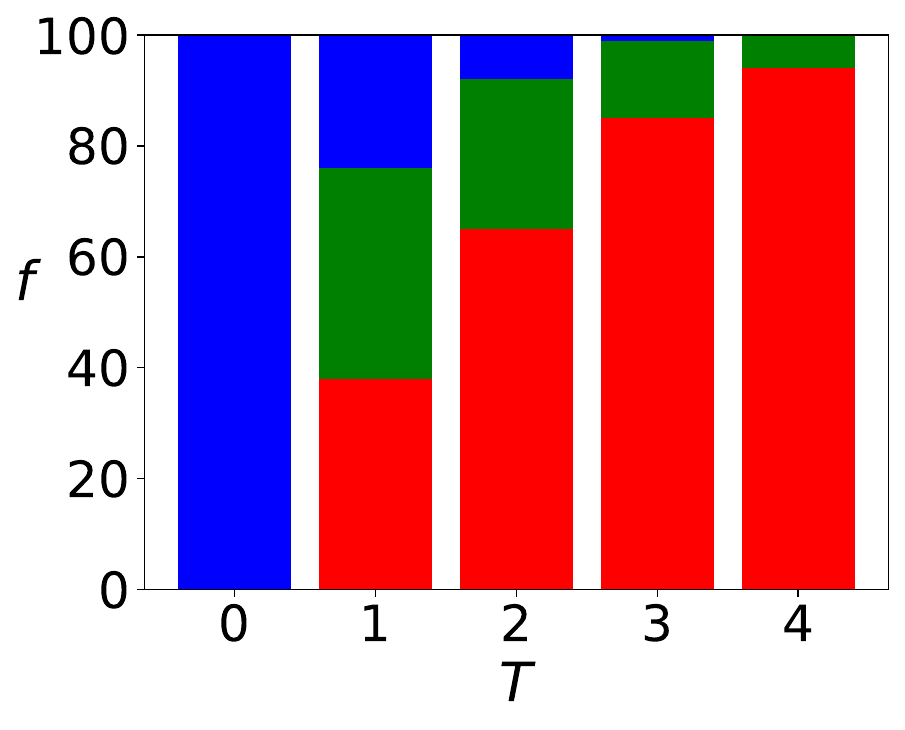}}\hfill
\subcaptionbox{$\alpha = 3$, $K = 4$\label{sfig:th12a3K4}}{\includegraphics[width=.24\textwidth]{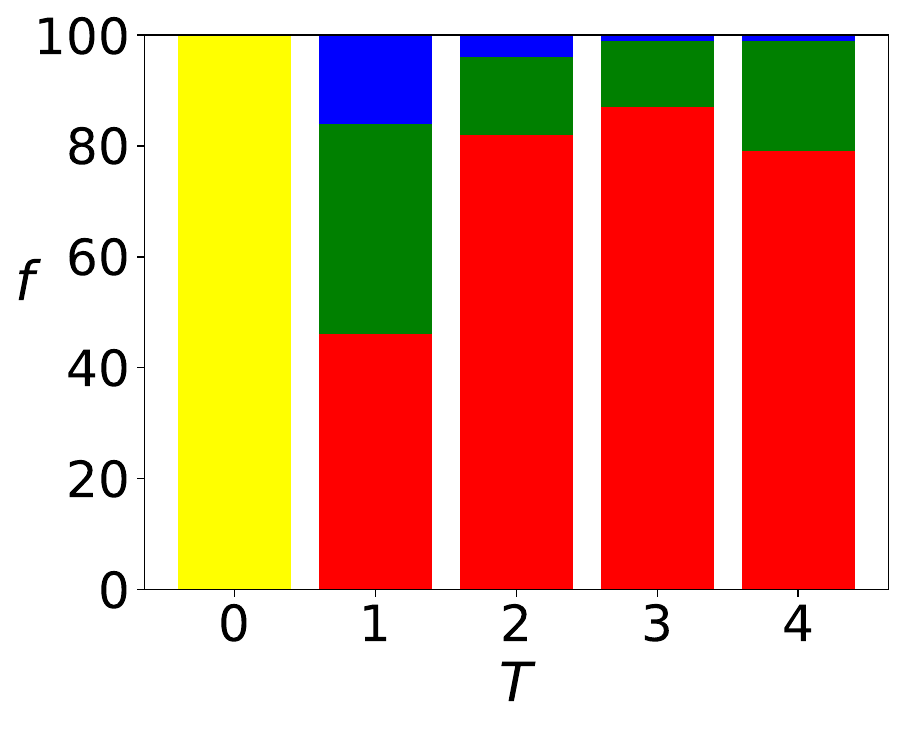}}\hfill
\subcaptionbox{$\alpha = 3$, $K = 5$\label{sfig:th12a3K5}}{\includegraphics[width=.24\textwidth]{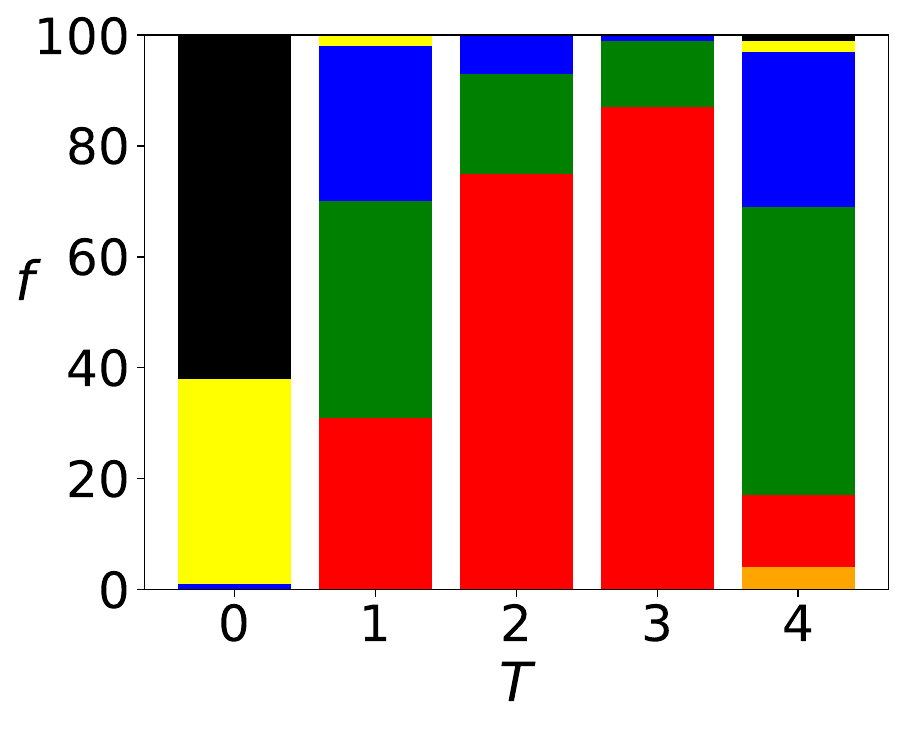}}\\
\subcaptionbox{$\alpha = 4$, $K = 2$\label{sfig:th12a4K2}}{\includegraphics[width=.24\textwidth]{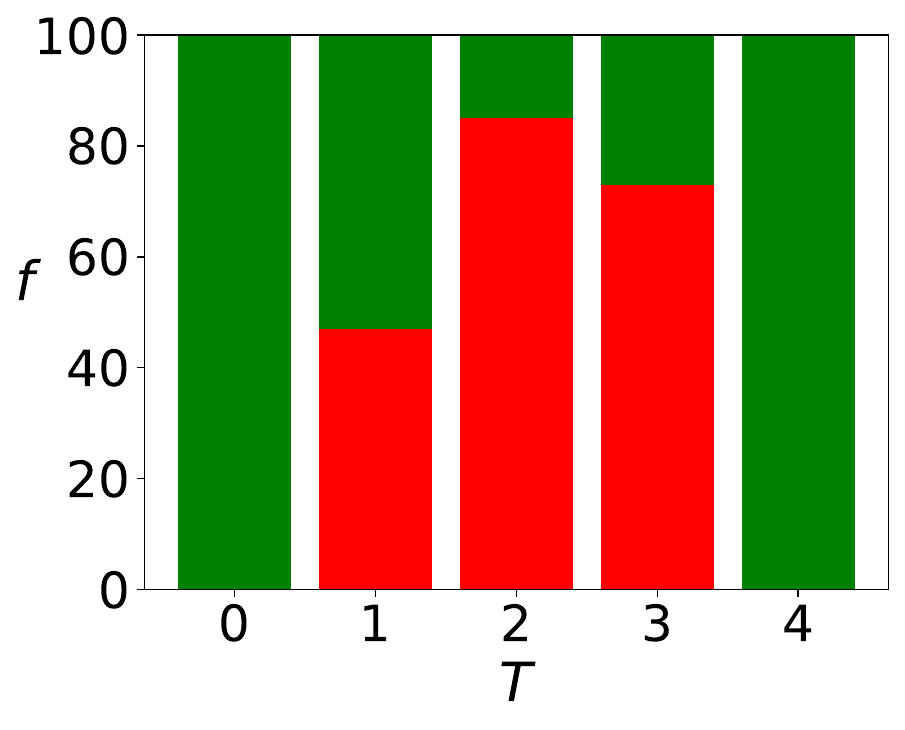}}\hfill
\subcaptionbox{$\alpha = 4$, $K = 3$\label{sfig:th12a4K3}}{\includegraphics[width=.24\textwidth]{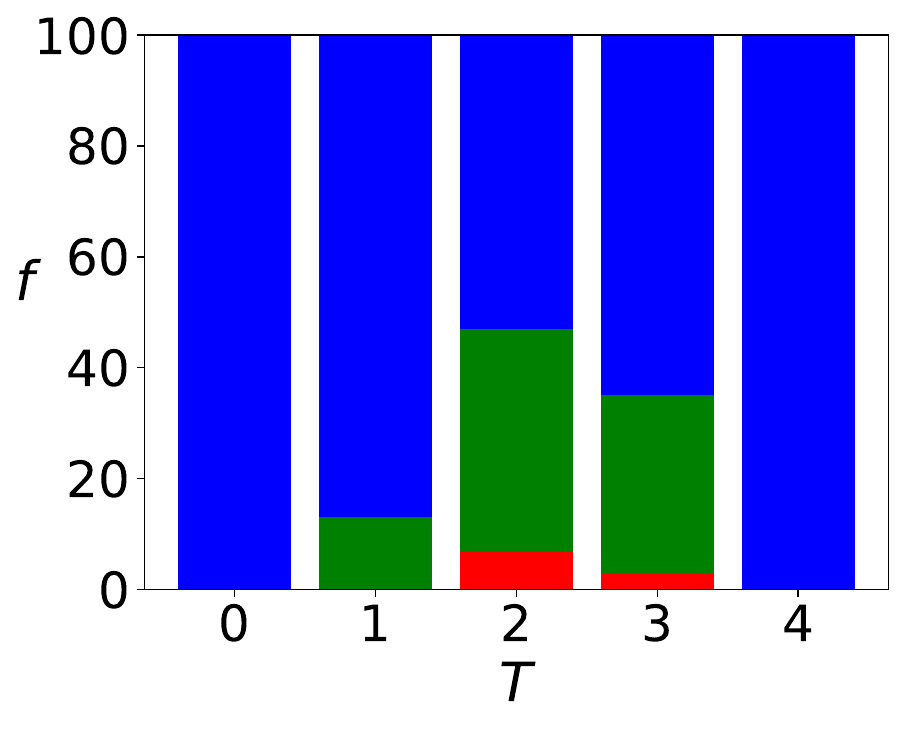}}\hfill
\subcaptionbox{$\alpha = 4$, $K = 4$\label{sfig:th12a4K4}}{\includegraphics[width=.24\textwidth]{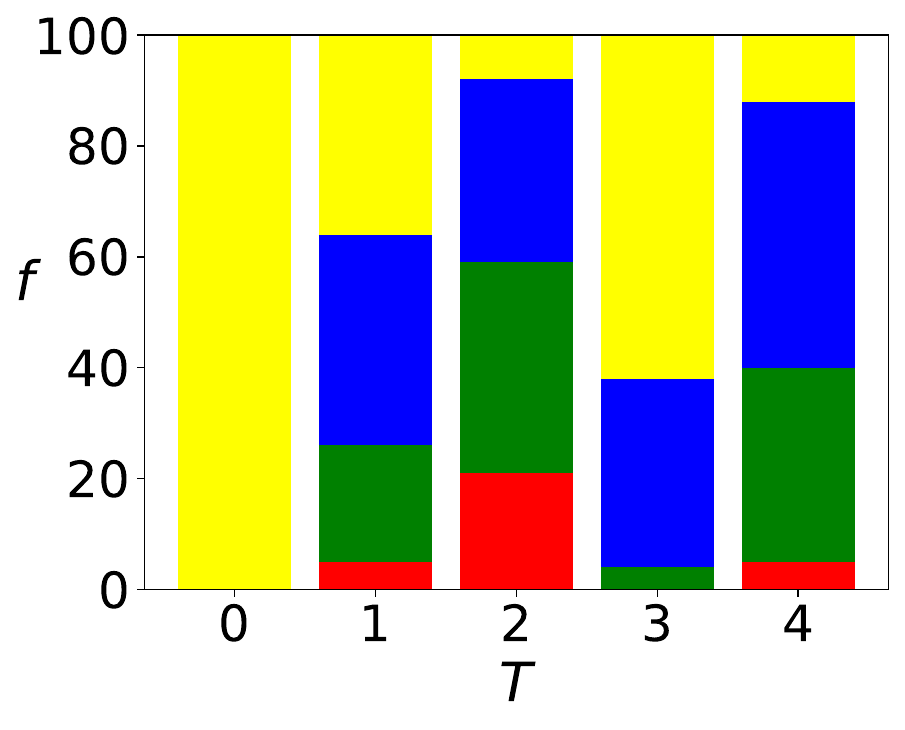}}\hfill
\subcaptionbox{$\alpha = 4$, $K = 5$\label{sfig:th12a4K5}}{\includegraphics[width=.24\textwidth]{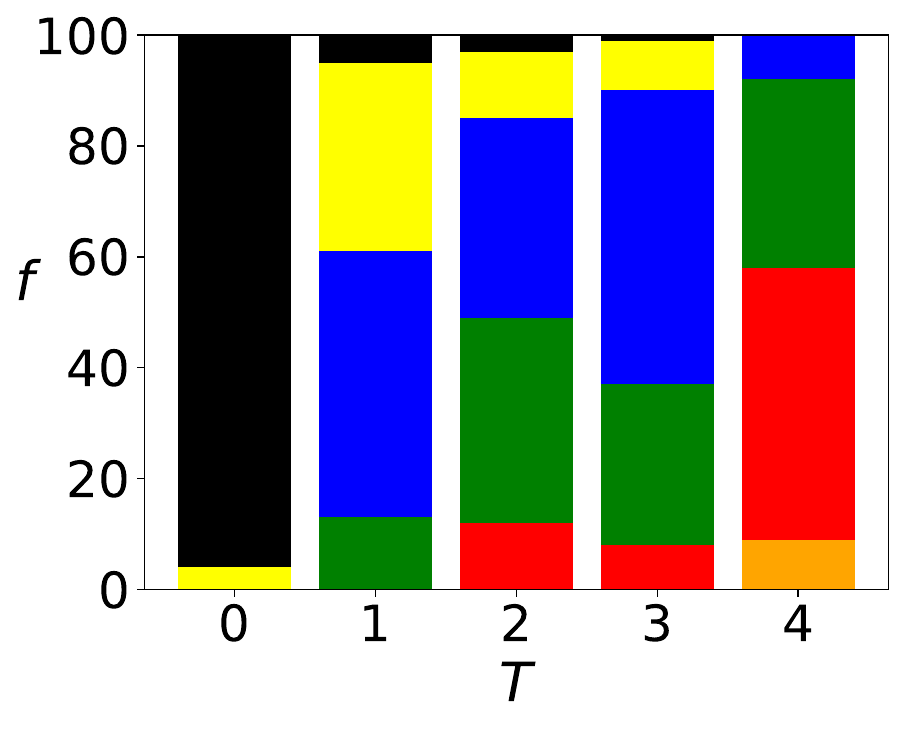}}\\
\subcaptionbox{$\alpha = 6$, $K = 2$\label{sfig:th12a6K2}}{\includegraphics[width=.24\textwidth]{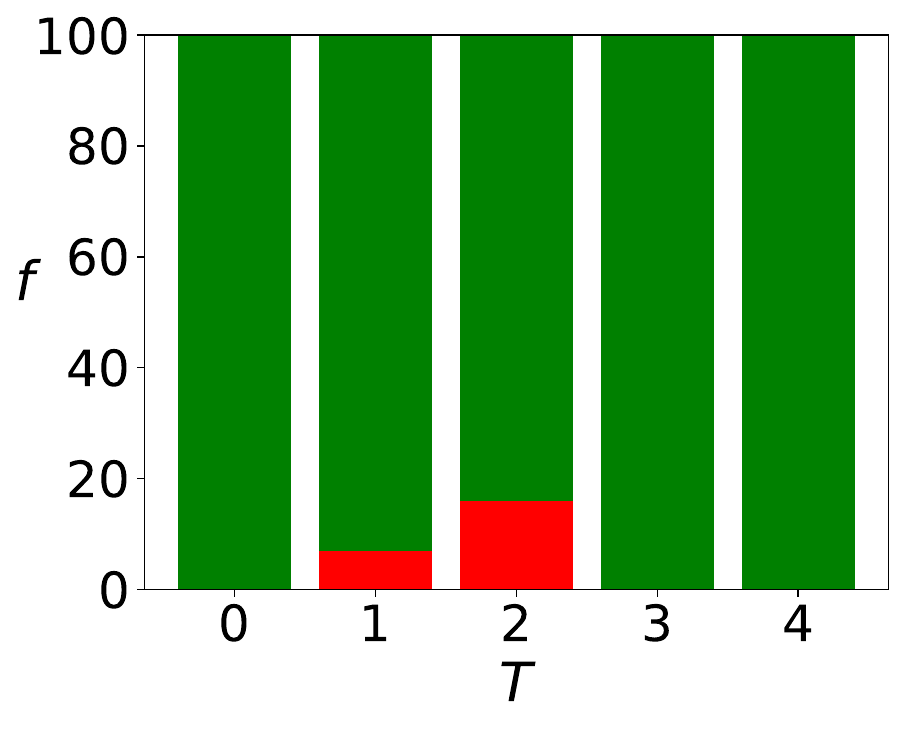}}\hfill
\subcaptionbox{$\alpha = 6$, $K = 3$\label{sfig:th12a6K3}}{\includegraphics[width=.24\textwidth]{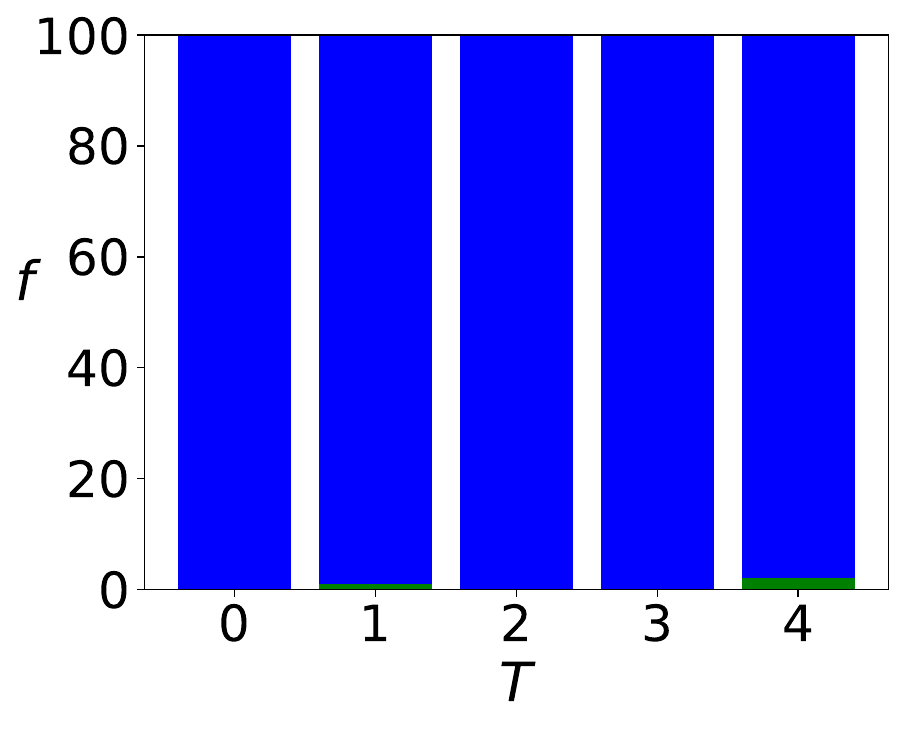}}\hfill
\subcaptionbox{$\alpha = 6$, $K = 4$\label{sfig:th12a6K4}}{\includegraphics[width=.24\textwidth]{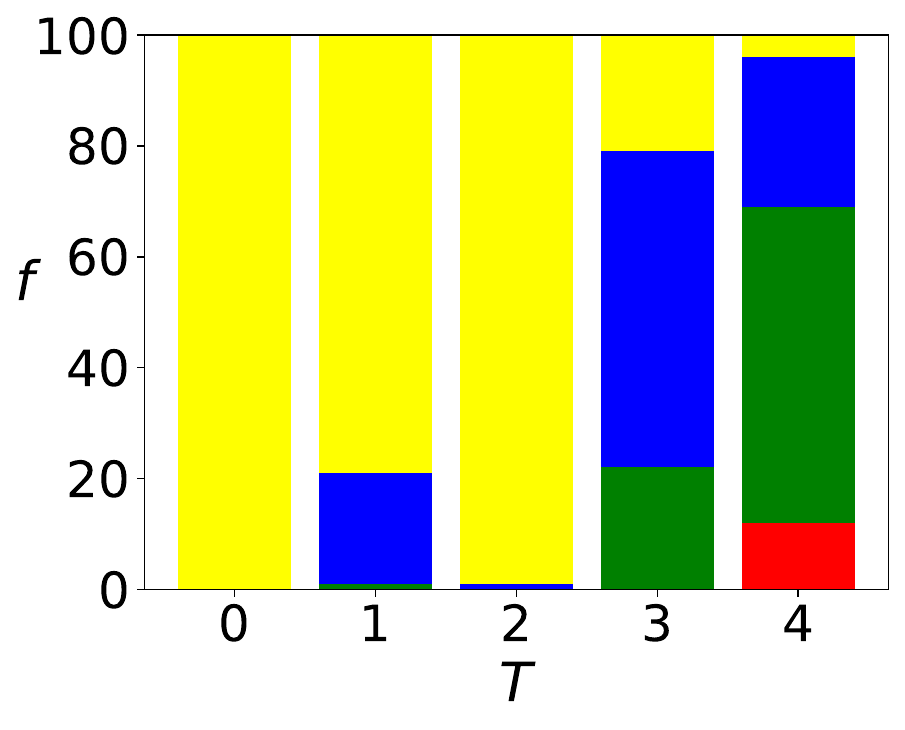}}\hfill
\subcaptionbox{$\alpha = 6$, $K = 5$\label{sfig:th12a6K5}}{\includegraphics[width=.24\textwidth]{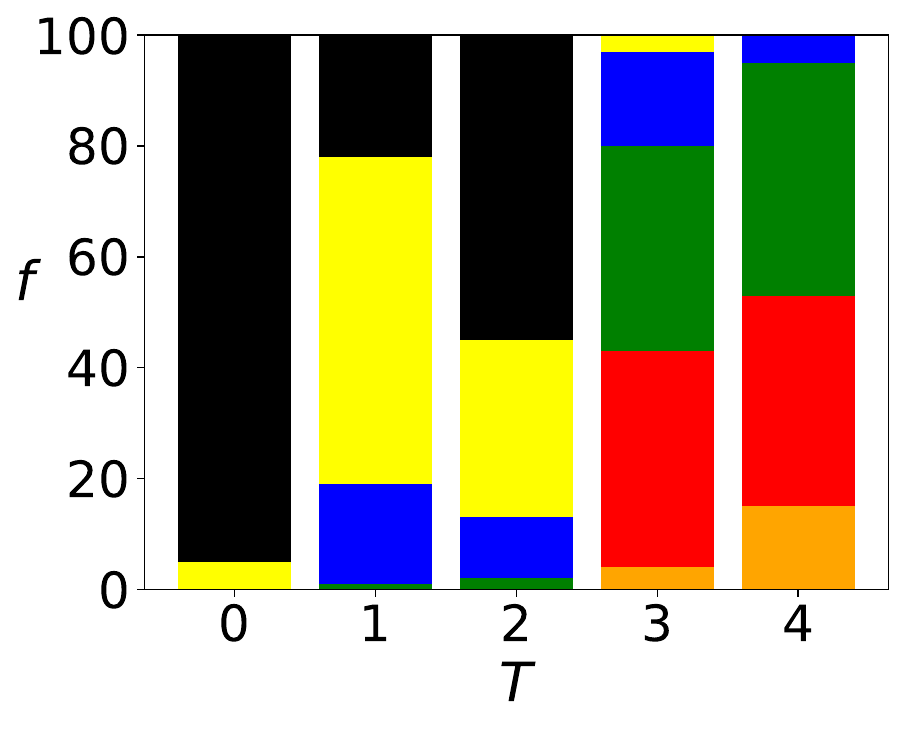}}\\
\caption{\label{fig:th12_sur}The histograms of frequencies $f$ of the number $\Phi$ of surviving opinions after $=10^3$ time steps and for various values of the distance scaling function exponent $\alpha$ and various values of the number of available opinions $K$. The system contains $L^2=41^2$ actors. The noise discrimination level $\theta=12$ and the results are averaged over $R=100$ independent simulations.}
\end{figure*}

\begin{figure*}[htbp]
\hfill\includegraphics[width=.12\textwidth]{img/legenda.png}\\
\subcaptionbox{$\alpha = 2$, $K = 2$\label{sfig:th25a2K2}}{\includegraphics[width=.24\textwidth]{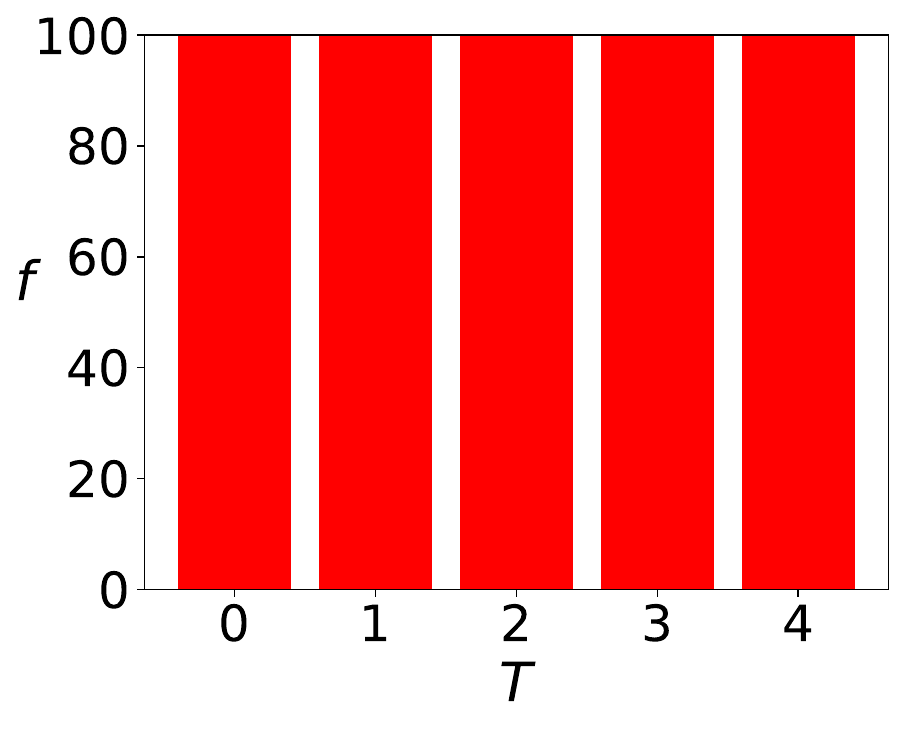}}\hfill
\subcaptionbox{$\alpha = 2$, $K = 3$\label{sfig:th25a2K3}}{\includegraphics[width=.24\textwidth]{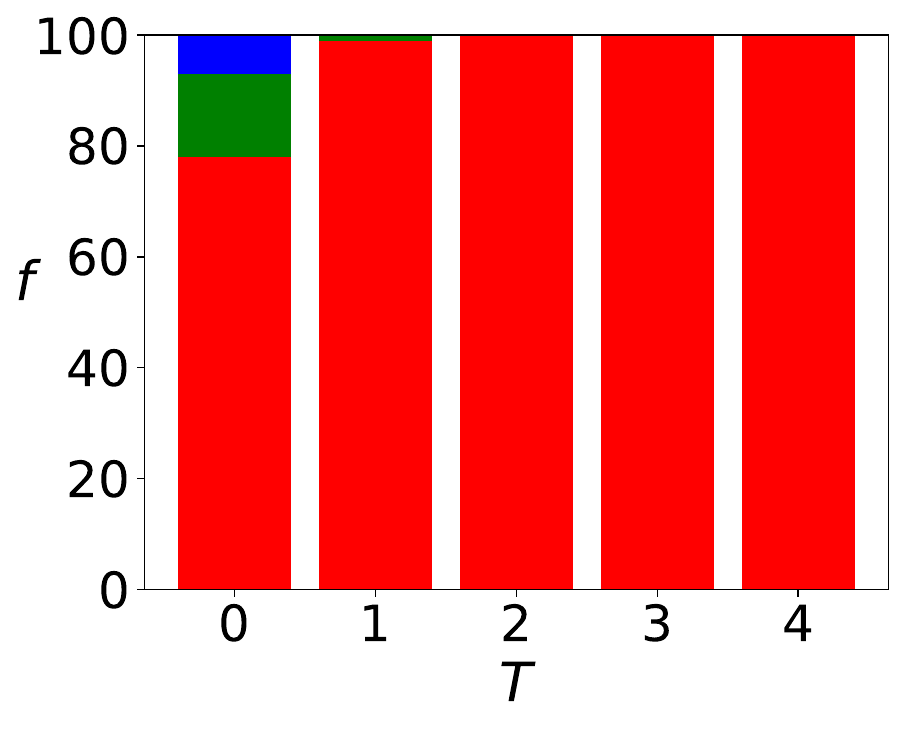}}\hfill
\subcaptionbox{$\alpha = 2$, $K = 4$\label{sfig:th25a2K4}}{\includegraphics[width=.24\textwidth]{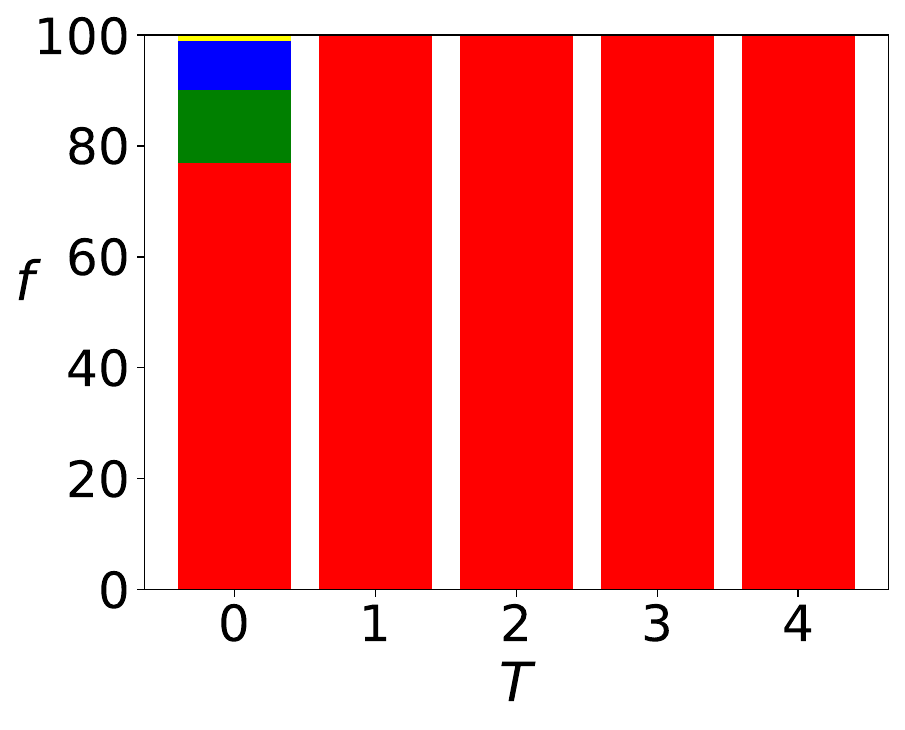}}\hfill
\subcaptionbox{$\alpha = 2$, $K = 5$\label{sfig:th25a2K5}}{\includegraphics[width=.24\textwidth]{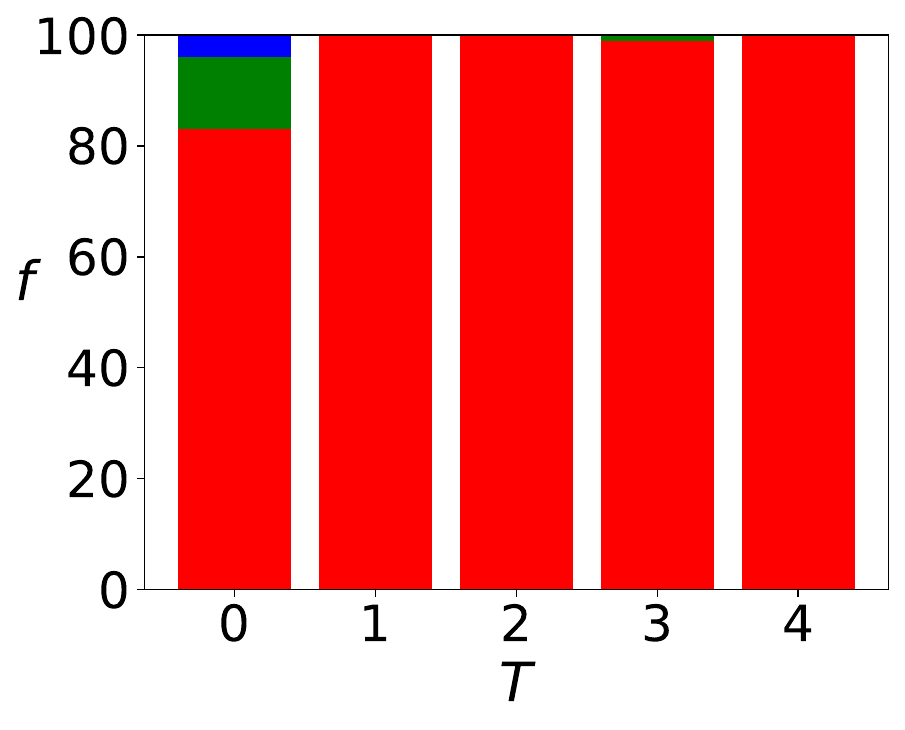}}\\
\subcaptionbox{$\alpha = 3$, $K = 2$\label{sfig:th25a3K2}}{\includegraphics[width=.24\textwidth]{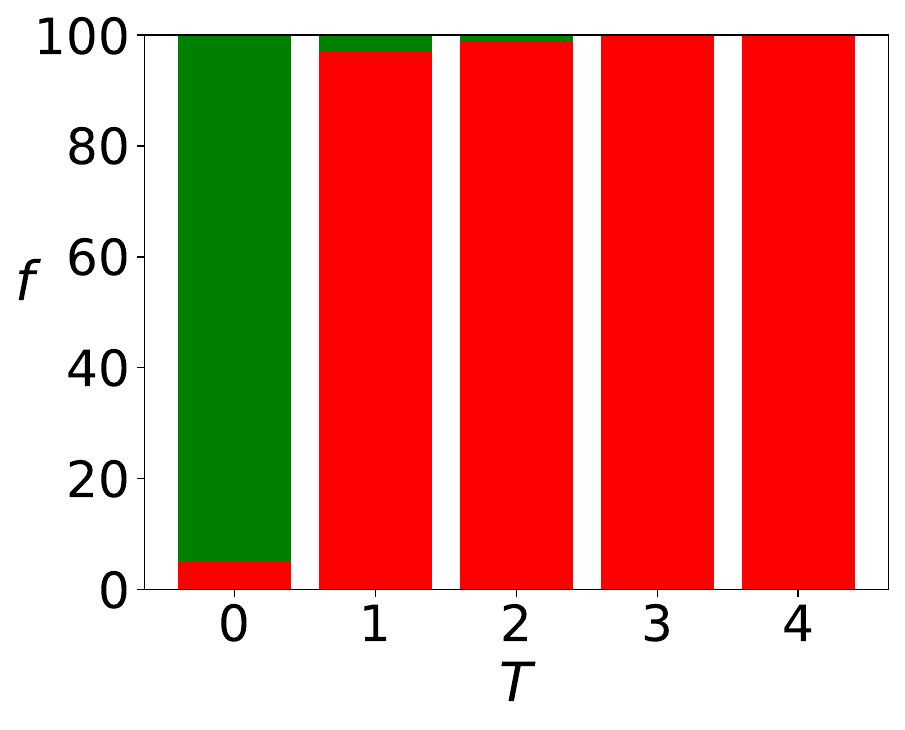}}\hfill
\subcaptionbox{$\alpha = 3$, $K = 3$\label{sfig:th25a3K3}}{\includegraphics[width=.24\textwidth]{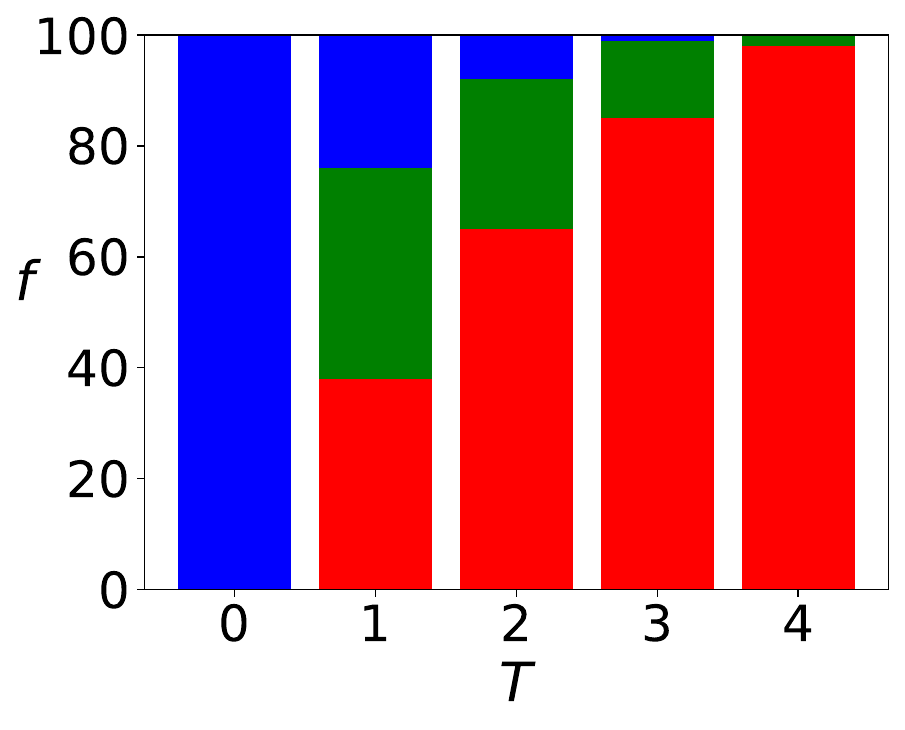}}\hfill
\subcaptionbox{$\alpha = 3$, $K = 4$\label{sfig:th25a3K4}}{\includegraphics[width=.24\textwidth]{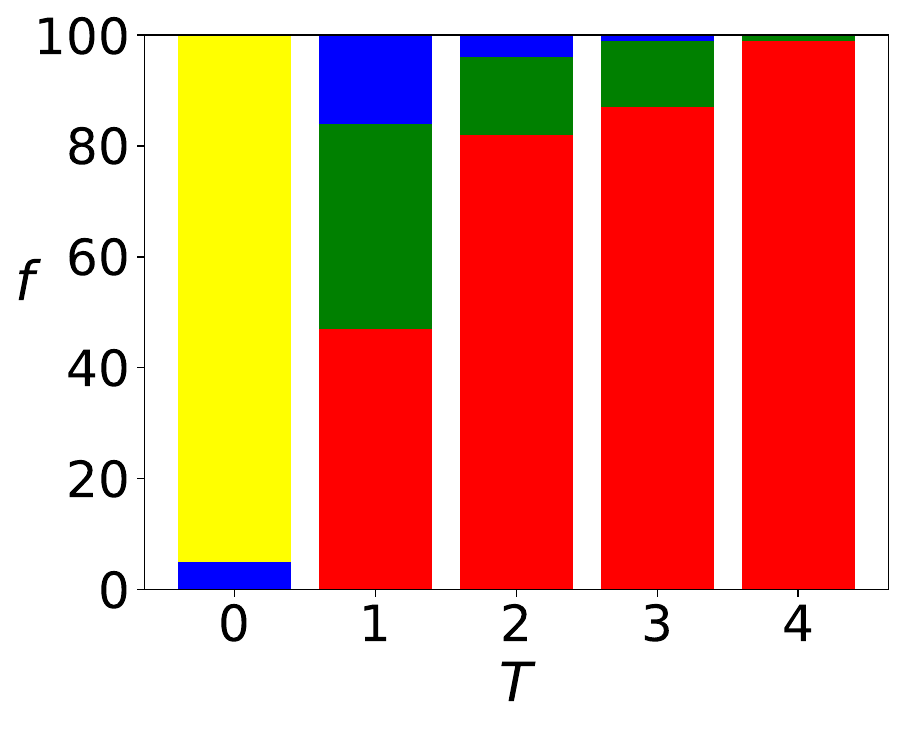}}\hfill
\subcaptionbox{$\alpha = 3$, $K = 5$\label{sfig:th25a3K5}}{\includegraphics[width=.24\textwidth]{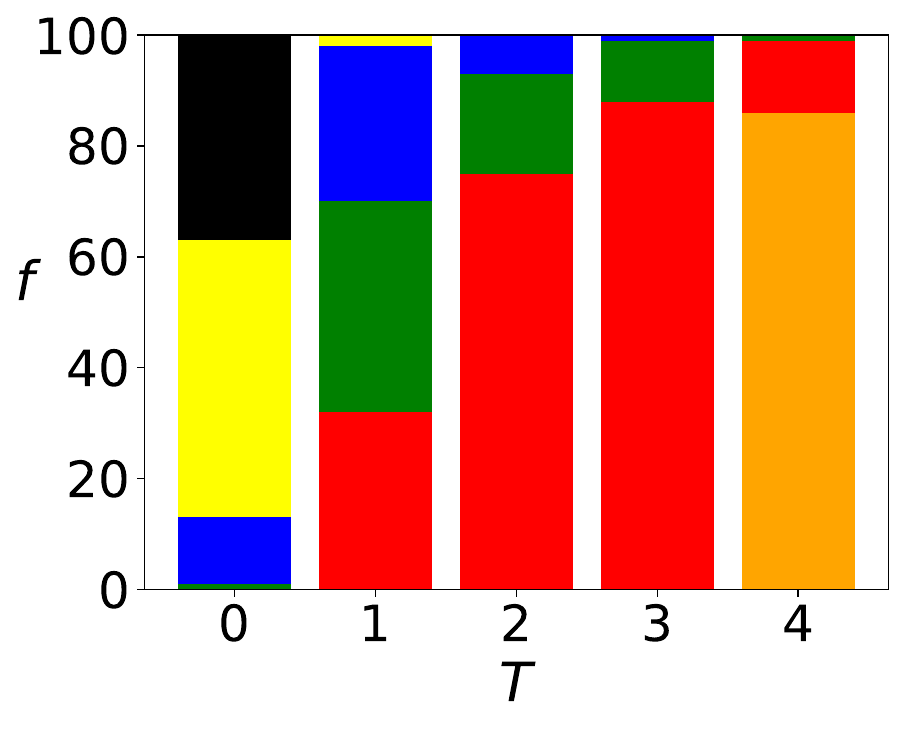}}\\
\subcaptionbox{$\alpha = 4$, $K = 2$\label{sfig:th25a4K2}}{\includegraphics[width=.24\textwidth]{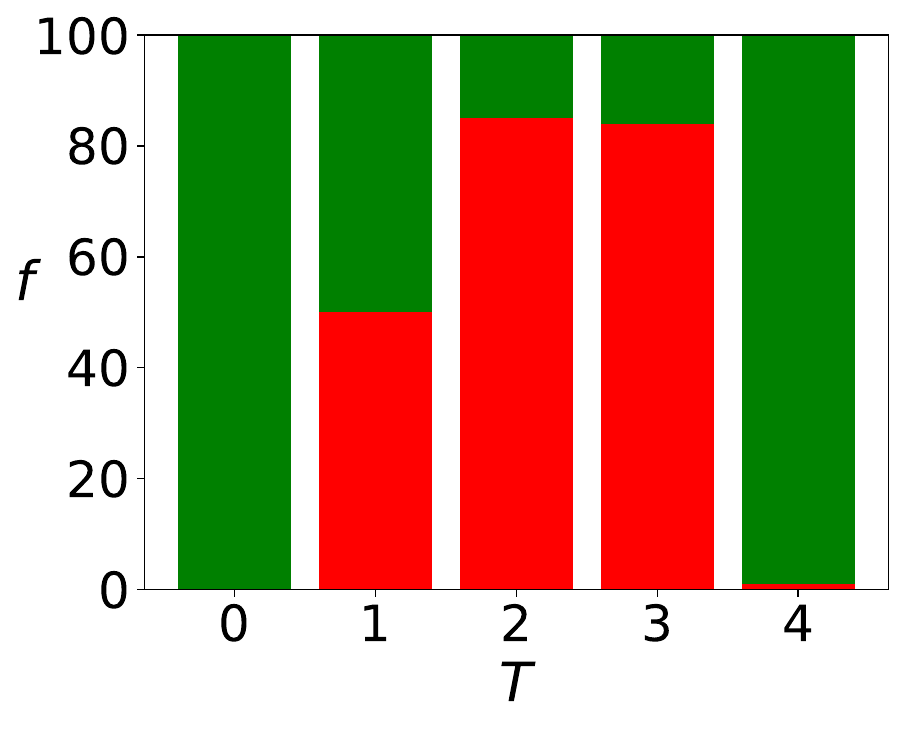}}\hfill
\subcaptionbox{$\alpha = 4$, $K = 3$\label{sfig:th25a4K3}}{\includegraphics[width=.24\textwidth]{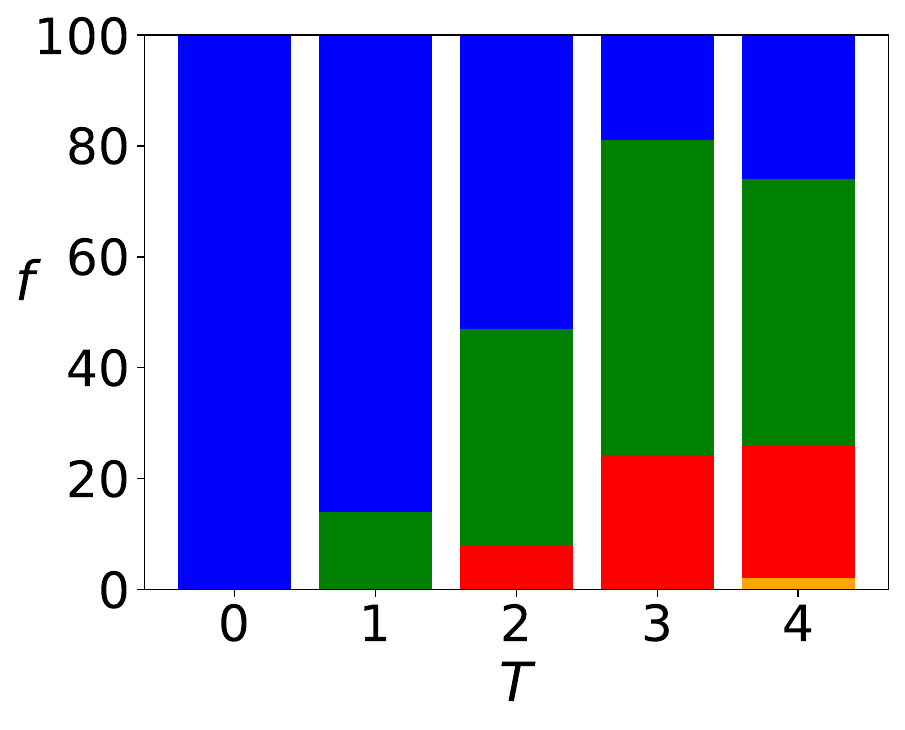}}\hfill
\subcaptionbox{$\alpha = 4$, $K = 4$\label{sfig:th25a4K4}}{\includegraphics[width=.24\textwidth]{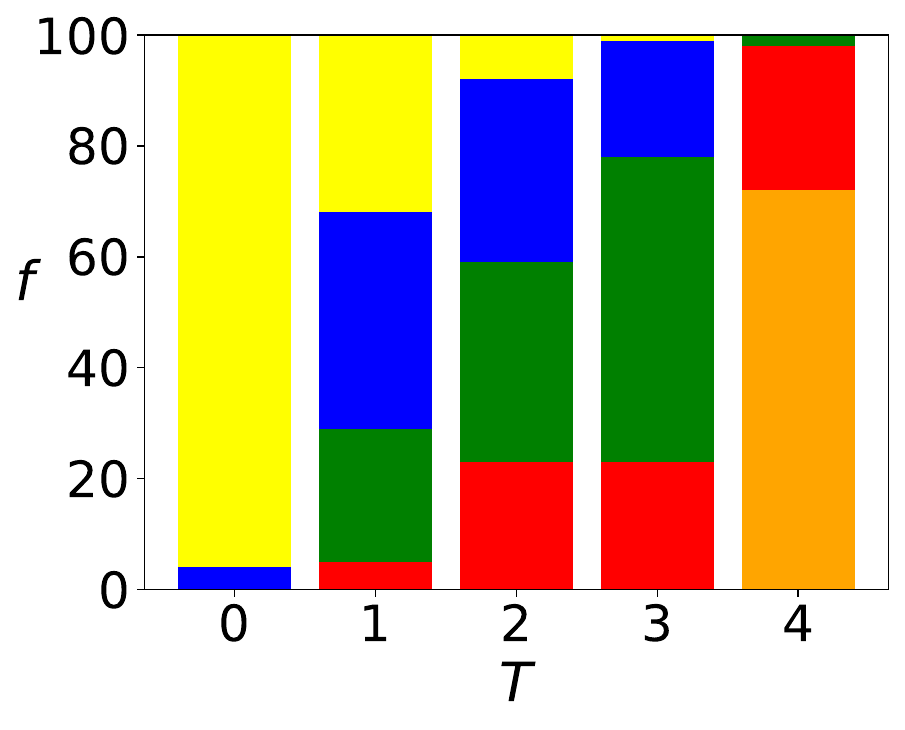}}\hfill
\subcaptionbox{$\alpha = 4$, $K = 5$\label{sfig:th25a4K5}}{\includegraphics[width=.24\textwidth]{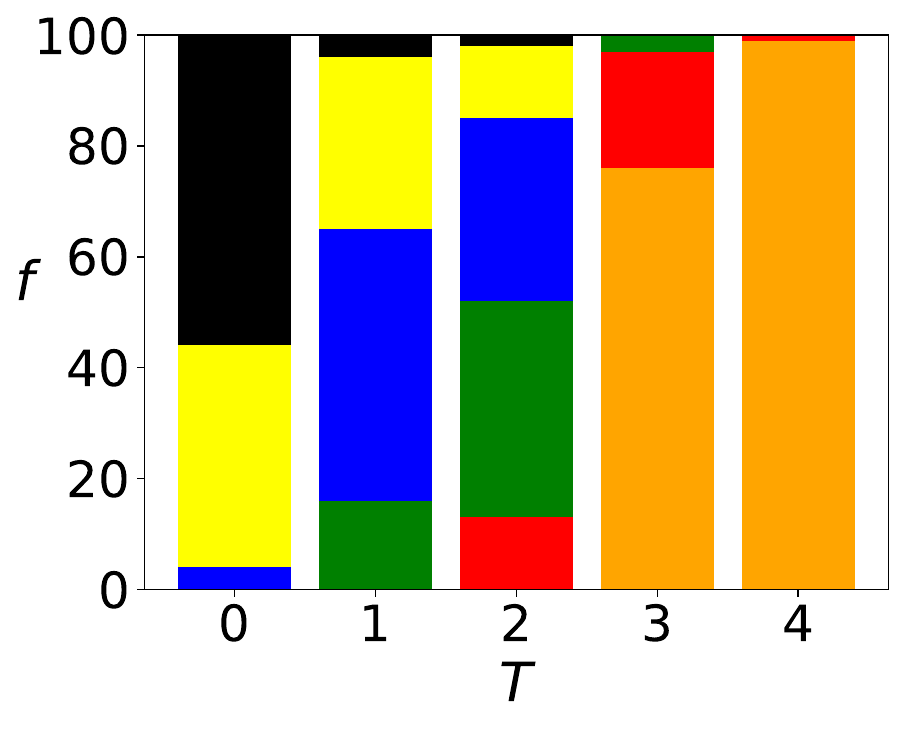}}\\
\subcaptionbox{$\alpha = 6$, $K = 2$\label{sfig:th25a6K2}}{\includegraphics[width=.24\textwidth]{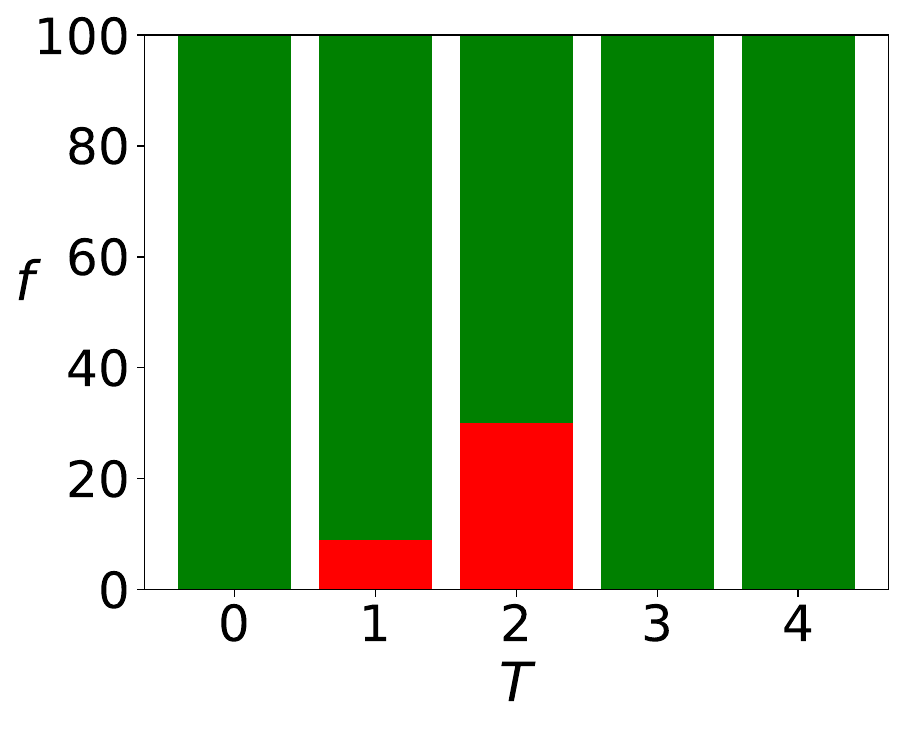}}\hfill
\subcaptionbox{$\alpha = 6$, $K = 3$\label{sfig:th25a6K3}}{\includegraphics[width=.24\textwidth]{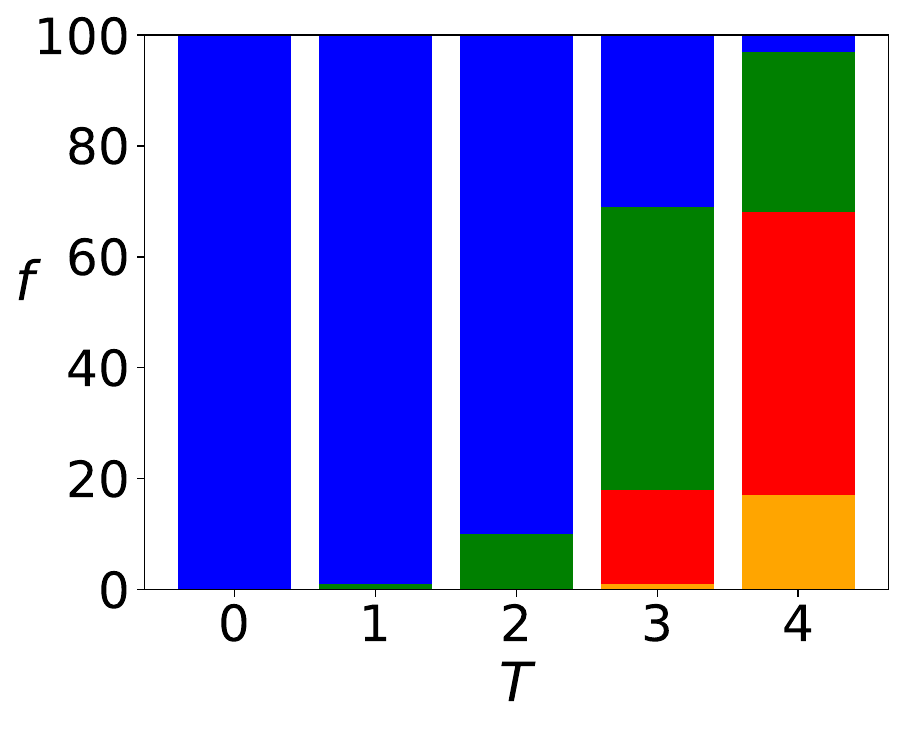}}\hfill
\subcaptionbox{$\alpha = 6$, $K = 4$\label{sfig:th25a6K4}}{\includegraphics[width=.24\textwidth]{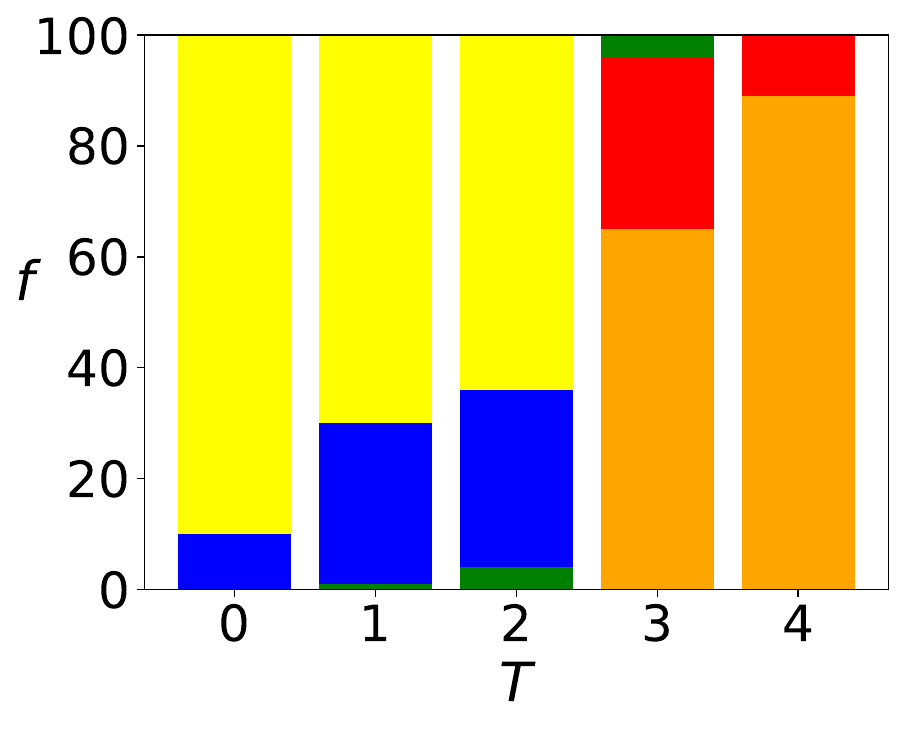}}\hfill
\subcaptionbox{$\alpha = 6$, $K = 5$\label{sfig:th25a6K5}}{\includegraphics[width=.24\textwidth]{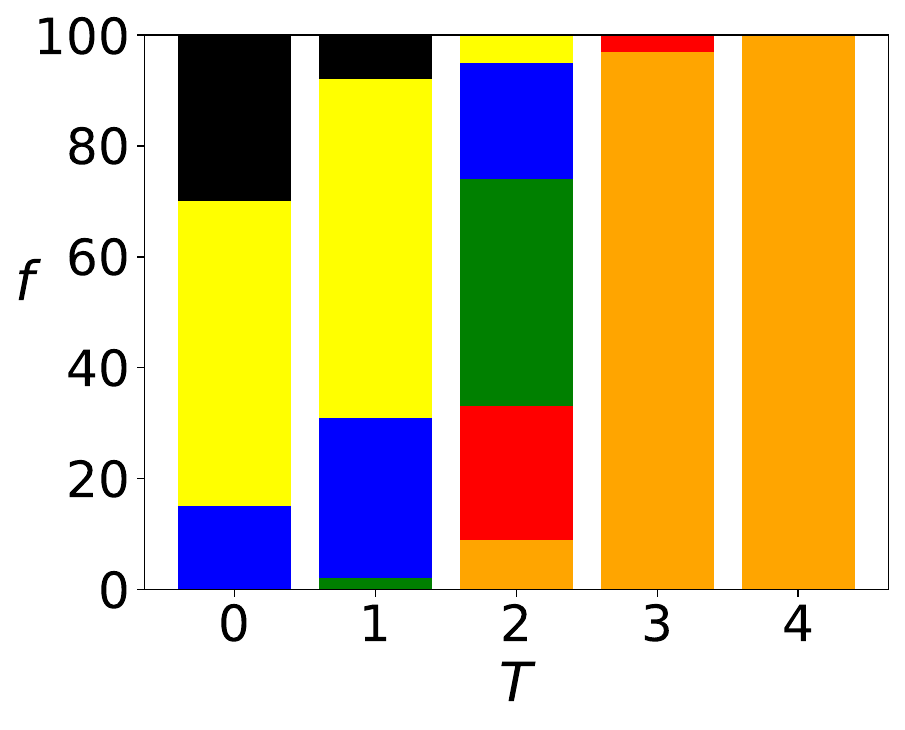}}\\
\caption{\label{fig:th25_sur}The histograms of frequencies $f$ of the number $\Phi$ of surviving opinions after $=10^3$ time steps and for various values of the distance scaling function exponent $\alpha$ and various values of the number of available opinions $K$. The system contains $L^2=41^2$ actors. The noise discrimination level $\theta=25$ and the results are averaged over $R=100$ independent simulations.}
\end{figure*}

\begin{figure*}[htbp]
\hfill\includegraphics[width=.12\textwidth]{img/legenda.png}\\
\subcaptionbox{$\alpha = 2$, $K = 2$\label{sfig:th50a2K2}}{\includegraphics[width=.24\textwidth]{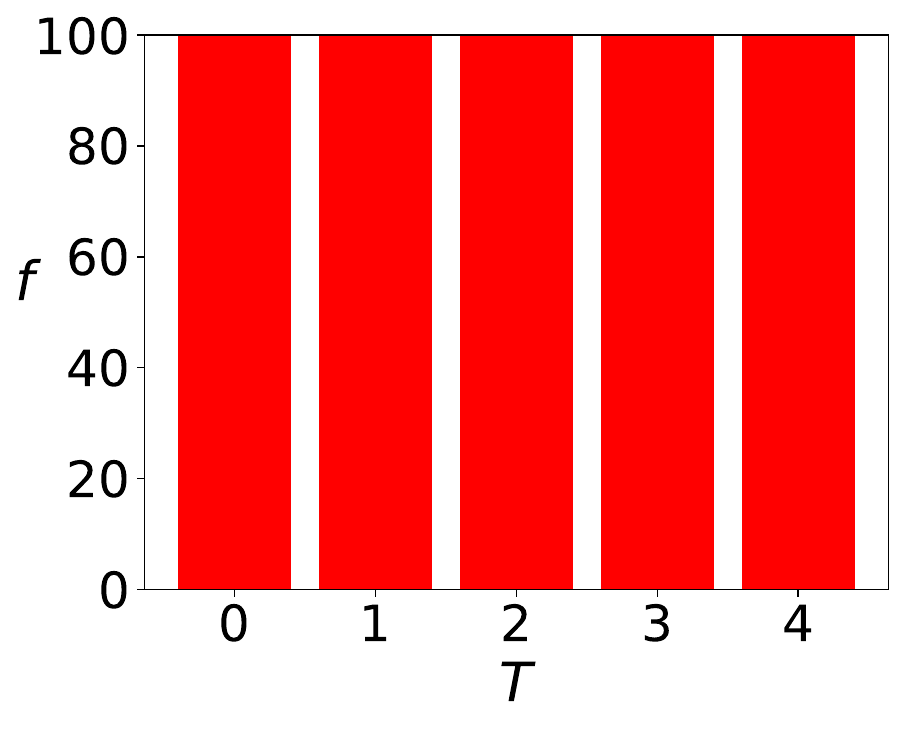}}\hfill
\subcaptionbox{$\alpha = 2$, $K = 3$\label{sfig:th50a2K3}}{\includegraphics[width=.24\textwidth]{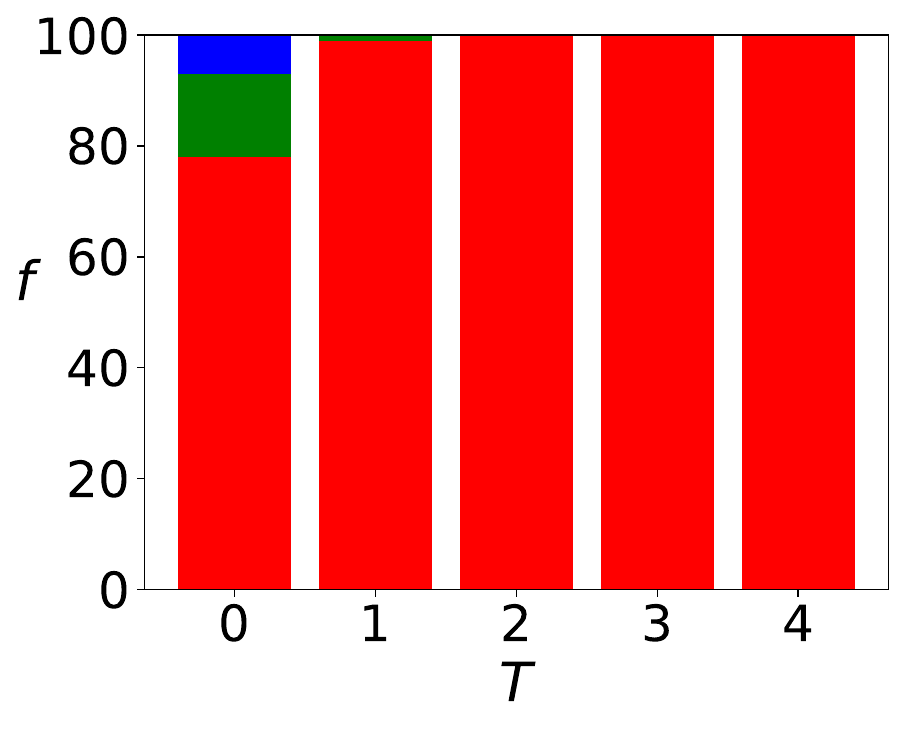}}\hfill
\subcaptionbox{$\alpha = 2$, $K = 4$\label{sfig:th50a2K4}}{\includegraphics[width=.24\textwidth]{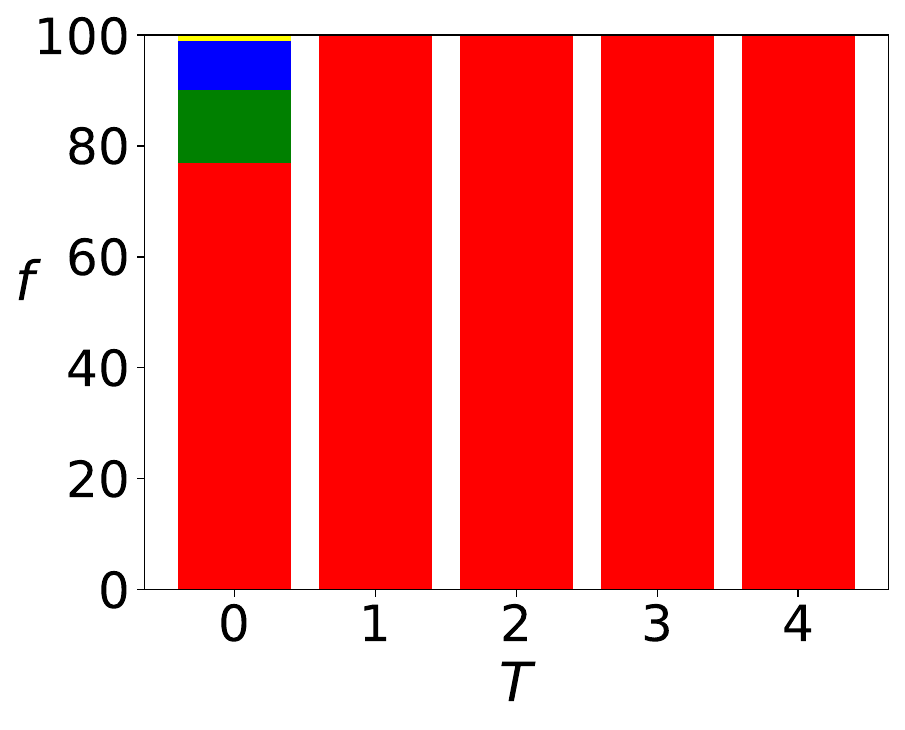}}\hfill
\subcaptionbox{$\alpha = 2$, $K = 5$\label{sfig:th50a2K5}}{\includegraphics[width=.24\textwidth]{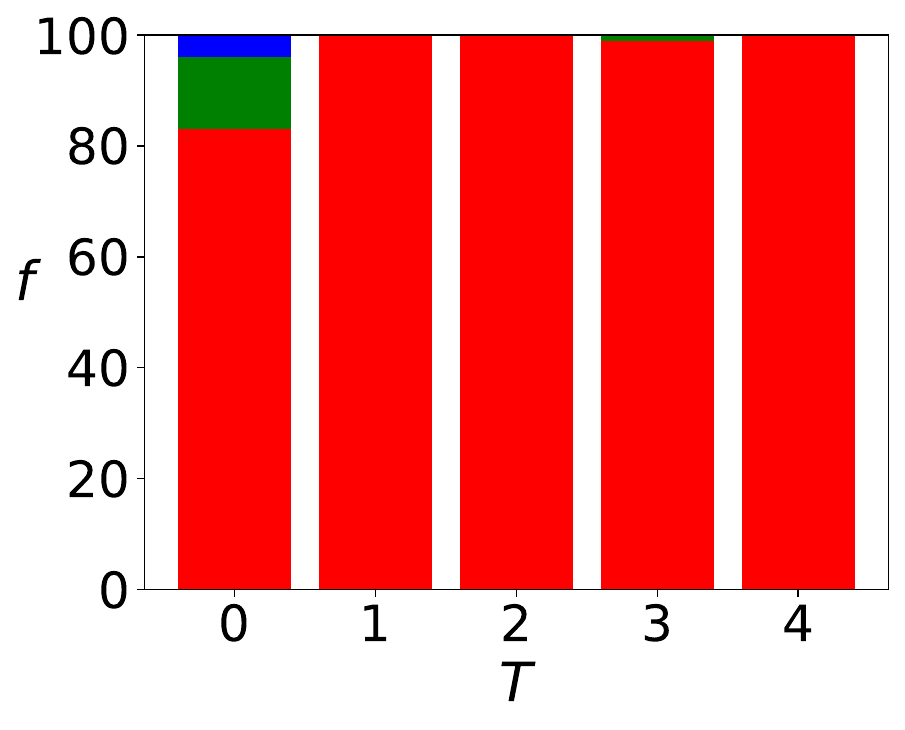}}\\
\subcaptionbox{$\alpha = 3$, $K = 2$\label{sfig:th50a3K2}}{\includegraphics[width=.24\textwidth]{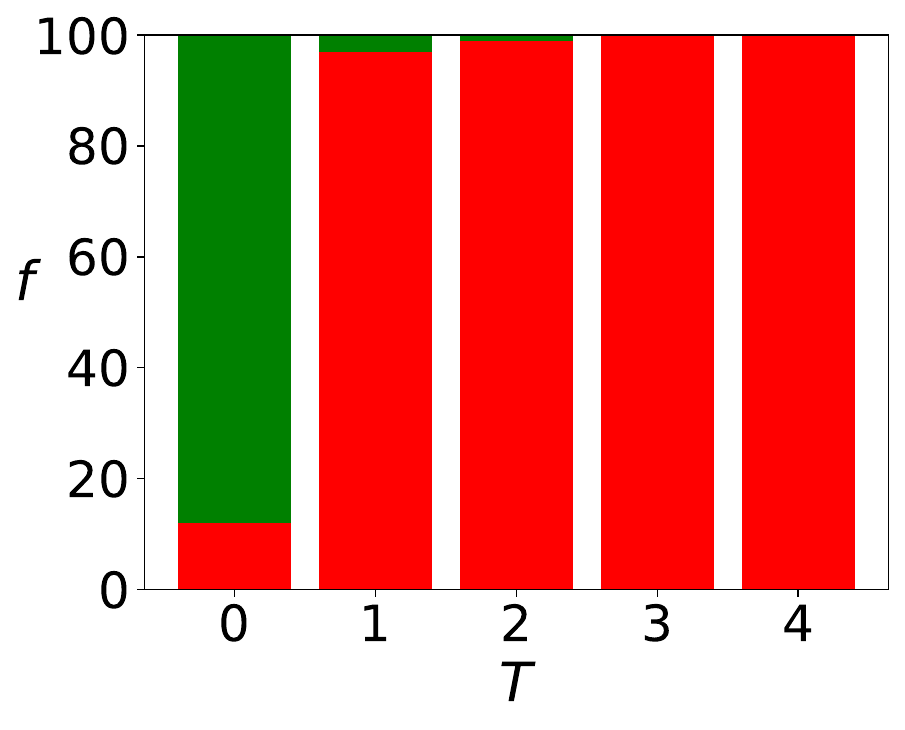}}\hfill
\subcaptionbox{$\alpha = 3$, $K = 3$\label{sfig:th50a3K3}}{\includegraphics[width=.24\textwidth]{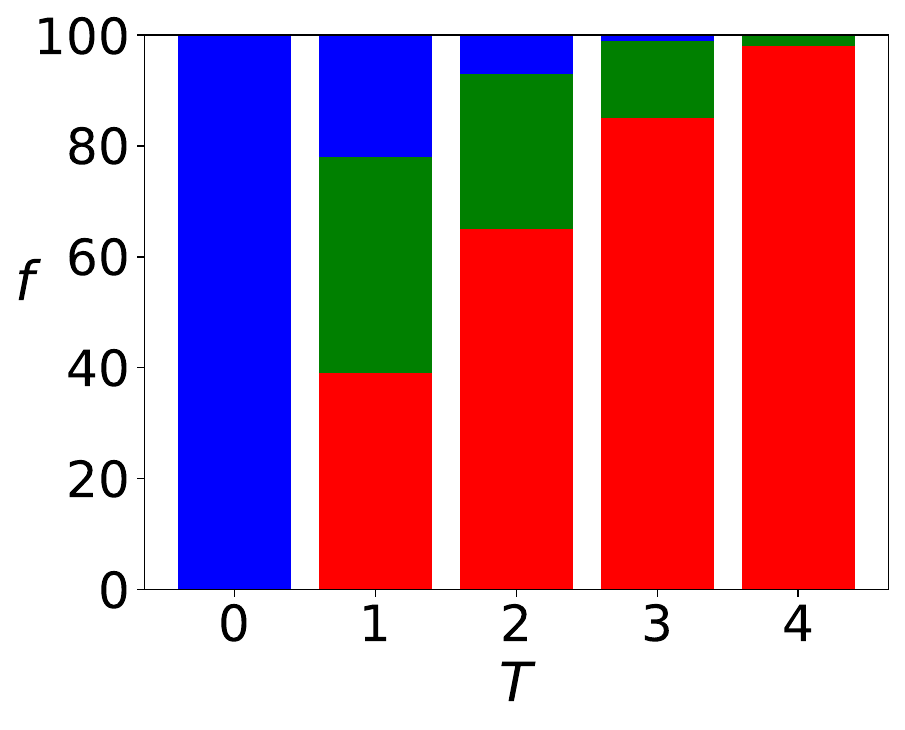}}\hfill
\subcaptionbox{$\alpha = 3$, $K = 4$\label{sfig:th50a3K4}}{\includegraphics[width=.24\textwidth]{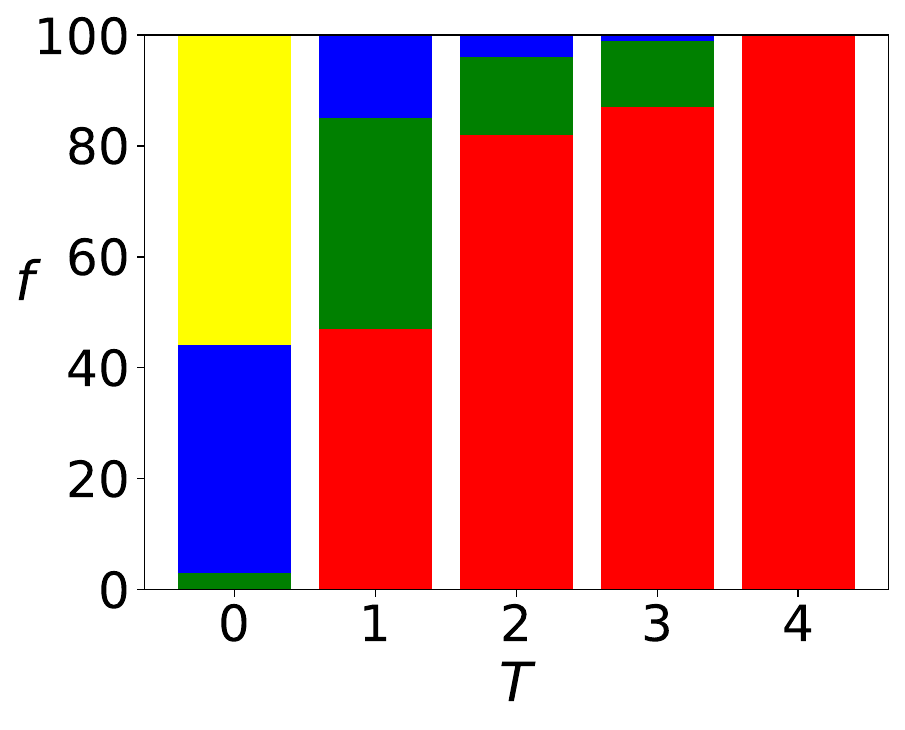}}\hfill
\subcaptionbox{$\alpha = 3$, $K = 5$\label{sfig:th50a3K5}}{\includegraphics[width=.24\textwidth]{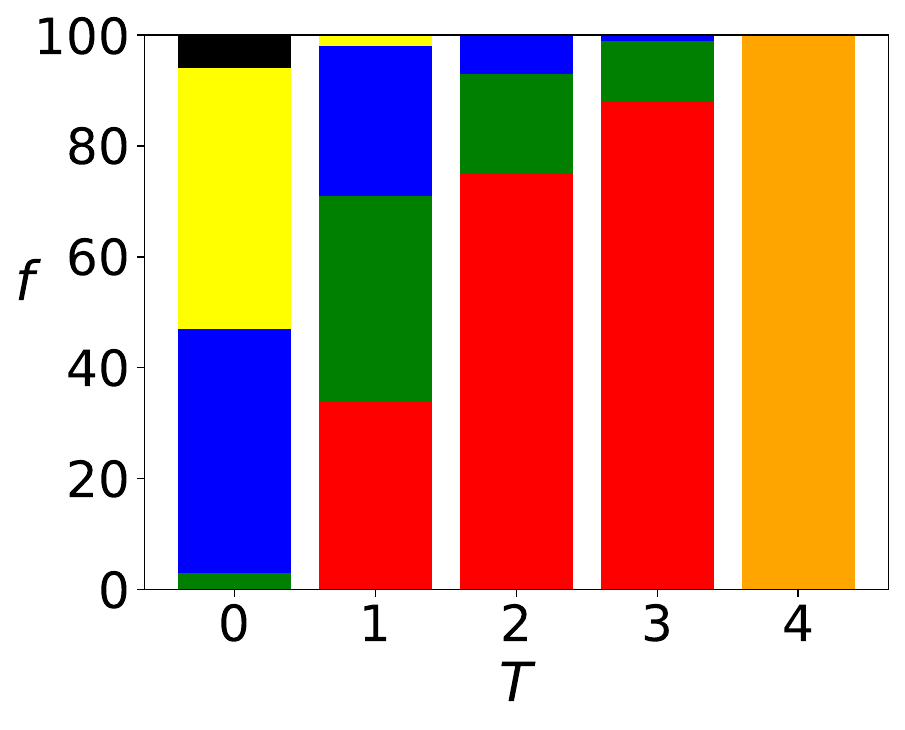}}\\
\subcaptionbox{$\alpha = 4$, $K = 2$\label{sfig:th50a4K2}}{\includegraphics[width=.24\textwidth]{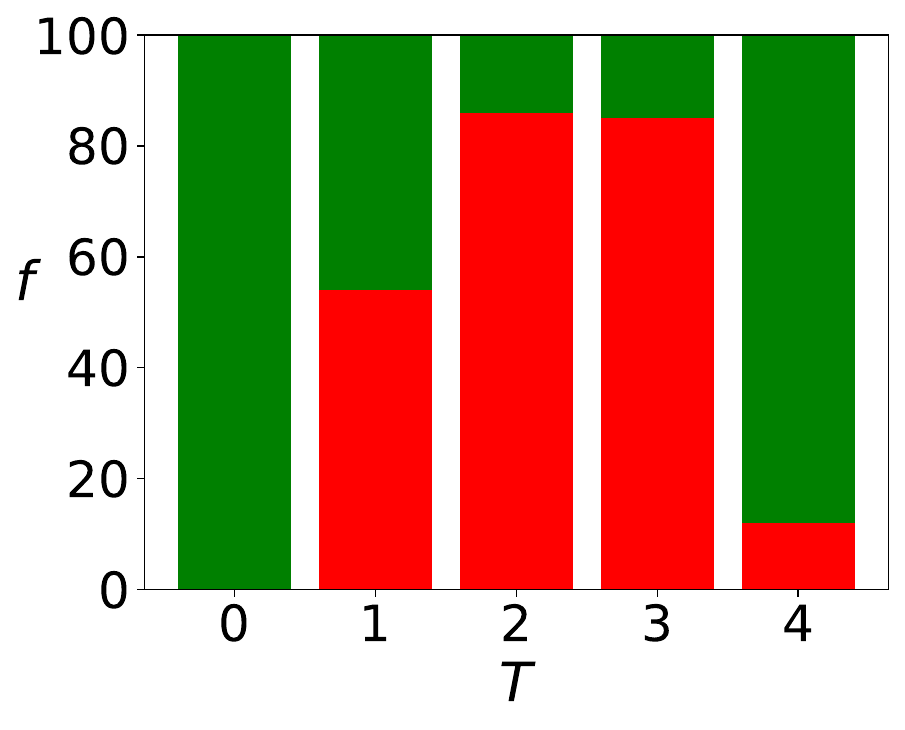}}\hfill
\subcaptionbox{$\alpha = 4$, $K = 3$\label{sfig:th50a4K3}}{\includegraphics[width=.24\textwidth]{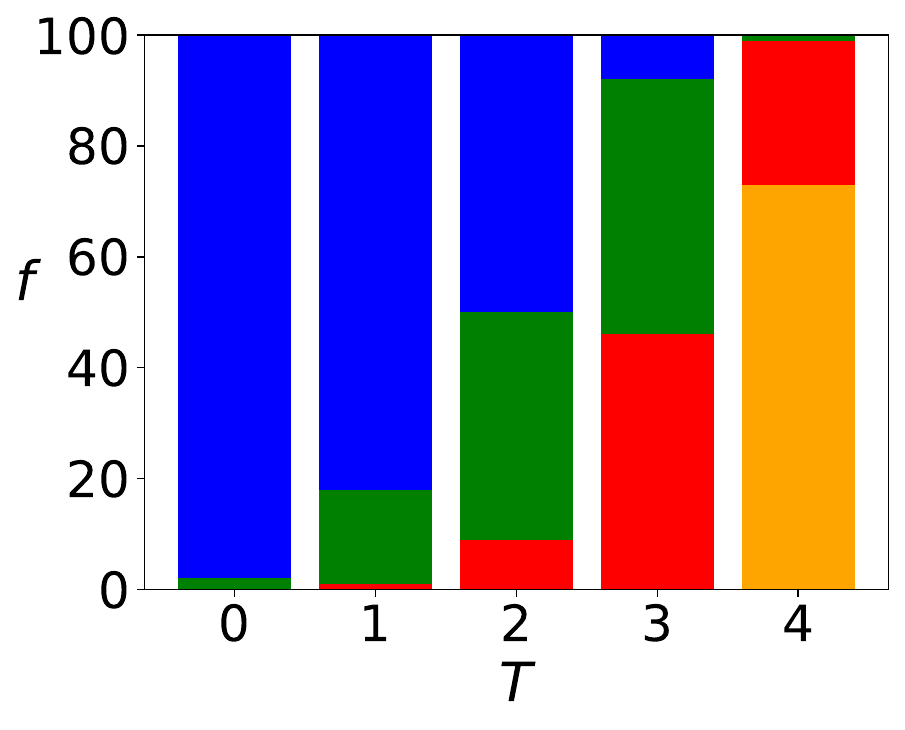}}\hfill
\subcaptionbox{$\alpha = 4$, $K = 4$\label{sfig:th50a4K4}}{\includegraphics[width=.24\textwidth]{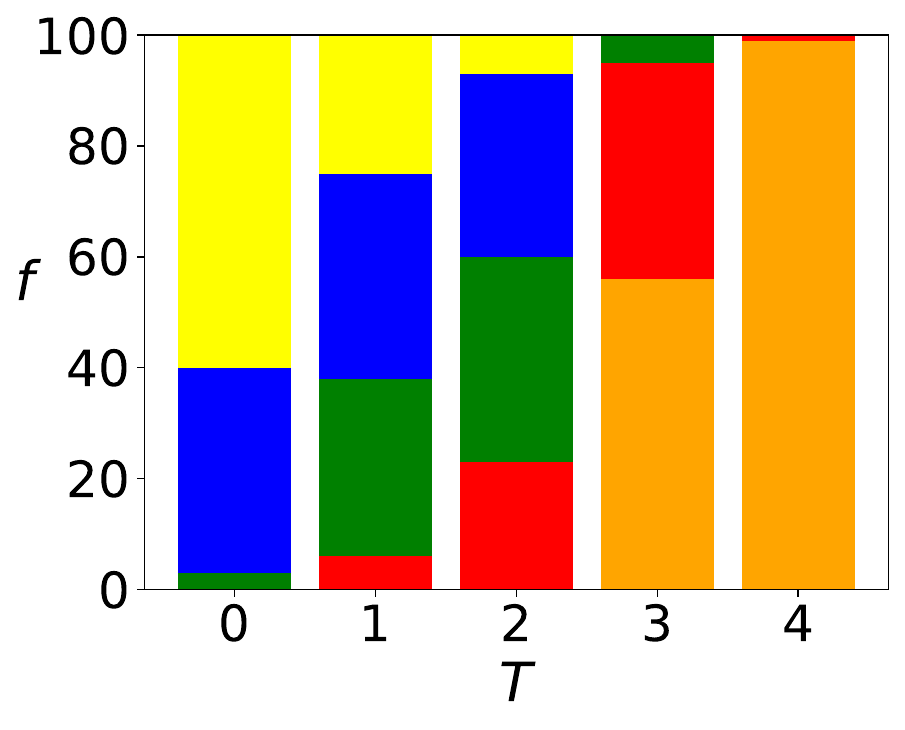}}\hfill
\subcaptionbox{$\alpha = 4$, $K = 5$\label{sfig:th50a4K5}}{\includegraphics[width=.24\textwidth]{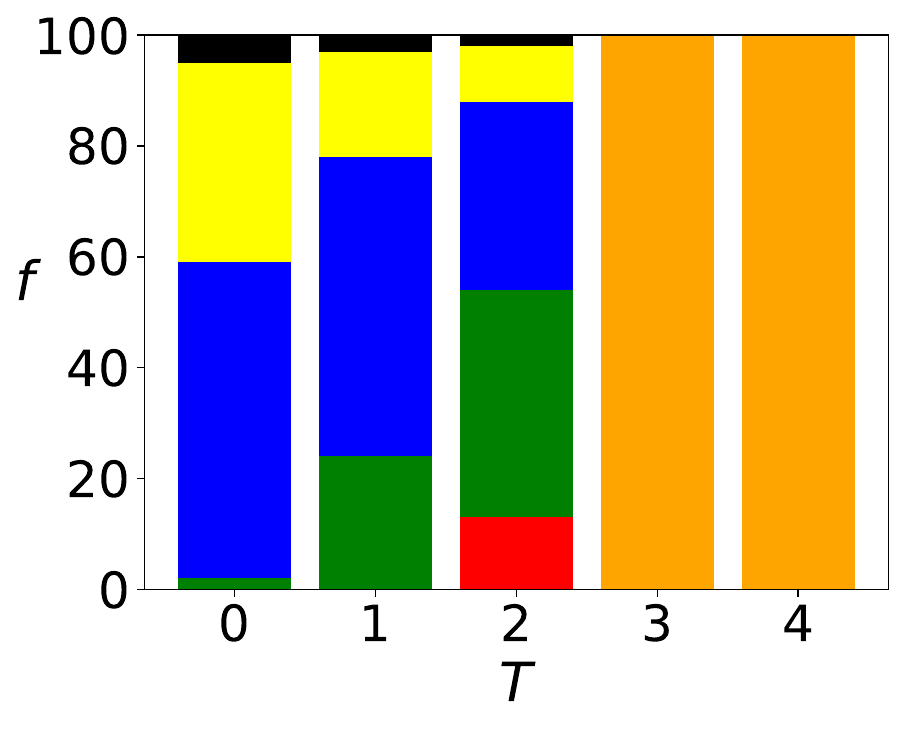}}\\
\subcaptionbox{$\alpha = 6$, $K = 2$\label{sfig:th50a6K2}}{\includegraphics[width=.24\textwidth]{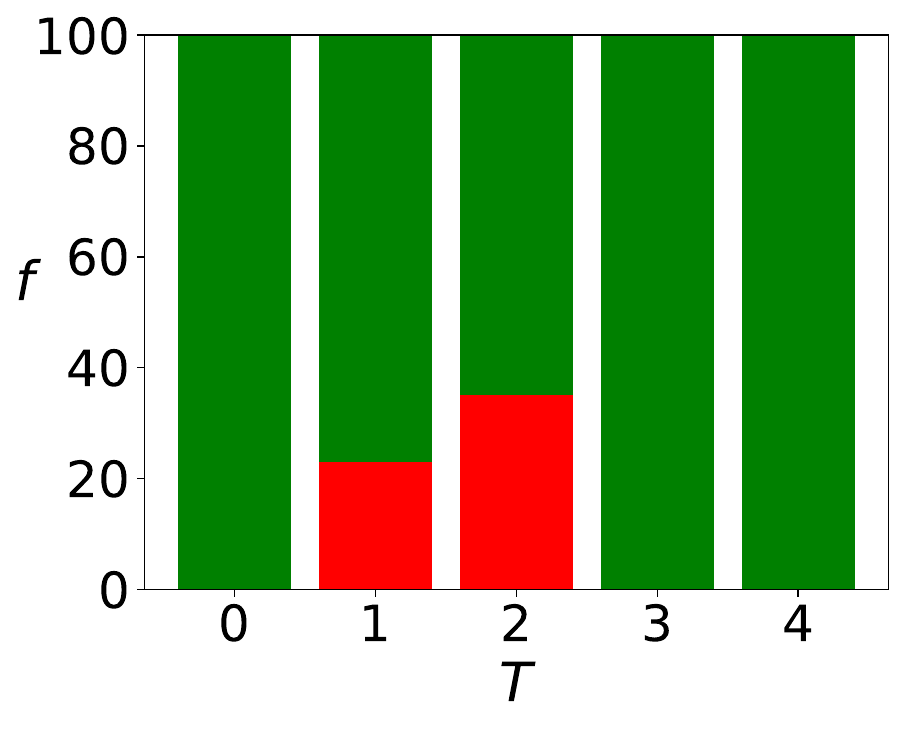}}\hfill
\subcaptionbox{$\alpha = 6$, $K = 3$\label{sfig:th50a6K3}}{\includegraphics[width=.24\textwidth]{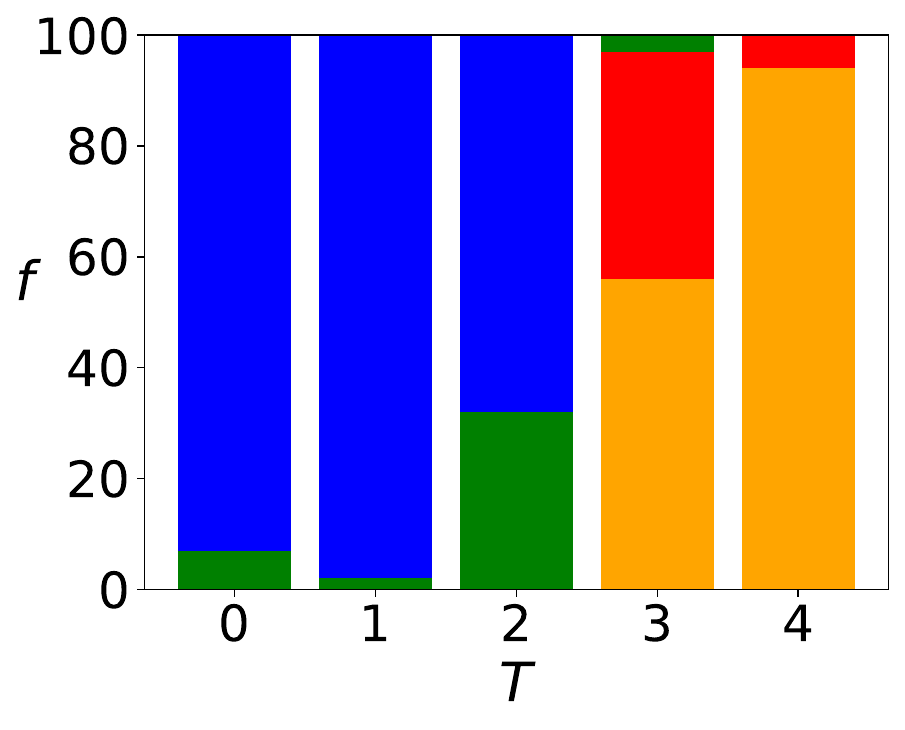}}\hfill
\subcaptionbox{$\alpha = 6$, $K = 4$\label{sfig:th50a6K4}}{\includegraphics[width=.24\textwidth]{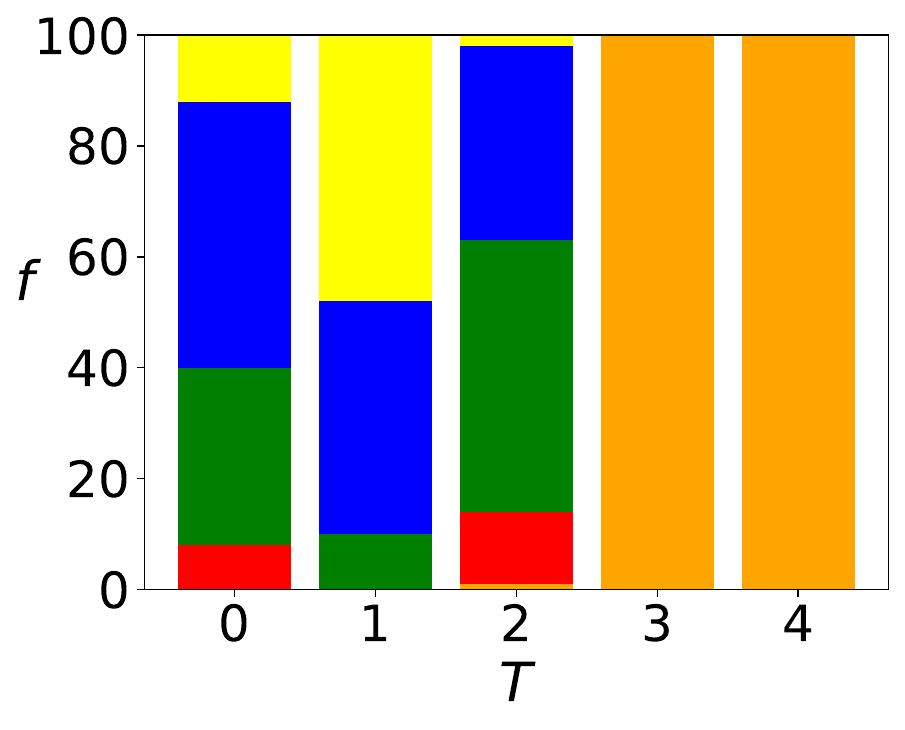}}\hfill
\subcaptionbox{$\alpha = 6$, $K = 5$\label{sfig:th50a6K5}}{\includegraphics[width=.24\textwidth]{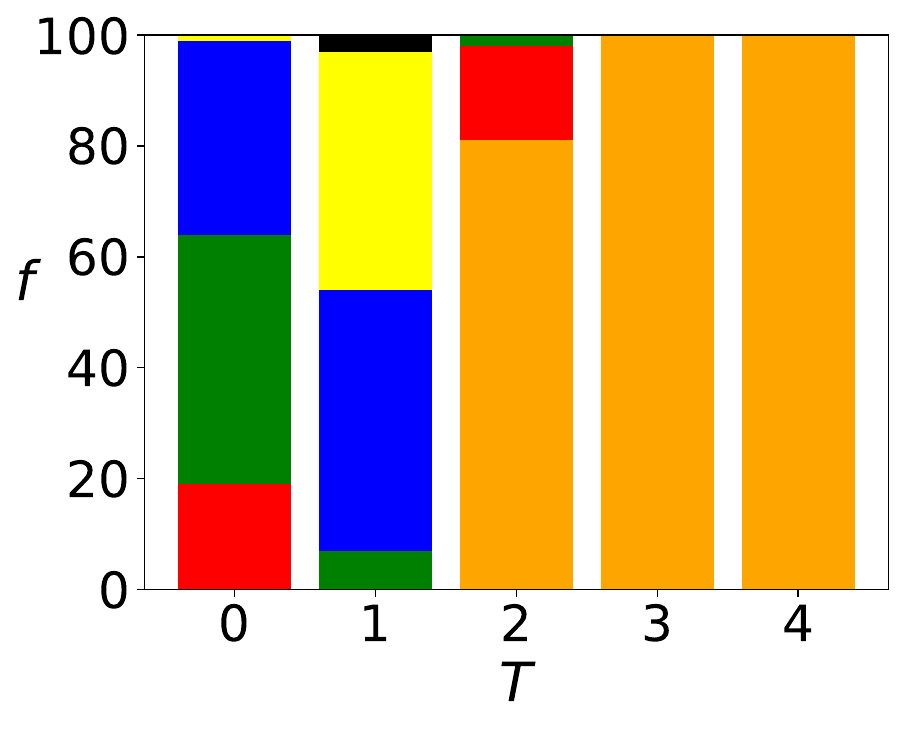}}\\
\caption{\label{fig:th50_sur}The histograms of frequencies $f$ of the number $\Phi$ of surviving opinions after $=10^3$ time steps and for various values of the distance scaling function exponent $\alpha$ and various values of the number of available opinions $K$. The system contains $L^2=41^2$ actors. The noise discrimination level $\theta=50$ and the results are averaged over $R=100$ independent simulations.}
\end{figure*}

\section{\label{apx:num_clu}Average number of clusters}

Average number $\langle\mathcal C\rangle$ of opinion clusters after $t=10^3$ time steps for various exponents of the distance scaling function $\alpha$ and various numbers $K$ of opinions available in the system. Noise discrimination threshold 
$\theta=12$ (\Cref{fig:th12_clust}),
$\theta=25$ (\Cref{fig:th25_clust}),
$\theta=50$ (\Cref{fig:th50_clust}).
The system contains $L^2=41^2$ actors. The results are averaged over $R=100$ independent system realisations.

\section{\label{apx:num_sur}The number of surviving opinions}

Histograms of the frequencies $f$ of the number $\Phi$ of surviving opinions after $=10^3$ time steps and for various values of the distance scaling function exponent $\alpha$ and various values of the number of available opinions $K$. The system contains $L^2=41^2$ actors. 
The noise discrimination level $\theta=12$ (\Cref{fig:th12_sur}), $\theta=25$ (\Cref{fig:th25_sur}), $\theta=50$ (\Cref{fig:th50_sur}) and the results are averaged over $R=100$ independent simulations.

\end{document}